\documentclass[12pt,preprint]{aastex}

\usepackage{epsf}
\usepackage{graphicx}

\def\deg{\ifmmode {^\circ}\else {$^\circ$}\fi}
\def\degree{\ifmmode {^\circ}\else {$^\circ$}\fi}
\def\mum{\ifmmode {\rm \,\mu {\rm m}}\else $\rm \,\mu {\rm m}$\fi}
\def\arcsec{\ifmmode ^{\prime \prime}\else $^{\prime \prime}$\fi}

\def\degpoint{\mbox{$\degree\mskip-7.6mu.\,$}}
\def\inch{\ifmmode ^{\prime \prime}\else $^{\prime \prime}$\fi}
\def\gs{\ifmmode {{\rm g~s^{-1}}}\else ${\rm g~s^{-1}}$\fi}
\def\msunyr{\ifmmode {M_{\odot}~{\rm yr^{-1}}}\else $M_{\odot}~{\rm yr^{-1}}$\fi}
\def\msun{\ifmmode {M_{\odot}}\else $M_{\odot}$\fi}
\def\rsun{\ifmmode {R_{\odot}}\else $R_{\odot}$\fi}
\def\lsun{\ifmmode {L_{\odot}}\else $L_{\odot}$\fi}
\def\mstar{\ifmmode {M_{\star}}\else $M_{\star}$\fi}
\def\rstar{\ifmmode {R_{\star}}\else $R_{\star}$\fi}
\def\tstar{\ifmmode {T_{\star}}\else $T_{\star}$\fi}
\def\lstar{\ifmmode {L_{\star}}\else $L_{\star}$\fi}
\def\mwd{\ifmmode {M_{wd}}\else $M_{wd}$\fi}
\def\rwd{\ifmmode {R_{wd}}\else $R_{wd}$\fi}
\def\twd{\ifmmode {T_{wd}}\else $T_{wd}$\fi}
\def\lwd{\ifmmode {L_{wd}}\else $L_{wd}$\fi}
\def\md{\ifmmode {M_d}\else $M_d$\fi}
\def\ld{\ifmmode {L_d}\else $L_d$\fi}
\def\ad{\ifmmode A_d\else $A_d$\fi}
\def\ldlwd{\ifmmode L_d / L_{wd}\else $L_d / L_{wd}$\fi}
\def\ldlstar{\ifmmode L_d / L_\star\else $L_d / L_{\star}$\fi}
\def\rearth{\ifmmode {\rm R_{\oplus}}\else $\rm R_{\oplus}$\fi}
\def\mearth{\ifmmode {\rm M_{\oplus}}\else $\rm M_{\oplus}$\fi}
\def\qdstar{\ifmmode Q_D^\star\else $Q_D^\star$\fi}
\def\vsqd{\ifmmode v^2 / Q_D^\star\else $v^2 / Q_D^\star$\fi}
\def\kms{\ifmmode {\rm km~s^{-1}}\else $\rm km~s^{-1}$\fi}
\def\ms{\ifmmode {\rm m~s^{-1}}\else $\rm m~s^{-1}$\fi}
\def\vrel{\ifmmode v_{rel}\else $v_{rel}$\fi}
\def\mdot{\ifmmode \dot{M}\else $\dot{M}$\fi}
\def\mdots{\ifmmode \dot{M}_\star\else $\dot{M}_\star$\fi}
\def\mdotv{\ifmmode \dot{M}_v\else $\dot{M}_v$\fi}
\def\mdotz{\ifmmode \dot{M}_0\else $\dot{M}_0$\fi}
\def\mesc{\ifmmode m_{esc}\else $m_{esc}$\fi}
\def\rmin{\ifmmode r_{min}\else $r_{min}$\fi}
\def\rmax{\ifmmode r_{max}\else $r_{max}$\fi}
\def\xmax{\ifmmode x_{max}\else $x_{max}$\fi}
\def\mmin{\ifmmode m_{min}\else $m_{min}$\fi}
\def\mmax{\ifmmode m_{max}\else $m_{max}$\fi}
\def\rmind{\ifmmode r_{min,d}\else $r_{min,d}$\fi}
\def\rmaxd{\ifmmode r_{max,d}\else $r_{max,d}$\fi}
\def\mmaxd{\ifmmode m_{max,d}\else $m_{max,d}$\fi}
\def\vrad{\ifmmode v_{rad}\else $v_{rad}$\fi}
\def\qz{\ifmmode q_{0}\else $q_{0}$\fi}
\def\qi{\ifmmode q_{i}\else $q_{i}$\fi}
\def\ql{\ifmmode q_{l}\else $q_{l}$\fi}
\def\qs{\ifmmode q_{s}\else $q_{s}$\fi}
\def\vhill{\ifmmode v_H\else $r_H$\fi}
\def\rhill{\ifmmode r_H\else $r_H$\fi}
\def\Rhill{\ifmmode R_H\else $R_H$\fi}
\def\rbrk{\ifmmode r_{brk}\else $r_{brk}$\fi}
\def\rdamp{\ifmmode r_{damp}\else $r_{damp}$\fi}
\def\rin{\ifmmode r_{in}\else $r_{in}$\fi}
\def\rout{\ifmmode r_{out}\else $r_{out}$\fi}
\def\tin{\ifmmode t_{in}\else $t_{in}$\fi}
\def\tout{\ifmmode t_{out}\else $t_{out}$\fi}
\def\ain{\ifmmode a_{in}\else $a_{in}$\fi}
\def\aout{\ifmmode a_{out}\else $a_{out}$\fi}
\def\r0{\ifmmode r_{0}\else $r_{0}$\fi}
\def\R0{\ifmmode R_{0}\else $R_{0}$\fi}
\def\m0{\ifmmode m_{0}\else $m_{0}$\fi}
\def\M0{\ifmmode M_{0}\else $M_{0}$\fi}
\def\xm{\ifmmode x_{m}\else $x_{m}$\fi}
\def\sigz{\ifmmode \Sigma_0\else $\Sigma_0$\fi}
\def\ergg{\ifmmode {\rm erg~g^{-1}}\else ${\rm erg~g^{-1}}$\fi}
\def\gyr{\ifmmode {\rm g~yr^{-1}}\else ${\rm g~yr^{-1}}$\fi}
\def\cmss{\ifmmode {\rm cm~s^{-2}}\else ${\rm cm~s^{-2}}$\fi}
\def\cms{\ifmmode {\rm cm~s^{-1}}\else ${\rm cm~s^{-1}}$\fi}
\def\gcms{\ifmmode {\rm g~cm^{-2}}\else $\rm g~cm^{-2}$\fi}
\def\gcmc{\ifmmode {\rm g~cm^{-3}}\else $\rm g~cm^{-3}$\fi}
\def\atil{\ifmmode {\tilde{a}}\else $\tilde{a}$\fi}
\def\ttil{\ifmmode {\tilde{t}}\else $\tilde{t}$\fi}
\def\sqrttt{\ifmmode {\tilde{t}^{1/2}}\else $\tilde{t}^{1/2}$\fi}

\def\rsunv{\ifmmode \vec{r}_\odot\else $\vec{r}_\odot$\fi}
\def\vsunv{\ifmmode \vec{v}_\odot\else $\vec{v}_\odot$\fi}
\def\vsun{\ifmmode v_\odot\else $v_\odot$\fi}
\def\xsun{\ifmmode x_\odot\else $x_\odot$\fi}
\def\vej{\ifmmode v_{ej}\else $v_{ej}$\fi}
\def\vejmin{\ifmmode v_{ej,min}\else $v_{ej,min}$\fi}
\def\veja{\ifmmode v_{ej,H}\else $v_{ej,H}$\fi}
\def\vejb{\ifmmode v_{ej,S}\else $v_{ej,S}$\fi}
\def\vejc{\ifmmode v_{ej,D}\else $v_{ej,D}$\fi}
\def\rv0{\ifmmode \vec{r}_0\else $\vec{r}_0$\fi}
\def\rvf{\ifmmode \vec{r}_f\else $\vec{r}_f$\fi}
\def\rfinal{\ifmmode r_f\else $r_f$\fi}
\def\rstarv{\ifmmode \vec{r}_\star\else $\vec{r}_\star$\fi}
\def\vv0{\ifmmode \vec{v}_0\else $\vec{v}_0$\fi}
\def\v0{\ifmmode v_0\else $v_0$\fi}
\def\b0{\ifmmode b_0\else $b_0$\fi}
\def\thz{\ifmmode \theta_0\else $\theta_0$\fi}
\def\phiz{\ifmmode \phi_0\else $\phi_0$\fi}
\def\phif{\ifmmode \phi_f\else $\phi_f$\fi}
\def\vesc{\ifmmode v_{esc}\else $v_{esc}$\fi}
\def\vvf{\ifmmode \vec{v}_f\else $\vec{v}_f$\fi}
\def\vfinal{\ifmmode v_f\else $v_f$\fi}
\def\bfinal{\ifmmode b_f\else $b_f$\fi}
\def\vstarv{\ifmmode \vec{v}_\star\else $\vec{v}_\star$\fi}
\def\xdot{\ifmmode \dot{x}\else $\dot{x}$\fi}
\def\ydot{\ifmmode \dot{y}\else $\dot{y}$\fi}
\def\zdot{\ifmmode \dot{z}\else $\dot{z}$\fi}
\def\xdots{\ifmmode \dot{x}\else $\dot{x}$\fi}
\def\ydots{\ifmmode \dot{y}\else $\dot{y}$\fi}
\def\zdots{\ifmmode \dot{z}\else $\dot{z}$\fi}
\def\facR{\ifmmode f_{R}\else $f_R$\fi}
\def\abin{\ifmmode a_{bin}\else $a_{bin}$\fi}
\def\amin{\ifmmode a_{min}\else $a_{min}$\fi}
\def\amax{\ifmmode a_{max}\else $a_{max}$\fi}
\def\ma{\ifmmode M_1\else $M_1$\fi}
\def\mb{\ifmmode M_2\else $M_2$\fi}
\def\mbh{\ifmmode M_{bh}\else $M_{bh}$\fi}
\def\rclose{\ifmmode r_{close}\else $r_{close}$\fi}
\def\masyr{\ifmmode {\rm milliarcsec~yr^{-1}}\else milliarcsec~yr$^{-1}$\fi}

\def\vrad{v$_{\rm rad}$}
\def\GC{Galactic Center}

\begin{document}

\title{Impact of the Galactic Disk and Large Magellanic Cloud on the
Trajectories of Hypervelocity Stars Ejected from the Galactic Center}

\author{Scott J. Kenyon}
\affil{Smithsonian Astrophysical Observatory,
 60 Garden St., Cambridge, MA 02138}
\email{skenyon@cfa.harvard.edu}

\author{Benjamin C. Bromley}
\affil{Department of Physics \& Astronomy, University of Utah,
 115 S 1400 E, Rm 201, Salt Lake City, UT 84112}
\email {bromley@physics.utah.edu}

\author{Warren R. Brown}
\affil{Smithsonian Astrophysical Observatory,
 60 Garden St., Cambridge, MA 02138}
\email{wbrown@cfa.harvard.edu}

\author{Margaret J. Geller}
\affil{Smithsonian Astrophysical Observatory,
 60 Garden St., Cambridge, MA 02138}
\email{mgeller@cfa.harvard.edu}

\clearpage

\begin{abstract}

We consider how the gravity of the Galactic disk and the Large Magellanic 
Cloud (LMC) modifies the radial motions of hypervelocity stars (HVSs) 
ejected from the Galactic Center. For typical HVSs ejected towards 
low (high) Galactic latitudes, the disk bends trajectories by up to 
30\deg\ (3\deg\ to 10\deg).  For many lines-of-sight through the Galaxy, 
the LMC produces similar and sometimes larger deflections.  Bound HVSs suffer 
larger deflections than unbound HVSs.  Gravitational focusing by the 
LMC also generates a factor of two overdensity along the line-of-sight 
towards the LMC.  With large enough samples, observations can detect the 
non-radial orbits and the overdensity of HVSs towards the LMC. For any 
Galactic potential model, the Galactic rest-frame tangential velocity 
provides an excellent way to detect unbound and nearly bound HVSs within 
10~kpc of the Sun. Similarly, the rest-frame radial velocity isolates 
unbound HVSs beyond 10--15~kpc from the Sun. Among samples of unbound HVSs, 
measurements of the radial and tangential velocity serve to distinguish
Galactic Center ejections from other types of high velocity stars.

\end{abstract}

\keywords{
        Galaxy: kinematics and dynamics ---
	Galaxy: structure ---
        Galaxy: halo ---
        Galaxy: stellar content ---
        stars: early-type
}

\section{INTRODUCTION}
\label{sec: intro}

Over the past decade, observations have revealed stars with space velocities 
sufficient to escape the Galaxy \citep{brown2005,edel2005,hirsch2005,brown2006a,
brown2006b,brown2007a,koll2007,heber2008,koll2009,till2009,brown2009b,
irrg2010,li2012,brown2012a,pereira2012,brown2014,zheng2014,zhong2014,
hawkins2015,brown2015a,brown2015b,geier2015,li2015,vickers2015,ziegerer2015,
favia2015,zhang2016,ziegerer2017,lennon2017,huang2017,marchetti2018b,
hattori2018b,shen2018,raddi2018,hawkins2018,li2018}.  
Many are apparently normal main sequence stars; some are hot subdwarfs or 
white dwarfs.  In the simplest examples, the radial velocity exceeds the 
local escape velocity. For other stars, a combination of radial velocity 
and proper motion provides evidence for a high total velocity.

Currently popular ejection mechanisms for these stars include
(i) tidal disruption of a binary by the supermassive black hole (SMBH) 
in the Galactic Center or ejection of a single star by a black hole binary 
somewhere in the Galaxy \citep[e.g.,][]{hills1988,yu2003,baumgardt2006,
bromley2006,kenyon2008,perets2009,zhang2010,bromley2012,zhang2013,
dremova2014,kenyon2014,rossi2017,marchetti2018a,wang2018},
(ii) close interactions among massive stars in a dense star cluster
\citep[e.g.,][]{pov1967,leon1991,bromley2009,pflamm2010,perets2012,kenyon2014,
tauris2015,ryu2017}, and
(iii) ejection of the low mass companion in a close binary during the
supernova explosion of a massive primary star \citep[e.g.,][]{blaauw1961,
dedonder1997,port2000,bromley2009,wang2009,pflamm2010,eldridge2011,napi2012,
kenyon2014,geier2015,tauris2015,renzo2018}.  
Disrupted dwarf galaxies \citep{abadi2009}, interacting galaxies \citep{piffl2014}, 
and star clusters disrupted by an SMBH \citep{capuzzo2015,fragione2016} may also 
contribute to the population of high velocity stars.

Aside from serving as possible probes of the mass and 3D gravitational 
potential of the Milky Way \citep[MW; e.g.,][]{gnedin2005,bromley2006,
yu2007,kenyon2008,brown2010a,gnedin2010,rossi2017,fragione2017b}, high 
velocity stars in the Galactic halo can also constrain the physical 
properties of the LMC and other local group galaxies 
\citep[e.g.,][]{laporte2017,laporte2018}. As one example, \citet{boubert2016} 
and \citet{boubert2017} demonstrate that high velocity stars ejected from the 
LMC might contribute to the population of high velocity stars observed in the 
thick disk and the halo of the Milky Way.

Here, we consider the impact of the gravitational potentials of the Galactic disk 
and the LMC on the space distribution of 3~\msun\ hypervelocity stars (HVSs) 
ejected from the \GC\ (GC). For HVSs with small ejection angles relative to the 
Galactic midplane ($\lesssim$ 30\deg), the gravity of the disk bends trajectories 
by up to 30\deg\ with respect to a purely radial trajectory.  Bound HVSs suffer 
larger deflections than unbound HVSs. When HVSs are ejected towards the Galactic 
pole, they maintain somewhat more radial trajectories.  Aside from generating 
similar deflections, the gravity of the LMC produces a factor of two overdensity 
of ejected stars along the line-of-sight towards the LMC.

With large enough samples, the non-radial orbits and the overdensity of ejected 
stars along specific lines-of-sight through the Galaxy are observable. In 
particular, the Galactic rest-frame tangential velocity provides an excellent way
to detect unbound and nearly bound stars within 10~kpc of the Sun. Similarly, the
rest-frame radial velocity isolates unbound stars at larger distances.

We begin with a discussion of the theoretical background (\S\ref{sec: back})
and numerical procedures (\S\ref{sec: sims}). After describing results 
quantifying the changing trajectories of HVSs (\S\ref{sec: res}), we develop 
several observational diagnostics for the shape of the potential (\S\ref{sec: obs}),
identify robust tools to isolate unbound HVSs from other high velocity stars
(\S\ref{sec: id}), and discuss the implications of the analysis for future Galactic 
surveys (\S\ref{sec: disc}).  We conclude with a brief summary (\ref{sec: summ}).



\section{BACKGROUND}
\label{sec: back}

When a binary system crosses the tidal radius of an SMBH,
it becomes unbound \citep{hills1988}. One component takes up an eccentric orbit
around the SMBH; to conserve energy, the other is ejected at high velocity
\citep[see also][]{gould2003,gual2005,gins2006,sari2010}. Hills suggested calling the 
ejecta HVSs. For HVSs that escape the SMBH, the ejection velocity depends on the 
physical properties of the binary and the SMBH and the distance of closest approach 
\citep[e.g.,][]{hills1988,gould2003,gual2005,bromley2006,gins2006,sari2010}. If 
the source of binaries is isotropic, the outward flow of HVSs is also isotropic
\citep{bromley2006,kenyon2008,kenyon2014,rossi2014,rossi2017}. An anisotropic
source of binaries or a BH binary companion to the SMBH generate anisotropies in 
the outflow \citep{yu2003,levin2006,sesana2006,lu2007,sesana2007,oleary2008,perets2009,
sesana2009,lu2010,zubovas2013,subr2016,coughlin2018}.

After HVSs travel a distance $r \approx$ 10--20~pc, the potential of the Galactic 
bulge acts as a high pass filter which prevents lower velocity stars from reaching the 
Galactic halo at distances $r \gtrsim$ 20~kpc from the GC \citep{kenyon2008,kenyon2014}. 
For a standard Galactic potential model (see below), HVSs that reach $r \approx$ 10--100~kpc 
require minimum ejection velocities $\v0\ \approx$ 800--925~\kms.  Many of these ejected 
stars remain bound to the Galaxy \citep{bromley2006,kenyon2008}.  Unbound HVSs require 
$\v0\ \gtrsim$ 925~\kms\ \citep{kenyon2008}.

When the Galactic potential is limited to a spherical bulge and a spherical halo, the 
radial distribution of HVSs about the GC is spherically symmetric \citep{bromley2006,
yu2007,kenyon2008,kenyon2014,rossi2014,rossi2017}. The purely radial trajectories of 
HVSs then provide a unique way to distinguish them from bulge, disk, and halo stars 
on more circular orbits around the GC \citep[see also][]{hattori2018a}.  Introducing a 
plausible amount of structure in the potential (e.g., a binary SMBH, mis-aligned 
circumnuclear disks in the GC, the Galactic bar, the Galactic disk, or a triaxial bulge 
or halo) eliminates spherical symmetry and may create observable asymmetries in the 3D 
distribution of HVSs \citep{gnedin2005,sesana2006,yu2007,sesana2009,subr2016,fragione2017a,
hamers2017}.

With a mass of roughly 10\% of the mass of the Galaxy, the LMC changes the trajectories 
of HVSs. Consider a single HVS ejected from the GC toward a fixed LMC located at $r$ = 
50~kpc.  When the HVS has $r \approx$ 35~kpc, the gravitational acceleration on the star 
from the LMC is roughly half the acceleration due to the Galaxy. Compared to a system 
with no LMC, this HVS decelerates more slowly relative to the GC, maintains a higher 
radial velocity, and travels farther out into the Galaxy. Compared to a Galaxy with 
no LMC, a Galaxy with the LMC then has fewer HVSs at 30--50~kpc.  If a star passes 
through the LMC and is at $r \approx$ 65~kpc, the radial acceleration from the LMC is 
comparable to the radial acceleration from the Galaxy. This extra deceleration causes an 
enhancement in the population of HVSs at 60--100~kpc relative to a Galaxy with no LMC.

HVSs ejected away from the LMC feel the extra acceleration from the LMC throughout 
their journey through the Galaxy. Per unit time, these stars must then reach smaller 
distances from the GC than their counterparts ejected towards the LMC.  The overall 
population of HVSs then has a larger space density away from the LMC than towards 
the LMC. Our goal is to learn whether the variation in HVS space density throughout 
the Galaxy is detectable with current observational tools.

Despite its somewhat lower mass, the Galactic disk can also bend the trajectories of 
stars ejected from the GC \citep{gnedin2005,yu2007}. Stars flowing radially outward 
at low Galactic latitude feel a larger acceleration from the disk than those at higher 
latitudes. Thus, HVSs at lower latitudes have a larger non-radial component of their 
motion than HVSs at higher latitudes. Numerical simulations of the space motions of 
HVSs will allow us to predict the non-radial motions of HVSs as a function of initial 
ejection velocity and Galactic latitude.

\section{NUMERICAL CALCULATIONS}
\label{sec: sims}

To explore the space motions of HVSs in a combined MW+LMC potential, we consider a 
set of numerical calculations.  As in previous papers \citep{bromley2006,kenyon2008,
bromley2009,kenyon2014}, we follow the dynamical evolution of an ensemble of HVSs 
throughout their main sequence lifetimes.  Snapshots yield predictions for the 3D 
distributions of space density, proper motion, and radial velocity.  With typical 
100--500~Myr travel times through the Galaxy, finite stellar lifetimes produce 
measurable differences in these observables for stars with a range of masses.  

For stars with main sequence lifetime $t_{ms}$, we generate initial position \rv0\ and 
velocity \vv0\ vectors, an ejection time $t_{ej}$, and an observation time $t_{obs}$, 
with $t_{ej} \le t_{obs} \le t_{ms}$.  For a flight time $t_f = t_{obs} - t_{ej}$, we 
integrate the orbit of each star in the MW+LMC potential and record the final position 
\rvf\ and velocity \vvf\ vectors at $t_{obs}$.  For an adopted position and velocity for 
the Sun, we derive a catalog of predicted observables $d$ (distance), $v_r$ (radial
velocity), $v_t$ (tangential velocity), $\mu_l$ (proper motion in Galactic longitude), 
and $\mu_b$ (proper motion in Galactic latitude). 

\subsection{Gravitational Potential of the Milky Way}
\label{sec: sims-pot}

As in \citet{kenyon2014}, we work in coordinate systems with an origin at the \GC\ (see 
Table~\ref{tab: vars}). Stars have cartesian positions $(x, y, z)$ and velocities 
$(v_x, v_y, v_z)$.  The distance from the GC to the star is $r$; the space velocity of 
the star relative to the GC is $v$.  The angle of the position vector of the star 
relative to the $x$ axis is $\theta$ (the GC longitude); 
the angle relative to the $x$--$y$ plane is $\phi$ (the GC latitude).  With 
$\varrho^2 = x^2 + y^2$, we also specify stellar positions and velocities in spherical 
$(r, \theta, \phi)$ or cylindrical $(\varrho, \theta, z)$ systems.

To measure dynamical properties in heliocentric coordinates, we adopt a cartesian
position $(-\rsun, 0, 0)$ and velocity $(0, \vsun, 0)$ for the Sun, 
where \rsun\ = 8~kpc is the distance of the Sun from the GC \citep[e.g.,][]{bovy2012} 
and \vsun\ = 235~\kms\ is the space velocity of the Sun relative to the GC 
\citep[e.g.,][]{hogg2005,bovy2012,reid2014,reid2016,russeil2017}.
Stars have distances $d = ( (x+\rsun)^2 + y^2 + z^2)^{1/2}$ and relative velocities 
$v_{rel} = (v_x^2 + (v_y - \vsun)^2 + v_z)^{1/2}$.  The galactic longitude $l$ of the star is 
the angle -- measured counter-clockwise in the $x-y$ plane -- from a line connecting the Sun 
to the GC, $l = {\rm tan^{-1}} (x~{\rm tan}~\theta / (x + \rsun))$.  The galactic latitude 
measures the height of the star above the galactic plane, 
$b = {\rm sin^{-1}}(z/d) = {\rm sin^{-1}} (r~{\rm sin}~\phi / d)$.  For $r \gg \rsun$, 
$\theta \approx l$ and $\phi \approx b$.

In this heliocentric system, the radial velocity of an HVS is:
\begin{equation}
v_{r, \odot} = v_x~{\rm cos}~l ~ {\rm cos}~b + (v_y - \vsun)~{\rm sin}~l~{\rm cos}~b + v_z~{\rm sin}~b ~ .
\label{eq: vrad}
\end{equation}
The tangential velocity follows from the relative velocity, 
$v_{t, \odot}^2 = v_{rel}^2 - v_{r, \odot}^2$. 
In the GC frame, the radial velocity is
\begin{equation}
v_r = v_{r, \odot} + \vsun~{\rm sin}~l~{\rm cos}~b ~ .
\label{eq: vrad-gc}
\end{equation}
The tangential velocity is then $v_t^2 = v^2 - v_r^2$. For our discussion, 
we consider velocities in the GC frame. In an observational program, $v_t$
requires accurate measurements of both the proper motion and distance.

We adopt a three component model for the Galactic potential $\Phi_G$ 
\citep{kenyon2008,kenyon2014} with parameters listed in Table~\ref{tab: pars}:
\begin{equation}
\label{eq: phi}
\Phi_G = \Phi_b + \Phi_d + \Phi_h ~ ,
\end{equation}
where
\begin{equation}
\label{eq: phib}
\Phi_b(r) = - G M_b / (r + r_b) ~ 
\end{equation}
is the potential of the bulge,
\begin{equation}
\label{eq: phid}
\Phi_d(\varrho,z) = - G M_d /\sqrt{\varrho^2 + [a_d + (z^2+b_d^2)^{1/2}]^2}
\end{equation}
is the potential of the disk, and
\begin{equation}
\label{eq: phih}
\Phi_h(r) = - G M_h \ln(1+r/r_h) / r
\end{equation}
is the potential of the halo \citep[e.g.,][]{hern1990,miya1975,nav1997}. 

For the bulge and halo, we set $M_b = 3.75 \times 10^9 \msun$,
$M_h = 10^{12} \msun$, $r_b$ = 105~pc, and $r_h$ = 20~kpc (Table~\ref{tab: pars}).
These parameters match measurements of the mass and velocity dispersion 
inside 1~kpc and outside 50~kpc \citep[see \S2.2 of][]{kenyon2008,kenyon2014}
and are consistent with various independent measures of the mass of the Galaxy
\citep[e.g.,][]{watkins2010,gnedin2010,boylan2013,piffl2014,penarrubia2016,
mcmillan2017,patel2017a,patel2018,gaia2018b,watkins2018,posti2018,monari2018}.

In some applications, the potential of the halo is expressed in terms of the 
virial mass $M_{vir}$ and the concentration parameter $c$ \citep[e.g.,][and 
references therein]{nav1997,zentner2003,gomez2015,bullock2017}. For a virial 
radius $r_{vir} = c r_h$, $M_{vir} = M_h [ {\rm ln}~(1+c) - c/((1+c) ] $.
Specifying $c$ and $M_{vir}$ is then equivalent to setting $r_h$ and $M_h$. 
Mass models for the Milky Way typically have $c \approx$ 10--15
\citep[e.g.,][]{dehnen2006,boylan2013,patel2017b,monari2018}, yielding 
$M_{vir} \approx 1.5 - 1.8 M_h$. 

To match the adopted 235~\kms\ circular velocity of the Sun, we adopt parameters 
for the disk potential: $M_d = 6 \times 10^{10} \msun$, $a_d$ = 2750~pc, and 
$b_d$ = 300~pc.  The complete set of parameters for the bulge, disk, and halo 
yields a flat rotation curve from 3--50~kpc.

Although the formal escape velocity for the halo is unbounded, we adopt a 
convenient definition based on the outward velocity required for a star 
to reach $r$ = 250~kpc with zero velocity. To place this reference point 
in context, a halo potential with a concentration parameter $c$ = 12.5 has a 
virial radius $r_{vir}$ = 250~kpc for the adopted $r_h$ = 20~kpc. We derive 
$v_{esc}(r)$ numerically by tracking the position and speed of particles 
dropped into the MW from rest at $r$ = 250~kpc. For the adopted MW potential,
the escape velocity in the $x-y$ plane is roughly 1~\kms\ larger than the 
escape velocity along the $z$-axis. We ignore this difference. 

Following \citet{gomez2015}, we assume a spherical potential for the LMC:
\begin{equation}
\label{eq: phil}
\Phi_L = -G M_L / \sqrt{r_s^2 + r_L^2} ~ ,
\end{equation}
where $M_L = 10^{11} \msun$, $r_L$ = 15~kpc, and $r_s$ is the distance from
the center of the LMC. The adopted mass and scale length are roughly in the 
middle of the range measured/proposed in the literature 
\citep[e.g.,][]{vandermarel2002,gomez2015,penarrubia2016,patel2017a}.  
Viewed from the \GC, the LMC scale length subtends an angle 
$\theta_L \approx$ 16\degpoint7 at a distance $d_L$ = 50~kpc.

Adding in the central SMBH, the total potential is
\begin{equation}
\Phi = \Phi_G + \Phi_L - G \mbh\ / r ~ ,
\end{equation}
where $ \mbh\ = 3.6 \times 10^6$ \msun\ is the mass of the central black hole.
Although 10\% to 20\% lower than current best values
\citep[e.g.,][]{boehle2016,gillenssen2017,eckart2017}, this value maintains
consistency with previous studies \citep{bromley2006,kenyon2008,kenyon2014}.
Adopting a larger value has little impact on the results.

With our adopted $(M_h, r_h)$ and $(M_L, r_L)$, the total acceleration from 
the Galaxy always dominates the acceleration from the LMC. Thus, there is no 
equivalent to a `Hill sphere,' a volume where the gravity of the LMC overcomes 
the gravity of the MW\footnote{Equivalently, a particle with velocity $v$ = 0
placed at $r_s$ = 0 falls into the GC after 9.8~Gyr.  Increasing $M_L$ or 
reducing $r_L$ allows the LMC to have a region where its gravity dominates.}.  
For stars ejected at an angle $\theta_0$ relative to the line-of-centers, 
however, the LMC produces an acceleration tangential to the velocity vector. 
When $d_L$ = 50~kpc, 0\deg\ $< \theta_0 \lesssim$ 25\deg, and $r \approx$ 
35--65~kpc, the LMC acceleration is a significant fraction of the deceleration 
from the MW. In this regime, trajectories are gravitationally focused and bend 
around the LMC.  In \S\ref{sec: res-toy}, we quantify this gravitational focusing.

When stars pass `through' the LMC, we ignore the possibility that ejected stars 
collide with gas or stars within the LMC.  The trajectories of HVSs follow paths 
dictated solely by the potential of the MW and the potential of the LMC.

\subsection{Initial Conditions}
\label{sec: sims-init}

To select \rv0\ and \vv0 for HVSs, we rely on published calculations. In our
approach, a single SMBH at the \GC\ disrupts close binary systems with semimajor 
axes \abin\ between \amin\ and \amax\ \citep{hills1988,kenyon2008,sari2010,rossi2014}.  
Our choice of the minimum semimajor axis \amin\ minimizes the probability of a 
collision between the two binary components during the encounter with the black hole 
\citep{gins2007,kenyon2008}.  Setting the maximum semimajor axis $\amax \approx$ 4~AU 
limits the number of low velocity ejections which cannot travel more than 10--100~pc 
from the \GC.  
To select stars capable of reaching $\gtrsim$ 10~kpc from the GC, we also set a minimum 
ejection velocity \vejmin\ = 750~\kms\ \citep{kenyon2008}.  Choosing smaller \amax\ and 
larger \vejmin\ precludes moderate velocity ejections that barely reach the Galactic 
halo and remain bound to the Galaxy.

These choices for \amax\ and \vejmin\ are consistent with expectations for ensembles of 
close binaries within 1--2~pc of the SMBH. For the selection procedure outlined below, 
results with \vejmin\ = 750~\kms\ and a somewhat smaller \amax\ between 0.6~AU and 4~AU 
are fairly similar to those with \vejmin\ = 750~\kms\ and \amax\ = 4~AU. In the dense
stellar system within a few pc of the GC \citep[e.g.,][]{tremaine2002,genzel2003}, 
binaries with equal mass components and \amax\ = 0.6~AU (4~AU) evaporate in roughly 2~Gyr 
(250~Myr) \citep{perets2009}.  Defining $\zeta = G m / 2 a \sigma^2$ where $m$ is the mass
of the binary and $\sigma$ is the stellar velocity dispersion, \citet{fragione2018} divide 
binaries into `soft' ($\zeta \ll 1$) and `hard' ($\zeta \gg 1$).  For the measured 
$\sigma \approx$ 60~\kms\ at a distance of 1--2~pc from the GC \citep{tremaine2002},
$\zeta \approx 1.1 ~ (m / 6~\msun) ~ ({\rm 1~AU} / a)$. As implied by the calculations 
of \citet{perets2009}, the adopted upper limit on \amax\ coupled with the lower limit on 
\vejmin\ is consistent with binaries that are hard enough to survive for up to 1~Gyr in 
the vicinity of the GC.

Numerical simulations of binary encounters with a single black hole demonstrate
that the probability of an ejection velocity \vej\ is a gaussian,
\begin{equation}
\label{eq: pvej}
p_H(\vej) \propto e^{(-(\vej - \veja)^2 / \sigma_v^2)}~ ,
\end{equation}
where the average ejection velocity is 
\begin{equation}\label{eq:vej} 
\veja = 1760 
\left(\frac{\abin}{\rm 0.1\ AU}\right)^{-1/2} 
       \left(\frac{\ma + \mb}{2~\msun}\right)^{1/3}
\left(\frac{\mbh}{3.5 \times 10^6~\msun}\right)^{1/6}
       \facR \ \   {\rm km~s^{-1}} ~ , ~
\end{equation}
and $\sigma_v \approx$ 0.2 \veja\ \citep{bromley2006}. Here
\ma\ (\mb) is the mass of the primary (secondary) star and \mbh\ is the mass 
of the central black hole.  The normalization factor $\facR$ depends on 
$r_{close}$, the distance of closest approach to the black hole:
\begin{eqnarray}\label{eq:facR}
\nonumber
    \facR & = & 0.774+(0.0204+(-6.23\times 10^{-4}
           +(7.62\times 10^{-6}+ \\
& &
(-4.24\times 10^{-8}
        +8.62\times 10^{-11}D)D)D)D)D,
\end{eqnarray}
where
\begin{equation}\label{eq:D}
D = D_0
\left(\frac{\rclose}{\abin}\right) ~ 
\end{equation}
and
\begin{equation} 
\label{eq: D0}
D_0 = \left[\frac{2 \mbh}{10^6 (\ma + \mb)}\right]^{-1/3}.
\end{equation}
This factor also sets the probability for an ejection, $P_{ej}$:
\begin{equation}
\label{eq:PE}
P_{ej} \approx 1 - D/175
\end{equation}
for $0 \le D \le 175$. For $D > 175$, $\rclose \gg \abin$; the binary
does not get close enough to the black hole for an ejection and
$P_{ej} \equiv 0$. 

To establish initial conditions, we select each HVS from a random 
distribution of \abin, \rclose, and \vej.  The binaries have semimajor 
axes uniformly distributed in log \abin\ \citep[e.g.,][]{abt1983,duq1991,
hea1998,ducati2011,dosSantos2017}. 
For binaries with $a$ = \amax, the maximum distance of closest approach is 
$r_{close,max} = 175 ~ \amax\ / D_0$. We adopt a minimum distance of closest 
approach $r_{close,min}$ = 1~AU.  Within this range, the probability of any 
\rclose\ grows linearly with $r$. Choosing two random deviates thus yields 
\abin\ and \rclose; \veja, $D$, and $P_{ej}$ follow from 
eqs.~(\ref{eq:vej}--\ref{eq:PE}).  Selecting a third random deviate 
from a gaussian distribution yields the ejection velocity. Two additional 
random deviates drawn from a uniform distribution spanning the main sequence
lifetime of the star fix $t_{ej}$ and $t_{obs}$.  To see whether this 
combination of parameters results in an ejection, we select a sixth random 
deviate, $P$. When $P_{ej} \ge P$, $\vej \ge v_{ej,min}$, and $t_{ej} < t_{obs}$, 
the star is ejected from the GC. Otherwise, we select new random numbers. 
We place each ejected star at a random location on a sphere with a radius of 
1.4~pc centered on the \GC\ and assign velocity components appropriate for a 
radial trajectory from the \GC.  These stars have initial Galactic longitude 
$l_0$, Galactic latitude $b_0$, GC longitude $\thz$, and GC latitude $\phiz$.  

\subsection{Numerical Technique}
\label{sec: sims-tech}

To integrate the motion of each ejected star through the MW+LMC potential,
we use an adaptive fourth-order integrator with Richardson extrapolation
\citep[e.g.,][]{press1992,bk2006,bromley2009}. Starting from an initial 
position \rv0\ with velocity \vv0, the code integrates the full three-dimensional
orbit through the Galaxy, allowing us to track position and velocity as a 
function of time. We integrate the orbit for a time $t_f = t_{obs} - t_{ej}$,
smaller than the main sequence lifetime of the ejected star.  This procedure 
allows us to integrate millions of orbits fairly rapidly. Several tests 
demonstrate that our approach yields typical errors of 0.01\% in position 
and velocity after 1--10~Gyr of evolution.  At $t_f$, stars have positions 
$(x_f, y_f, z_f)$, equivalently $(r_f, \theta_f, \phi_f)$ or 
$(\varrho_f, \theta_f, z_f)$, and velocities $(v_{xf}, v_{yf}, v_{zf})$.

To enable comparisons with other studies, we quote results for several simple 
calculations (Table~\ref{tab: tests}). In these tests, massless particles 
released at rest fall toward the GC from several locations along the $x$-axis 
(HVx models) or the $z$-axis (HVz models). The Table lists the time $t(r)$ to
reach several distances $r$ between the starting point $r_0$ and the GC.  
Time scales to fall in along the $x$-axis are 0.5--1.5 yr shorter than those
along the $z$-axis. Velocities at $r$ = 1~pc and $r$ = 100~kpc are independent
of the initial position. At 8~kpc, the velocity of particles falling through 
the disk is somewhat larger than $v(r)$ for infall perpendicular to the disk.
At the $\pm$1~\kms\ level, velocities from the numerical calculations agree 
with the analytic result $v = \sqrt{2 (\Phi_{G,0} - \Phi_G)}$, where $\Phi_{G,0}$ 
is the gravitational potential at $r$ = $r_0$ and $\Phi_G$ is the potential 
at $r$.  Calculations with these starting velocities at $z_0 \approx$ 1~pc (or 
$x_0 \approx$ 1~pc) achieve the appropriate maximum distance from the GC on 
the listed infall time scales.

\section{RESULTS: HVS TRAJECTORIES}
\label{sec: res}

\subsection{The Galactic Disk}
\label{sec: res-disk}

To illustrate the deflection of HVS trajectories by the Galactic disk, we consider 
stars with \v0\ = 900~\kms\ traveling in the $x-z$ plane for $t_f$ = 1~Gyr 
\citep[for other examples, see][]{gnedin2005,yu2007}.  When the initial angle 
relative to the $x-y$ plane is $\phi_0$ = 1\deg, stars reach a maximum height 
above the $x$-axis of $z \approx$ 60~pc at $x \approx$ 8400~pc ($t \approx$ 13~Myr). 
Acceleration from the disk then pulls the star through the midplane ($z$ = 0) 
at $t \approx$ 52--53~Myr when $x \approx$ 27--28~kpc.  Although $v_z$ is then 
only $-$2~\kms, the disk potential is too weak to pull the star back towards 
the disk \citep[see Fig. 1 of][]{kenyon2008}.  The star continues to move 
farther below the disk midplane, reaching $z \approx -750$~pc ($\phi \approx$ 
0\degpoint25) when $x \approx$ 175~kpc and $t \approx$ 1~Gyr.

\begin{figure}
\includegraphics[width=6.5in]{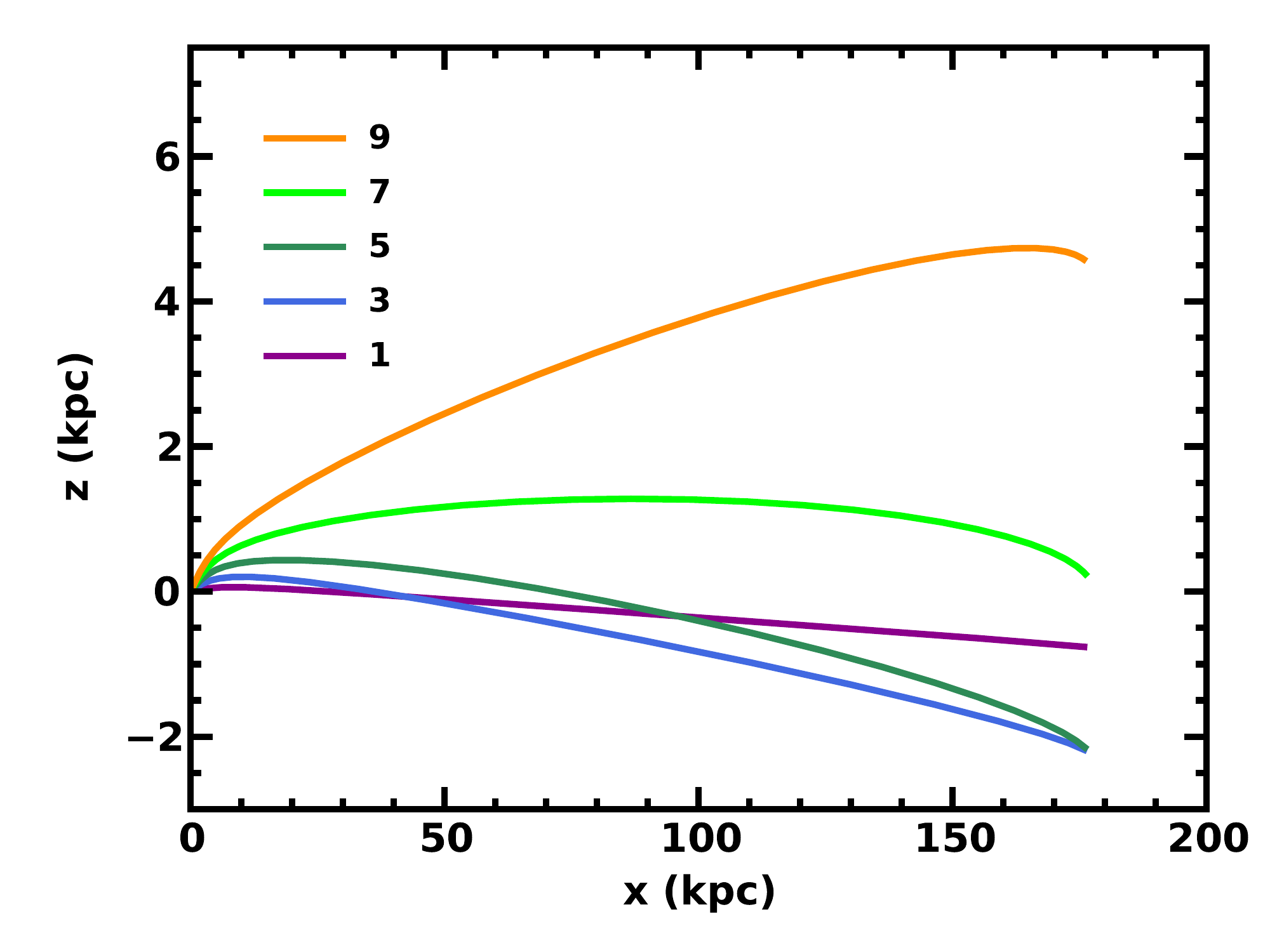}
\vskip 2ex
\caption{\label{fig: disk1}
Trajectories in the $x-z$ plane for HVSs ejected with \v0\ = 900~\kms\ at various 
angles \phiz\ (in deg, as indicated in the legend) relative to the Galactic plane.  
After travel times of 1~Gyr, stars with $\phiz\ \approx$ 1\deg--6\deg\ (purple,
blue, and dark green curves) lie significantly below the plane. Others ejected at 
somewhat larger \phiz\ (light green and orange curves) turn around and head back 
to the plane after 500~Myr to 1~Gyr.
}
\end{figure}

Stars ejected at somewhat larger angles end up farther below the disk midplane
after 1~Gyr (Fig.~\ref{fig: disk1}, blue and green curves).  When $\phi_0 \approx$ 
3\deg--5\deg, stars feel a larger gravitational force from the disk.  Despite 
their larger initial $v_z$, stars with larger $\phi_0$ decelerate more rapidly and 
have $v_z \approx -4$~\kms\ as they pass through the midplane. After 1~Gyr, these
stars almost reach the halo, $z = -2200$ pc ($\phi \approx$ 0\degpoint65). 

As the initial angle of ejection $\phi_0$ grows, it is harder and harder for the disk
to pull the star across the disk midplane (Fig.~\ref{fig: disk1}, lime and orange 
curves).  In these examples, the initial $v_z$ is too large for the disk gravity to 
overcome completely. Although the stars reach a peak $z$ distance and then begin to 
fall back toward the disk, they remain above the midplane at $t$ = 1~Gyr. 

Among all HVSs, larger ejection velocities lead to smaller deflections
(Fig.~\ref{fig: disk2}).  When \v0\ = 900~\kms\ and $t_f$ = 600~Myr (purple curve),
the difference between the initial and final values for the GC latitude, 
$\delta \phi = | \phi_0 - \phi_f|$, grows from zero at $\phi_0$ = 0\deg\ to nearly 
8\deg\ at $\phi_0 \approx$ 15\deg. Although the deflection then decreases at larger 
$\phi_0$, it is still significant -- roughly 1\deg\ -- for ejections towards the 
Galactic pole ($\phi_0$ $\approx$ 89\deg). For stars with larger \v0, the maximum 
deflection decreases to roughly 4\deg\ for \v0\ = 1050~\kms. It is 2\degpoint5 for 
\v0\ = 1200~\kms\ and 1\degpoint9 for \v0\ = 1350 ~\kms. Despite the variation in the 
maximum deflection with initial velocity \v0, the form of the $\delta \phi$--$\phi_0$ 
relation is independent of \v0, with the peak deflection always at $\phi_0 \approx$ 15\deg.

The magnitude of the deflection is sensitive to the disk parameters, $M_d$, $a_d$, 
and $b_d$. For modest changes in $M_d$, the maximum $\delta \phi$ scales approximately 
with disk mass. Larger (smaller) disk masses result in larger (smaller) deflections, 
with no shift in the $\phiz$ for maximum deflection.  The two scale factors -- $a_d$ 
and $b_d$ -- control the amplitude and shape of the $\delta \phi$--$\phi_0$ curve.  
Smaller (larger) $a_d$ and $b_d$ enable larger (smaller) $\delta \phi$. Larger $a_d$ 
($b_d$) shifts the maximum $\delta \phi$ to smaller (larger) $\phi_0$. 

For fixed $M_d$, it is easier to generate larger deflections with modest changes 
in $a_d$ and $b_d$ than it is to produce smaller deflections. With 
$M_d = 6 \times 10^{10}$~\msun, for example, setting $a_d$ = 2250~pc (3250~pc)
results in a maximum deflection of 10\deg\ (7\deg) instead of the 
$\delta \phi \approx$ 8\deg\ for our nominal $a_d$ = 2750~pc. Similarly, adopting
$b_d$ = 450~pc (150~pc) yields a maximum $\delta \phi \approx$ 7\deg\ (10\deg) 
instead of $\delta \phi \approx$ 8\deg\ for $b_d$ = 300~pc. 

Systematic deflection of HVS trajectories by the disk has a clear observational
consequence.  Without deflection, the fraction of all HVSs detected in a survey 
is proportional to the sky coverage; e.g., surveying 50\% of the sky should yield 
50\% of all HVSs. Because HVSs with \phiz\ somewhat larger than 30\deg\ end up 
with \phiz\ somewhat smaller than 30\deg, deflection reduces the ability of halo 
surveys to recover HVSs.  This reduction depends on the initial ejection velocity.  
In these examples, the fraction of stars with $\phi_f \le$~30\deg\ ranges from 
54\% for \v0\ = 1200~\kms\ to 57\% for \v0\ = 1050~\kms\ to 60\% for \v0\ = 900~\kms.  
The impact for unbound HVSs with \v0\ $\ge$ 925--950~\kms\ is smaller than for 
bound HVSs with \v0\ $\le$ 900--925~\kms. 

To explore these issues in more detail, we consider $10^7$ intermediate mass stars
(3~\msun, B spectral type, $t_{ms}$ = 350~Myr) with random ejection parameters as 
outlined in \S\ref{sec: sims-init}. The 
sample includes bound stars which barely make it out of the bulge and unbound stars 
ejected from the Galaxy. Aside from the larger lower velocity limit, \v0\ = 
750~\kms\ instead of \v0\ = 600~\kms, this set of calculations is identical to 
those in \citet{kenyon2014}. Outcomes are also similar. Although the higher minimum 
\v0\ precludes bound stars with maximum $r$ = 1--8~kpc, statistics for stars with
$r \gtrsim$ 10~kpc are nearly identical to those in \citet{kenyon2014}.

To establish the importance of bound and unbound stars in this sample, we derive the 
variation of the space density with distance from the GC \citep[see also][]{bromley2006}. 
For a set of radial bins, $r_i$, 
we define a space density, $\rho_i \propto r_i^2 N_i$, where $N_i$ is the number of 
stars in a bin extending from $r_i - 0.5 \delta r$ to $r_i + 0.5 \delta r$. Setting 
$\delta r$ = 5~kpc yields a reasonable number of stars per bin.  Within the full sample,
`halo-like' stars have a space velocity $v$ smaller than 75\% of the local escape 
velocity \vesc; `bound outliers' have $0.75 \vesc \le v \le \vesc$.  `Unbound' stars 
have velocities relative to the GC that exceed the local escape velocity. 
Table~\ref{tab: den-stats} summarizes the fraction of these three types of stars as a 
function of $r$.

Our choice for the boundary between halo-like stars and bound outliers is motivated by
radial velocity surveys of the halo \citep[e.g.,][]{battaglia2005,smith2007,brown2008,
xue2008,brown2010a,kafle2012,brown2014,loebman2014,king2015,cohen2017}. Within these 
surveys, the radial velocity distribution consists of a gaussian component and a small
set of outliers. The outliers have velocities roughly 2--3 times larger than the 
half-width of the gaussian component. For typical surveys, unambiguous outliers have 
velocities exceeding roughly 75\% of the local escape velocity.

At $r \le$ 70~kpc, bound stars dominate the population (Fig.~\ref{fig: rho1}). Nearly
all of the bound stars have $v \le$ 0.75\vesc; these stars have positions and velocities
similar to those of the indigenous population in the Galactic halo \citep[e.g.,][]{brown2006a,
brown2006b,brown2007a,kenyon2014}. Roughly 20\% have space velocities large enough 
($v >$ 0.75\vesc) to be identified as outliers in a halo radial velocity survey but they are 
still bound to the Galaxy.  Only a small fraction of the ejected stars at these distances 
(2\% at 10--20~kpc, 6\% at 20--40~kpc, and 21\% at 40--80~kpc; Table~\ref{tab: den-stats}) 
are unbound.  

The variation of $\rho_i$ with $r$ for the bound stars depends on the stellar lifetime 
and the initial ejection velocity from the GC \citep{bromley2006,kenyon2008,kenyon2014}. 
Bound HVSs ejected with \v0\ = 900~\kms\ take 100~Myr to reach $r$ = 45~kpc and another 
100~Myr to approach $r$ = 77~kpc.  In an ensemble of HVSs ejected from the GC at random 
times, a 100~Myr travel time is a modest fraction of the main sequence lifetime, 
$t_{ms}$ = 350~Myr.
Nearly all bound HVSs can travel to 50~kpc; the density is then roughly constant with $r$. 
Beyond 50~kpc, the travel time becomes a larger and larger fraction of $t_{ms}$; the 
density then begins to drop because the stars die. For HVSs ejected with \v0\ = 900~\kms\ as 
zero-age main sequence stars, the {\it maximum} distance from the \GC\ is roughly 110~kpc.  
At this point, the density of bound stars is zero.

When $r \gtrsim$ 80~kpc ($r \gtrsim$ 125~kpc), most (all) stars are unbound.  Unbound
stars ejected at high velocities, \v0\ $\approx$ 1100~\kms, reach 100~kpc (125~kpc) on 
time scales, $\sim$ 100~Myr (175~Myr). On these time scales, it is fairly easy for a B-type 
main sequence star to travel 100--110~kpc from the \GC; however, it is much more challenging 
to reach distances much beyond 150~kpc. Thus, the density gradually rises until 100--120~kpc 
and then begins to fall due to the finite stellar lifetime. Although stars ejected at the 
largest velocities, 1500~\kms, are still on the main sequence at $r \approx$ 200--300~kpc, 
these stars are rare. At these distances, the density of B-type HVSs is negligible.

Despite the smaller set of bound stars in a sample at $r \approx$ 80--160~kpc, there is a 
dramatic variation in $\delta \phi$ with $r$ and \phiz\ (Figs.~\ref{fig: disk3}--\ref{fig: disk4};
see also Yu \& Madau 2007).  At these distances, the highest velocity stars with 
\v0\ = 1400--1500~\kms\ feel a modest deceleration from the disk and have 
$\delta \phi \lesssim$ 1\deg. For lower velocity stars, typical deflections range 
from 5\deg\ ($\phi_0 \lesssim$ 30\deg) to 0\degpoint5 ($\phi_0 \gtrsim$ 70\deg; 
Fig.~\ref{fig: disk3}, lower panel).  Maximum deflections are roughly twice the typical 
deflections. Within the full set of stars, 57\% have $\phi_f \lesssim$ 30\deg, 
illustrating the dramatic impact of gravitational focusing by the disk.

Among less distant stars with $r$ = 40--80~kpc, the range of $\delta \phi$ is roughly 
50\% larger (Fig.~\ref{fig: disk3}, upper panel). Compared to the 80--160~kpc group,
unbound stars at these distances have somewhat smaller initial velocities and therefore 
experience somewhat larger overall deflections. However, most stars with the largest 
$\delta \phi$ are bound; with \v0\ $\approx$ 850--900~\kms, they spend more time at smaller
$r$ and undergo much larger deflections. As a result, more stars in this sample have
$\phi_f \lesssim$ 30\deg\ (60\%). 

For stars with $r$ = 10--40~kpc, the variation of $\delta \phi$ with $\phi_0$ is more 
complicated (Fig.~\ref{fig: disk4}). In this distance range, nearly all of the stars 
are bound (94\%; Table \ref{tab: den-stats} and Fig.~\ref{fig: rho1}). Among the bound 
stars, roughly 90\% have halo-like space velocities. On their first pass out through 
the Galaxy, these stars endure somewhat larger deflections than higher velocity stars 
at larger $r$. The maximum $\delta \phi$ is roughly 20\deg\ for stars with $r$ = 20--40~kpc 
and $\phi_0 \approx$ 15\deg; stars with $r$ = 10--20~kpc experience a maximum $\delta \phi$ of
30\deg\ at $\phi_0 \approx$ 20\deg\ to 30\deg. As in Fig.~\ref{fig: disk3}, the maximum
$\delta \phi$ is smaller for stars with $\phi_0 \lesssim$ 10\deg\ and $\phi_0 \gtrsim$ 30\deg;
the typical $\delta \phi$ is half the maximum. 

Some bound stars with $\phi_0 \gtrsim$ 30\deg\ and $r$ = 10--40~kpc travel out from the \GC, 
reach apogalacticon, and head back towards the Galactic disk. If they live long enough to 
pass through the midplane of the disk, they end up on the opposite side of the disk relative 
to their starting point and have $\delta \phi \gtrsim \phi_0$. This group produces the
concentrations of stars extending from ($\phi_0$, $\delta \phi$) $\approx$ (35\deg, 30\deg) 
to ($\phi_0$, $\delta \phi$) $\approx$ (90\deg, 150\deg) in each panel of Fig.~\ref{fig: disk4}. 
Stars with smaller \v0\ that reach smaller maximum $r$ are more likely to live long enough 
to pass through the disk plane than higher velocity stars at larger $r$. Thus, there are 
more stars with very large $\delta \phi$ at 10--20~kpc than at 20--40~kpc.

When bound stars follow purely radial orbits, they simply retrace their path after
reaching apogalacticon.  Within a real MW, however, the disk deflects trajectories 
for stars on their way out of the \GC\ and continues to deflect them as they try to 
return to the \GC\ (Fig.~\ref{fig: disk1}). In this situation, stars follow very 
non-radial orbits where the total deflection is roughly proportional to $\phi_0$: 
$\delta \phi \approx \alpha \phi_0 + \beta$ with $\alpha \approx$ 2.5 and 
$\beta$ = $-70$\deg\ for $\phi_0$ = 40\deg\ to 90\deg.  

We next consider the impact of the LMC on the trajectories of HVSs. With a mass almost
twice the mass of the Galactic disk, the LMC should generate larger deflections than
the disk. To quantify changes to HVS trajectories, we consider simple models with a 
stationary LMC and then examine results for an LMC on a more realistic orbit relative
to the \GC.

\subsection{Toy LMC Models}
\label{sec: res-toy}

As a first exploration of the impact of the LMC on HVSs ejected from the GC,
we consider a simple potential model where an LMC analog lies along the $+$z-axis 
at a distance of 49.01~kpc from the GC and 49.66~kpc from the Sun. Compared to a
system with no LMC, the extra mass in the MW+LMC potential changes $r(t)$ and 
$v(t)$ for HVSs ejected from the GC. After showing how the LMC modifies $v(r)$ 
for HVSs ejected along the $+z$-axis, we follow the structure of the previous 
subsection and quantify how the gravity of the LMC modifies the radial trajectories 
of individual stars ejected from the GC. We then examine the range of possible 
deflections for ensembles of $10^7$ HVSs selected with the standard prescription
outlined in \S\ref{sec: sims-init}.

Fig.~\ref{fig: vr1} compares $v(r)$ for calculations with (dashed lines) and
without (solid lines) an LMC on the $+z$-axis. In a pure MW potential, stars 
ejected with $\v0\ \lesssim$ 775~\kms\ towards the LMC have maximum $r \lesssim$ 
10~kpc \citep[see also][]{kenyon2008}. As \v0\ grows, ejected stars reach larger 
$r$.  Although the LMC produces negligible changes to $v(r)$ for low velocity 
ejections which never reach the LMC, there are clear changes in $v(r)$ for high 
velocity ejections.  When \v0\ $\approx$ 800~\kms, stars traveling toward the LMC 
achieve larger distances ($r \approx$ 40~kpc) than those trying to escape from 
a pure MW potential ($r \approx$ 30~kpc). When the ejection velocity is larger 
(\v0\ $\approx$ 900--1000~\kms), the LMC acceleration produces a clear `bump' in the 
$v(r)$ track centered on the distance of the LMC from the \GC\ ($r \approx$ 49~kpc).  

In this example, HVSs ejected towards the LMC reach larger distances than HVSs ejected 
into a potential with no LMC. When \v0\ = 900~\kms, HVSs ejected along the $+z$-axis 
reach $r$ = 105~kpc after 300~Myr of travel time. With no LMC, stars reach only 101~kpc. 
The smaller deceleration before the star reaches the LMC compensates for the larger 
deceleration after the star passes through the LMC.  Despite the larger $r$, HVSs 
traveling through the LMC have $v \approx$ 200~\kms\ at 300~Myr compared to 212~\kms\ for 
HVSs ejected into a pure MW potential. Although the LMC helps HVSs ejected along the 
$+z$-axis reach larger distances, these HVSs have 10\% smaller velocities.

Although changing $r_L$ results in negligible differences in $v(r,t)$, the amplitude 
of the bump in the $v(r)$ track responds to the adopted $M_L$. More (less) massive 
LMC analogs yield larger (smaller) bumps. In a system where the LMC has twice (half) 
the nominal mass, an HVS ejected with \v0\ = 900~\kms\ along the $+z$-axis reaches a
distance of 107~kpc (103~kpc) with a velocity of 185~\kms\ (205~\kms) after a 300~Myr
travel time. For ejections along the $-z$-axis, \v0\ = 900~\kms\ yields $(r, v)$ = 
(100~kpc, 202~\kms) for the light LMC analog, (97~kpc, 192~\kms) for the nominal LMC 
analog, and (94~kpc, 172~\kms) for the heavy LMC analog, 

Fig.~\ref{fig: lmc1} illustrates trajectories for HVSs ejected with \v0\ = 900~\kms\ at 
various angles $\phi_0$ relative to the Galactic plane. When $\phi_0$ = 90\deg, the axisymmetric
potential of the disk and the LMC simply speed up or slow down an HVS without changing its
overall path (Fig.~\ref{fig: vr1}). For smaller ejection angles, however, the LMC deflects 
stars more than the disk does (Fig.~\ref{fig: disk1}).
Over a travel time of 600~Myr, stars with $\phi_0 \approx$ 89\deg\ reach a maximum $x \approx$ 
910~pc at $z \approx$ 64~kpc and then bend back towards the $z$-axis, reaching 
$x \approx -200$~pc at $z \approx$ 146~kpc. Stars with smaller $\phi_0$ achieve larger 
maximum $x$ distances from the $z$-axis before bending around the LMC. The maximum $x$
distance and the $z$ distance for this maximum increase with decreasing $\phi_0$. 

Doubling (halving) the LMC mass increases (decreases) deflections (Fig.~\ref{fig: lmc1}). 
Over the first 90--100~Myr, HVS trajectories are fairly independent of the mass of the LMC. 
After 100~Myr, the heavier LMC analog sharply bends the path of an HVS towards the $x$-axis
(Fig.~\ref{fig: lmc1}, dark green dashed line). With the lighter LMC analog, the trajectory 
is much more radial (Fig.~\ref{fig: lmc1}, dark green dot-dashed line). Despite the different
magnitude of the deflections in this example, the overall speed of an HVS at 200~Myr is
nearly identical: 267~\kms\ (light LMC), 270~\kms\ (nominal LMC), 268~\kms\ (heavy LMC). 
Somewhat counterintuitively, an HVS traveling past the heavy LMC travels a larger distance
(85~kpc) than in the gravity well of the nominal LMC (81~kpc) or the light LMC (80~kpc).

The variation of $\delta \phi$ with ejection angle and LMC mass is a signature of 
gravitational focusing by the LMC. All stars ejected at some angle $\phi^{\prime}$ 
($= 90\deg\ - \phi$) relative to the $+z$-axis feel an acceleration towards 
the $+z$-axis. When $\phi^{\prime}$ is small, the acceleration in the $x-y$ plane 
is also small. At large $\phi^{\prime}$, the Galactic potential dominates. In 
both regimes, focusing is negligible. For $\phi^\prime \approx$ 15\deg--25\deg, acceleration 
from the LMC at $r \approx$ 35--65~kpc is large enough to bend trajectories by 
5\deg--10\deg. At these angles, the acceleration at $r \approx$ 35--50~kpc exceeds the
deceleration at $r \approx$ 50--65~kpc. Compared to stars with $\phi^\prime \lesssim$
15\deg\ or $\phi^\prime \gtrsim$ 25\deg, these stars end up with slightly larger velocities 
after passing by the LMC. Because the deceleration from the Galaxy is independent of
$\phi^\prime$, the stars maintain their larger velocities as they continue to speed through
the halo.

\begin{figure}
\includegraphics[width=6.5in]{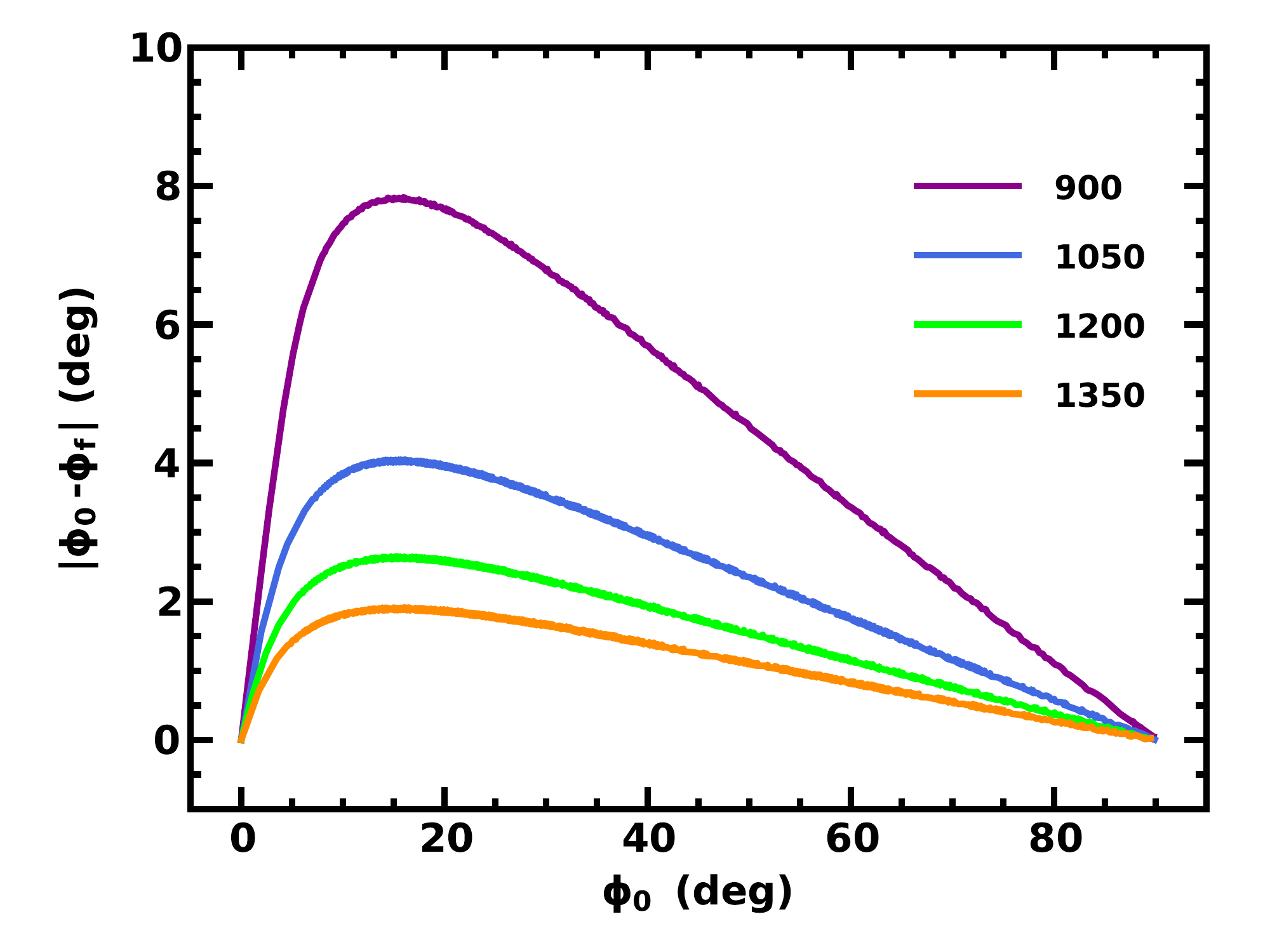}
\vskip 2ex
\caption{\label{fig: disk2}
Difference between the initial \phiz\ and final \phif\ GC latitude, 
$\delta \phi = | \phi_0 - \phi_f|$, as a function of \phiz\ for HVSs ejected at 
various \v0\ (in \kms) as listed in the legend. Independent of \v0, all 
ejected stars reach peak $\delta \phi$ when $\phi_0 \approx$ 15\deg. Stars with
large (small) \v0\ have smaller (larger) deflections.
}
\end{figure}

\begin{figure}
\includegraphics[width=6.5in]{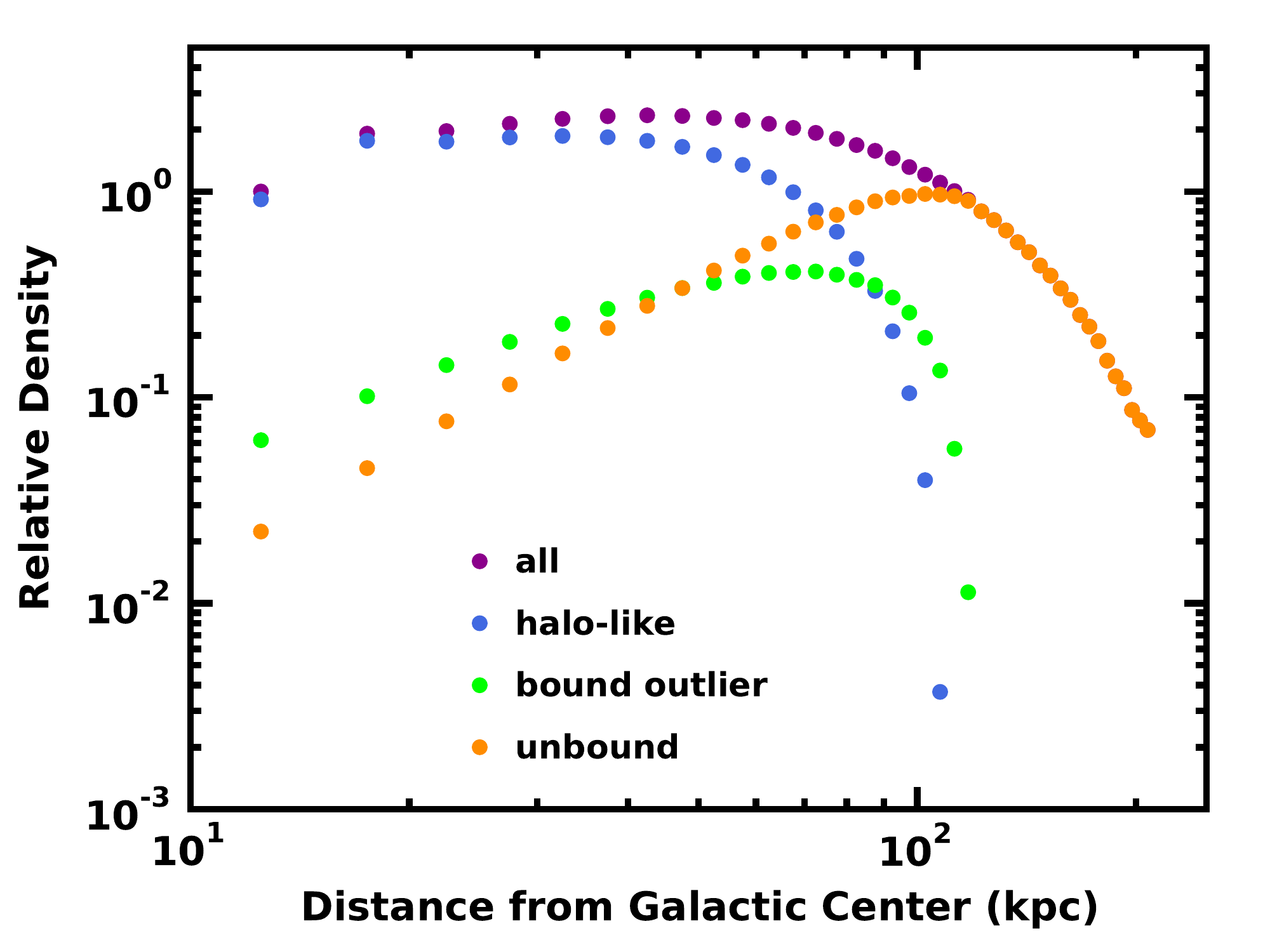}
\vskip 2ex
\caption{\label{fig: rho1}
Relative density -- $\rho(r) \propto r^2 N$ -- of an ensemble of ejected stars 
at 10--200~kpc.  The full sample ('all'; purple symbols) consists of 
`halo-like' stars with $v \le$ 0.75~\vesc\ (blue), `bound outliers' with 
$v >$ 0.75~\vesc\ and $v \le \vesc$ (green), and `unbound' stars with 
$v > \vesc$ (orange). At $r \le$ 50~kpc (80~kpc), halo-like stars and bound 
outliers (bound outliers) are more numerous than the population of unbound 
stars. At $r >$ 70~kpc, unbound stars dominate.
}
\end{figure}

\begin{figure}
\includegraphics[width=6.5in]{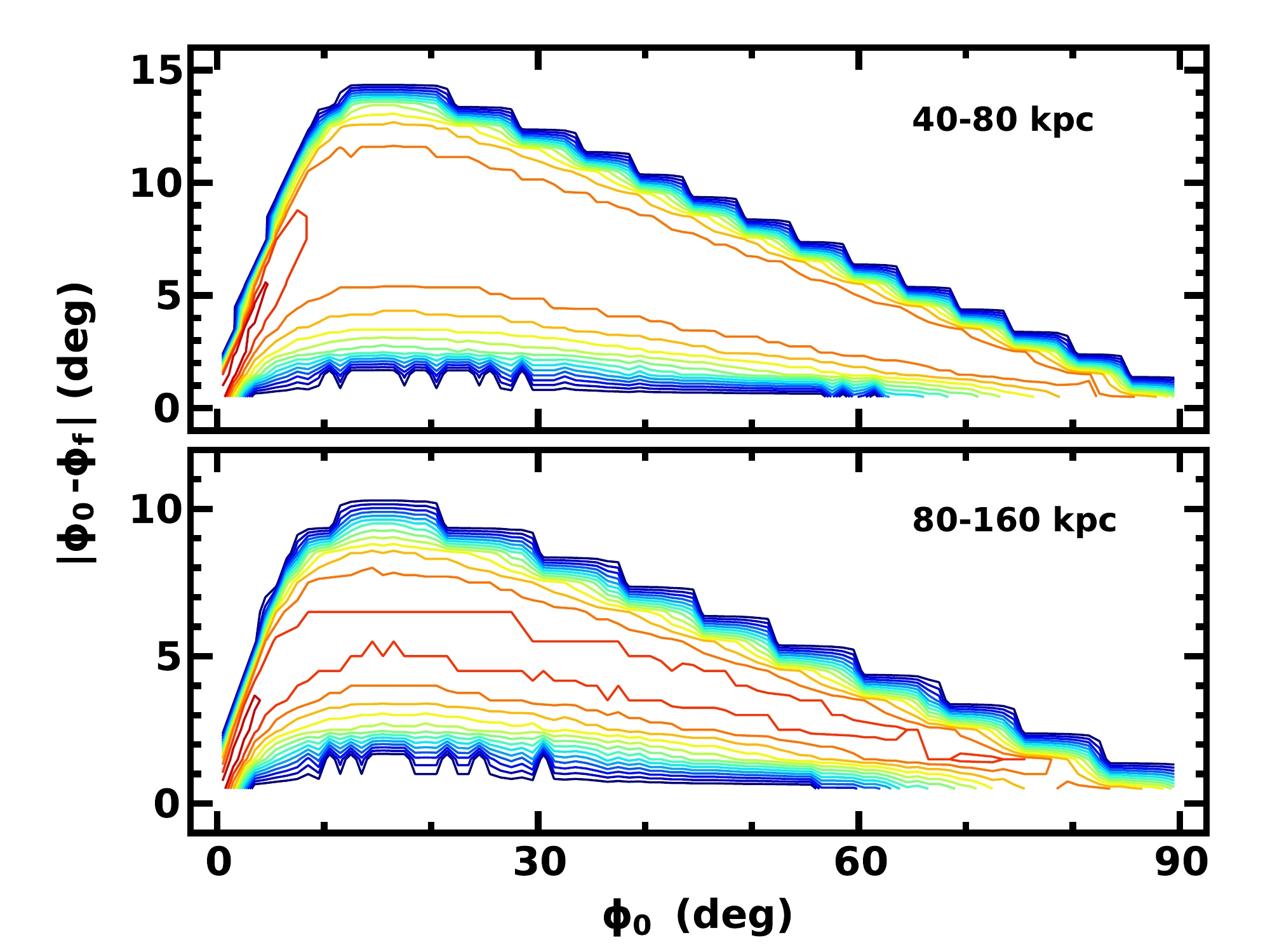}
\vskip 2ex
\caption{\label{fig: disk3}
Frequency distribution of $\delta \phi$ in an ensemble of $10^7$ HVSs with 
random \v0, \b0, and travel times. The density varies logarithmically from 
1 star per bin (blue contours) to $10^4$ stars per bin (red contours). 
Stars with $r$ = 80--160~kpc (lower panel) are deflected less than stars
with $r$ = 40--80~kpc
}
\end{figure}

\begin{figure}
\includegraphics[width=6.5in]{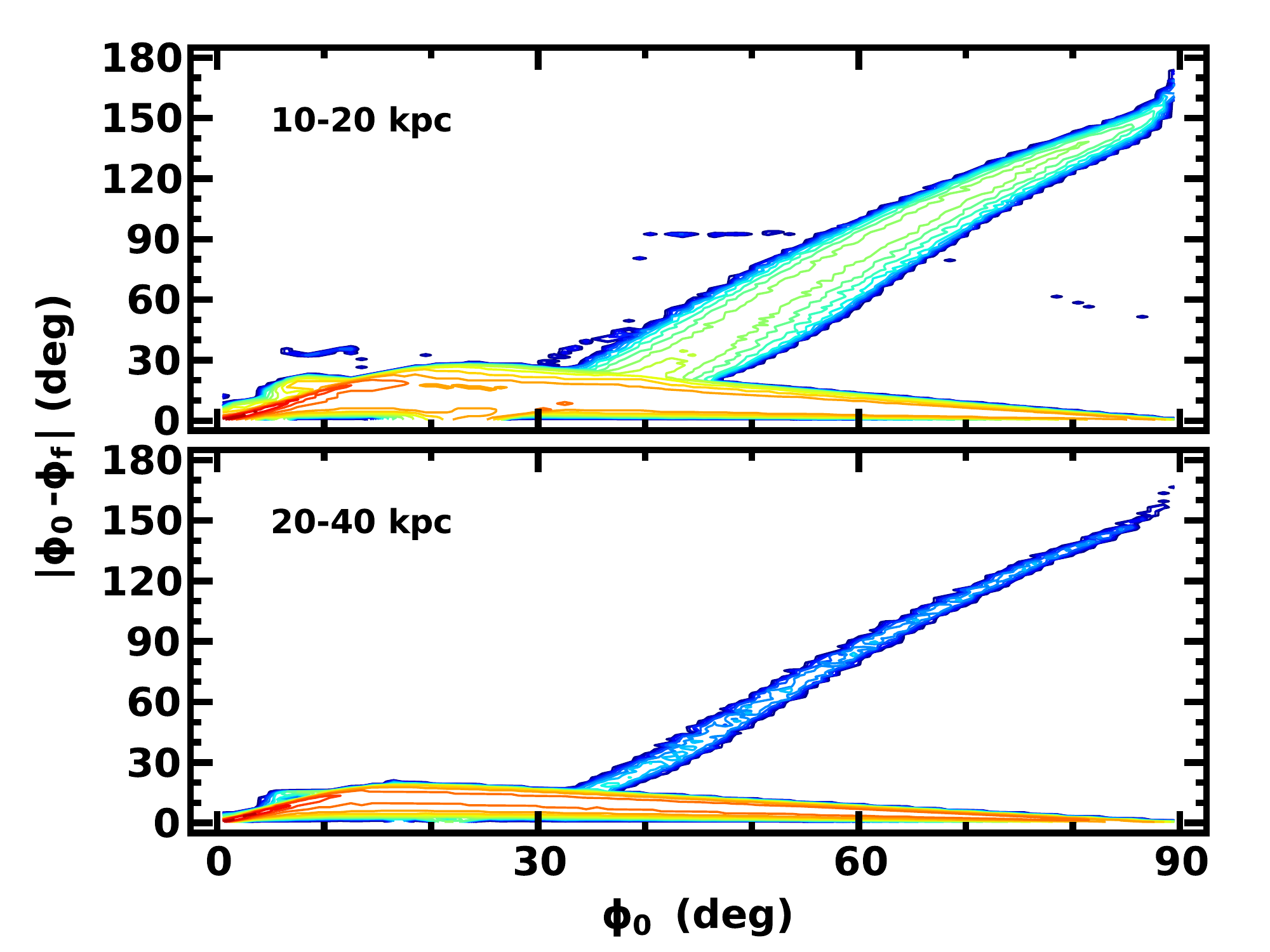}
\vskip 2ex
\caption{\label{fig: disk4}
As in Fig.~\ref{fig: disk3} for stars with $d$ = 20--40~kpc (lower panel)
and $d$ = 10--20~kpc (upper panel). Note the change in vertical scale 
(from 0\deg--15\deg\ to 0\deg--180\deg). HVSs making their first pass through 
the Galaxy lie in the dense population with $\delta \phi \lesssim$ 20\deg--30\deg. 
Lower velocity stars whose trajectories have been deflected back towards the 
Galactic disk lie in the low density region in each panel with 
$\phi_0 \gtrsim$ 30\deg--40\deg\ and $\delta \phi \gtrsim$ 30\deg--40\deg.
}
\end{figure}

\begin{figure}
\includegraphics[width=6.5in]{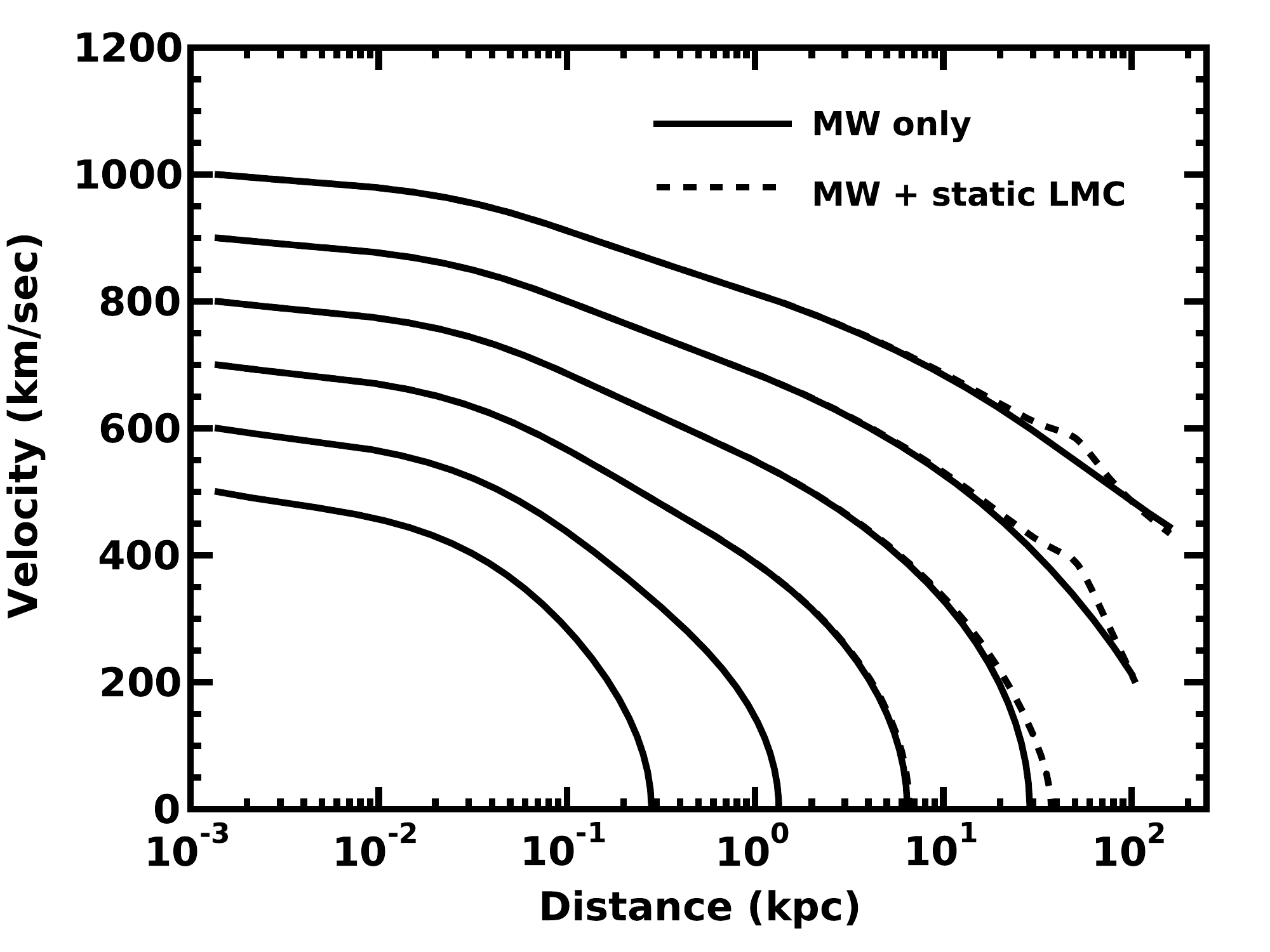}
\vskip 2ex
\caption{\label{fig: vr1}
Variation of space velocity with distance from the \GC\ for HVS in a Galactic 
potential with (dashed lines) and without (solid lines) an LMC-analog on the 
z-axis at a distance of 50~kpc from the \GC. With no LMC, reaching the halo of 
the Milky Way ($d \gtrsim$ 10--30~kpc) requires an ejection velocity $\v0 \gtrsim$ 
775--800~\kms\ \citep[solid lines;][]{kenyon2008}. For stars ejected in the 
direction of the LMC analog (dashed lines), the space velocity is roughly 5\%
larger at $d$ = 30~kpc and 15\% larger at $d$ = 50~kpc. When $d \gtrsim$ 100~kpc, 
stars traveling through the MW+LMC potential have smaller space velocity than 
stars in the MW-only potential. 
}
\end{figure}

\begin{figure}
\includegraphics[width=6.5in]{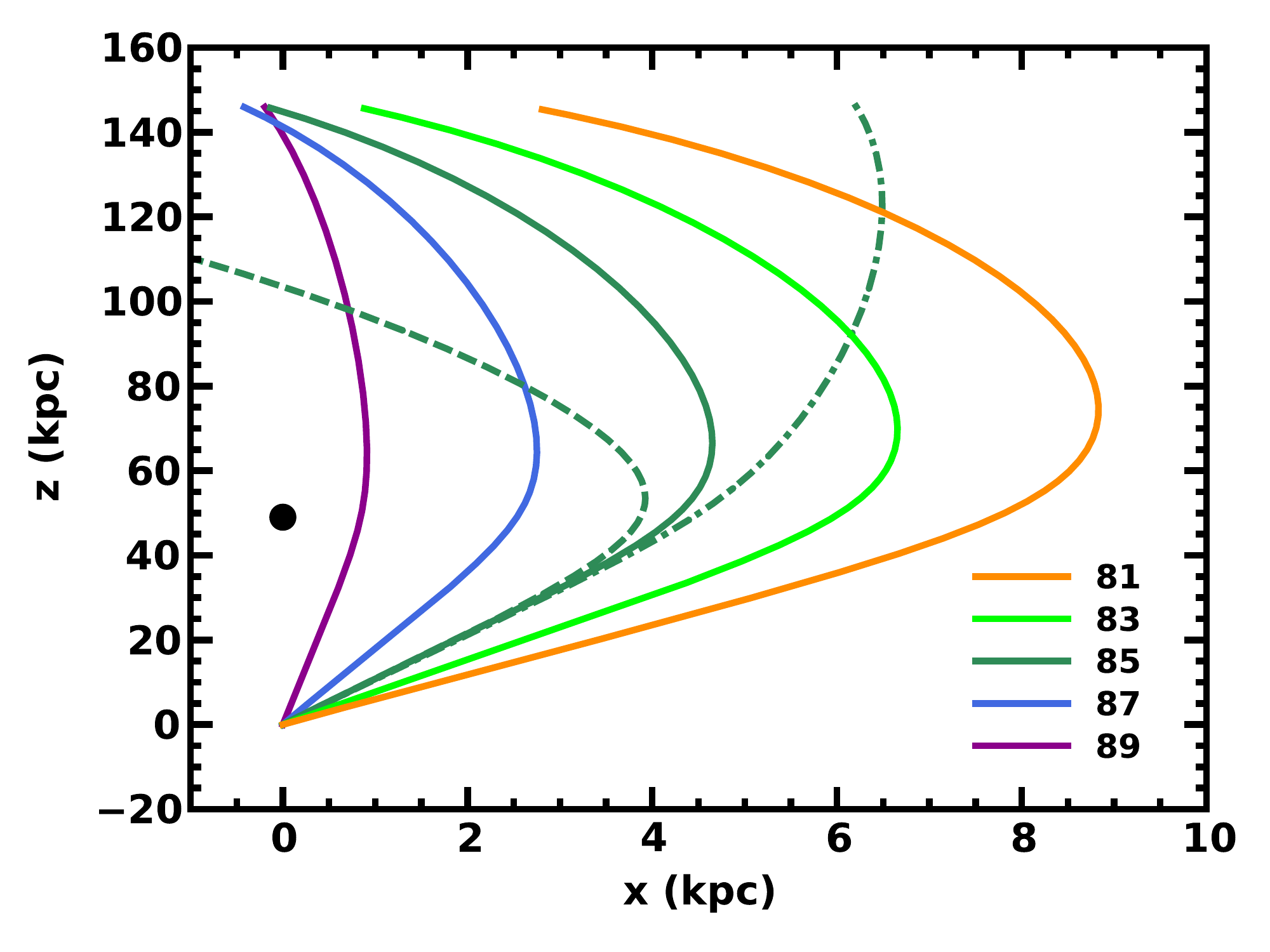}
\vskip 2ex
\caption{\label{fig: lmc1}
Trajectories in the $x-z$ plane for HVSs ejected with \v0\ = 900~\kms\ at various 
angles $\phi_0$ (in deg, as indicated in the legend) relative to the Galactic plane 
in a MW+LMC system.  The filled circle indicates the position of the LMC analog.  
Solid lines illustrate trajectories for the nominal mass of the LMC 
($M_L = 10^{11}~\msun$). The dashed (dot-dashed) dark green line shows a trajectory 
for $\phi_0$ = 85\deg\ and twice (half) the nominal LMC mass.  Over travel times of 
600~Myr, the gravity of the LMC bends trajectories towards the galactic pole. Stars 
with $\phi_0 \approx$ 85\deg--89\deg\ (dark green, blue, and purple curves) cross 
the $z$-axis; others ejected at somewhat smaller $\phi_0$ (orange and light green 
curves) approach the $z$-axis after 300--600~Myr. Heavier (lighter) LMC analogs bend 
trajectories more (less).
}
\end{figure}

The amplitude of the `bump' in $v$ in Fig.~\ref{fig: vr1} and the deflections of trajectories
in Fig.~\ref{fig: lmc1} are also functions of the ejection velocity from the GC. With the 
gravitational focusing factor $f_g \propto (v_{esc} / v_0)^2$, the magnitude of the bump or
deflection responds more to changes in \v0\ than to changes in $M_L$.  For our nominal $M_L$, 
ensembles of stars with \v0\ = 1200~\kms\ and various ejection angles \phiz\ have nearly 
identical \vfinal\ after a travel time of 300~Myr. Deflections from purely radial trajectories 
are minimal. Halving or doubling the mass of the LMC has a fairly minimal impact on the 
trajectories for these high velocity stars. However, halving (doubling) \v0\ leads to much 
smaller (larger) deflections (Fig.~\ref{fig: disk2}).

\begin{figure}
\includegraphics[width=6.5in]{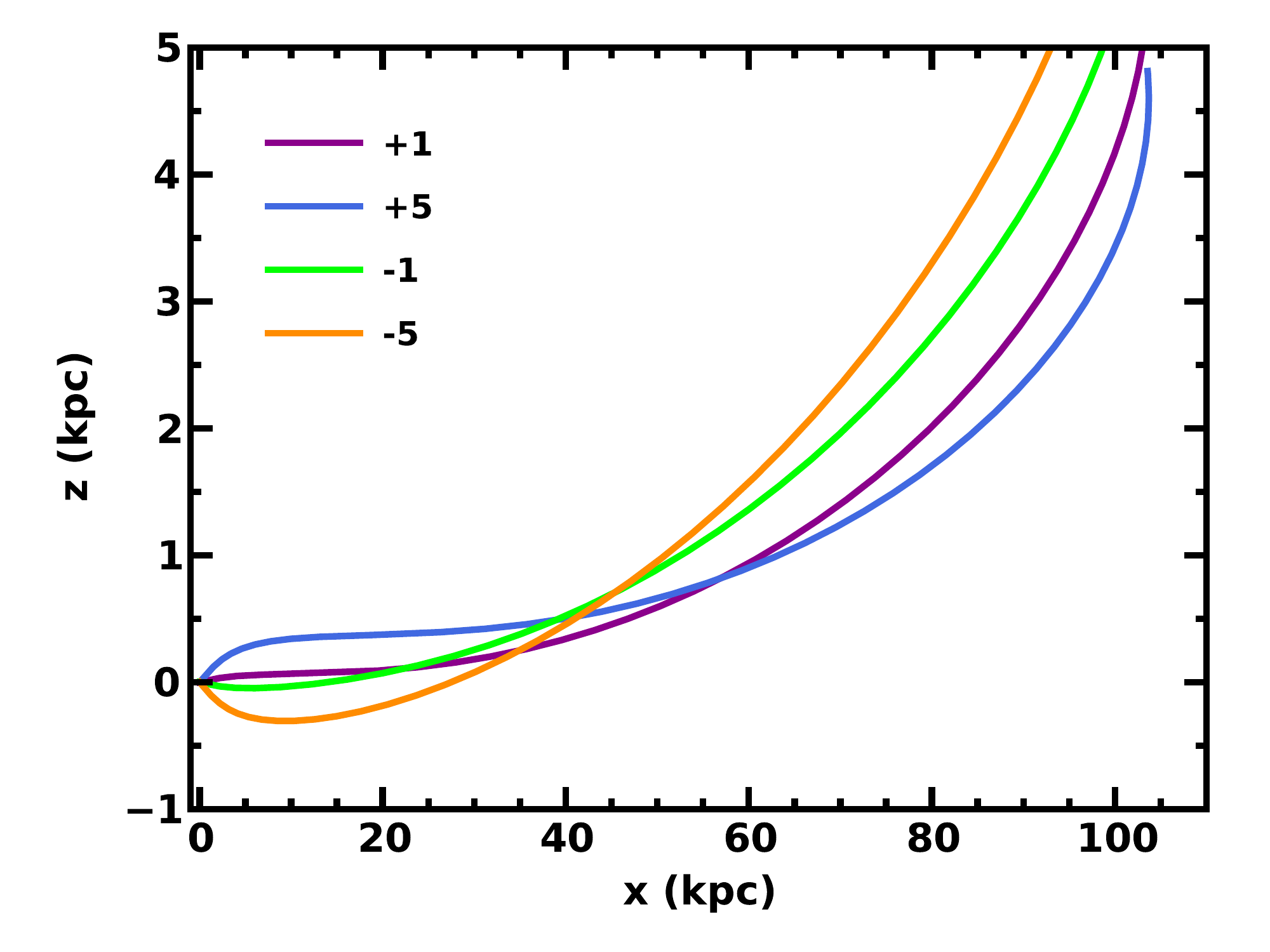}
\vskip 2ex
\caption{\label{fig: lmc2}
As in Fig.~\ref{fig: lmc1} for HVSs ejected with \v0\ = 900~\kms\ in the Galactic 
plane.  The legend indicates the ejection angle (in deg) relative to the Galactic 
plane. Compared to a system without the LMC (Fig.~\ref{fig: disk1}), trajectories 
are bent away from the plane in the direction of the LMC. Stars with $\phi_0 <$ 0 
(light green and orange curves) experience larger deflections than those with 
$\phi_0 > 0$ (blue and purple curves).
}
\end{figure}

Aside from these obvious gravitational focusing effects, the LMC potential also bends the 
trajectories of HVSs ejected into the Galactic plane (Fig.~\ref{fig: lmc2}). For stars with 
$\phi_0 \approx$ 1\deg--5\deg, the disk gravity works to deflect stars back toward the midplane 
as in Fig.~\ref{fig: disk1}. By the time stars reach $x$ = 40~kpc, the gravity of the more distant 
LMC overcomes the weaker disk gravity and pulls stars away from the plane and into the halo. 

When stars are ejected with $\phi_0$ = $-$1\deg\ to $-$5\deg, the impact of the LMC is more
pronounced. At small $x$, the disk gravity pulls stars towards the midplane; the z-component
of the velocity changes sign from negative to positive. After stars cross the midplane, the 
gravity of the disk is too weak compared to the LMC to pull them back. The gravity from the 
LMC continues to pull stars farther and farther above the midplane. Because these stars already 
have a positive $v_z$, they overtake HVSs ejected with $\phi_0 >$ 0. The trajectories of stars 
with $\phi_0 < 0$\deg\ are therefore bent more than those of stars with $\phi_0 > 0$\deg.

For HVSs ejected with a range of \phiz, the distance reached after a fixed time depends on 
\v0\ and \phiz. Stars ejected towards the fixed LMC ($\phiz\ \gtrsim$ 60\deg) with \v0\ =
900~\kms\ achieve 1\% to 5\% larger distances after a 300~Myr travel time. Stars ejected 
away from the LMC (\phiz\ $\lesssim -60$\deg) end up at 4\% smaller distances. 
Stars ejected approximately into the plane ($-60$\deg\ $\lesssim \phiz \lesssim$ 60\deg) 
are mainly slowed by the extra gravity from the LMC and have 2\% to 3\% smaller maximum
distances than stars ejected into a pure MW potential.

Independent of \phiz, HVSs in a MW+LMC potential have smaller space velocities. For 
\v0\ = 900~kms, speeds after 300~Myr range from 95\% (\phiz\ $\approx$ 75\deg) to 90\% 
(\phiz\ $\approx -90$\deg) of HVS speeds in the pure MW potential. The maximum final 
speeds occur for HVSs that pass within 1-2 LMC scale lengths of the LMC center. 

HVS ejected with larger (smaller) \v0\ have a smaller (larger) impact on their distances 
and speeds after 300~Myr traveling through the Galaxy. High (low) velocity stars spend less
(more) time near the LMC and thus experience smaller (larger) overall deceleration. For 
HVSs capable of escaping the Galaxy (\v0\ $\gtrsim$ 1000~\kms), final distances and speeds
are only somewhat less with the LMC than without the LMC. Bound stars with \v0\ $\lesssim$
900~\kms\ have much smaller distances and velocities, with trajectories modified significantly
by the LMC (Figs.~\ref{fig: lmc1}--\ref{fig: lmc2}).

\begin{figure}
\includegraphics[width=6.5in]{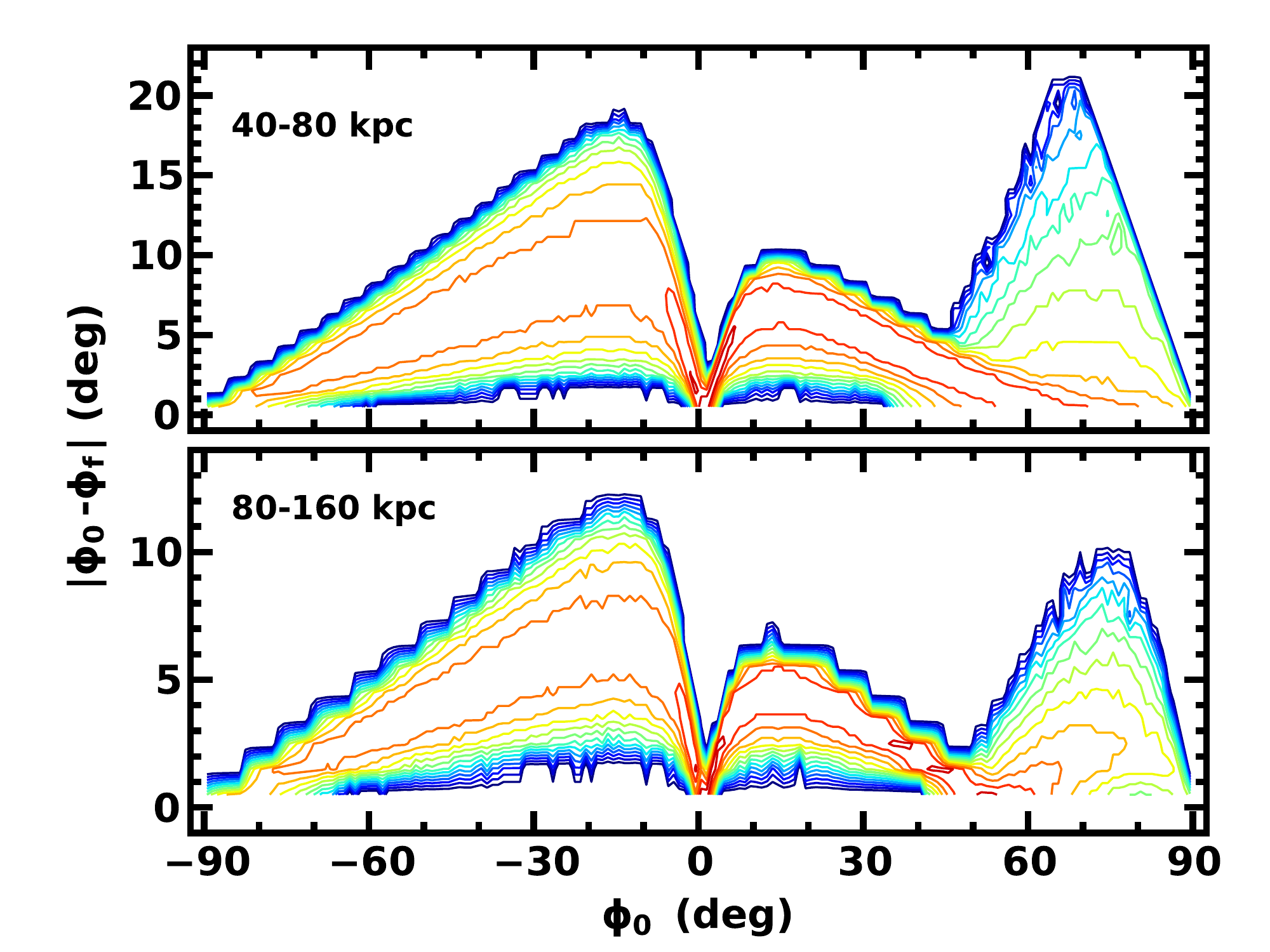}
\vskip 2ex
\caption{\label{fig: lmc3}
Frequency distribution of $\delta \phi$ in an ensemble of $10^7$ HVSs in a MW+LMC 
potential with random \v0, \phiz, and travel times. The LMC analog is on the $+z$-axis.
The density varies logarithmically from 1 star per bin (blue contours) to $10^4$ stars 
per bin (red contours).  Compared to a potential with no LMC (Fig.~\ref{fig: disk3}),
the typical $\delta \phi$ is much larger, with major peaks at 
(i) $\phi_0 \approx -15$\deg, where the gravity of the LMC draws stars across 
the disk midplane,
(ii) $ \phi_0 \approx +15$\deg, where the disk deflects stars across the midplane, 
and 
(iii) $\phi_0 \approx +75$\deg, where stars ejected into the halo bank around 
the LMC.  Trajectories of stars at 40--80~kpc (upper panel) bend more than 
those at 80--160~kpc (lower panel).
}
\end{figure}

To quantify the impact of the LMC on HVSs trajectories in more detail, we consider a second 
sample of $10^7$ stars ejected from the \GC\ with a stationary LMC analog positioned on the 
$+z$-axis at a distance of 49.66~kpc from the Sun.  Once again, the predicted $\delta \phi$ 
is a strong function of 
$\phi_0$ and $v_0$ (Fig.~\ref{fig: lmc3}). When an LMC analog lies along the $+z$-axis, there 
are three peaks in the $\phi_0$--$\delta \phi$ relation: 
(i) at $\phi_0 \approx -15$\deg, where stars ejected with a negative $v_z$ are pulled toward
the midplane by the LMC and the Galactic disk,
(ii) at $\phi_0 \approx$ 15\deg, where the gravity from the disk counters the gravity from the
LMC, and
(iii) at $\phi_0 \approx$ 75\deg, where the LMC gravity focuses stars around it.

With our adopted LMC mass, the three peaks have very different maximum deflections. In a pure
MW potential, the two peaks at $\phi_0 = \pm15$\deg\ are symmetric: stars ejected with 
positive or negative $\phi_0$ are equally drawn to the disk midplane. Adding in the LMC
potential creates an asymmetry. Stars ejected with negative $\phi_0$ are pulled towards 
the disk midplane by the LMC {\it and} the disk. For stars with positive $\phi_0$, the 
gravity of the LMC {\it counters} the gravity of the disk. When $|\phi_0| \lesssim$ 60\deg, 
stars with negative $\phi_0$ have larger $\delta \phi$ than those with positive $\phi_0$. 

Stars ejected along the $+z$-axis ($\phi_0 \gtrsim$ 60\deg) experience much larger deflections
from the gravity of the LMC than from the gravity of the disk. Among these stars, the disk gravity 
decelerates stars and produces a modest deflection (Fig.~\ref{fig: disk2}). All of these stars, 
however, pass within 2$r_L$ of the LMC and are focused towards the $+z$-axis.  The amount of 
focusing depends on \v0: stars with large (small) \v0\ spend less (more) time near the LMC and 
have smaller (larger) $\delta \phi$.

As in examples for the pure MW potential, the typical $\delta \phi$ is a strong function 
of \v0. Stars with the largest \v0\ reach the largest $r$ and experience the smallest 
deflections (Fig.~\ref{fig: lmc3}, lower panel). Compared with HVSs in a pure MW potential, 
HVSs in the MW+LMC potential with $r \gtrsim$ 80~kpc and $\phi_0 \approx -15$\deg\ ($+15$\deg) 
have larger (smaller) $\delta \phi$. Most of these stars are unbound; aside from deflecting
stars above the plane, the LMC has little impact on their escape from the Galaxy.

High velocity HVSs ejected towards the LMC ($\phi_0 \gtrsim$ 60\deg) fall into two groups.
Nearly all of these stars are unbound (Fig.~\ref{fig: rho1}). The LMC deflects these stars 
by a few deg, as indicated by the red contour in the lower right corner of the panel. A few
stars are bound; before falling back towards the GC, the LMC bends their trajectories by as
much as 10\deg.

Lower velocity stars that reach $r$ = 40--80~kpc have systematically larger $\delta \phi$
(Fig.~\ref{fig: lmc3}, upper panel). For nearly all of these stars, the typical $\delta \phi$
is roughly 50\% larger than for higher velocity stars at 80--160~kpc. However, the shape of
the $\delta \phi$--\phiz\ relation is mostly unchanged, with two peaks at $\phiz$ = 
$-15$\deg\ and $\phiz \approx +15$\deg, and a third peak at a somewhat larger 
$\phiz \approx$ $+75$\deg\ instead of $+65$\deg.

\begin{figure}
\includegraphics[width=6.5in]{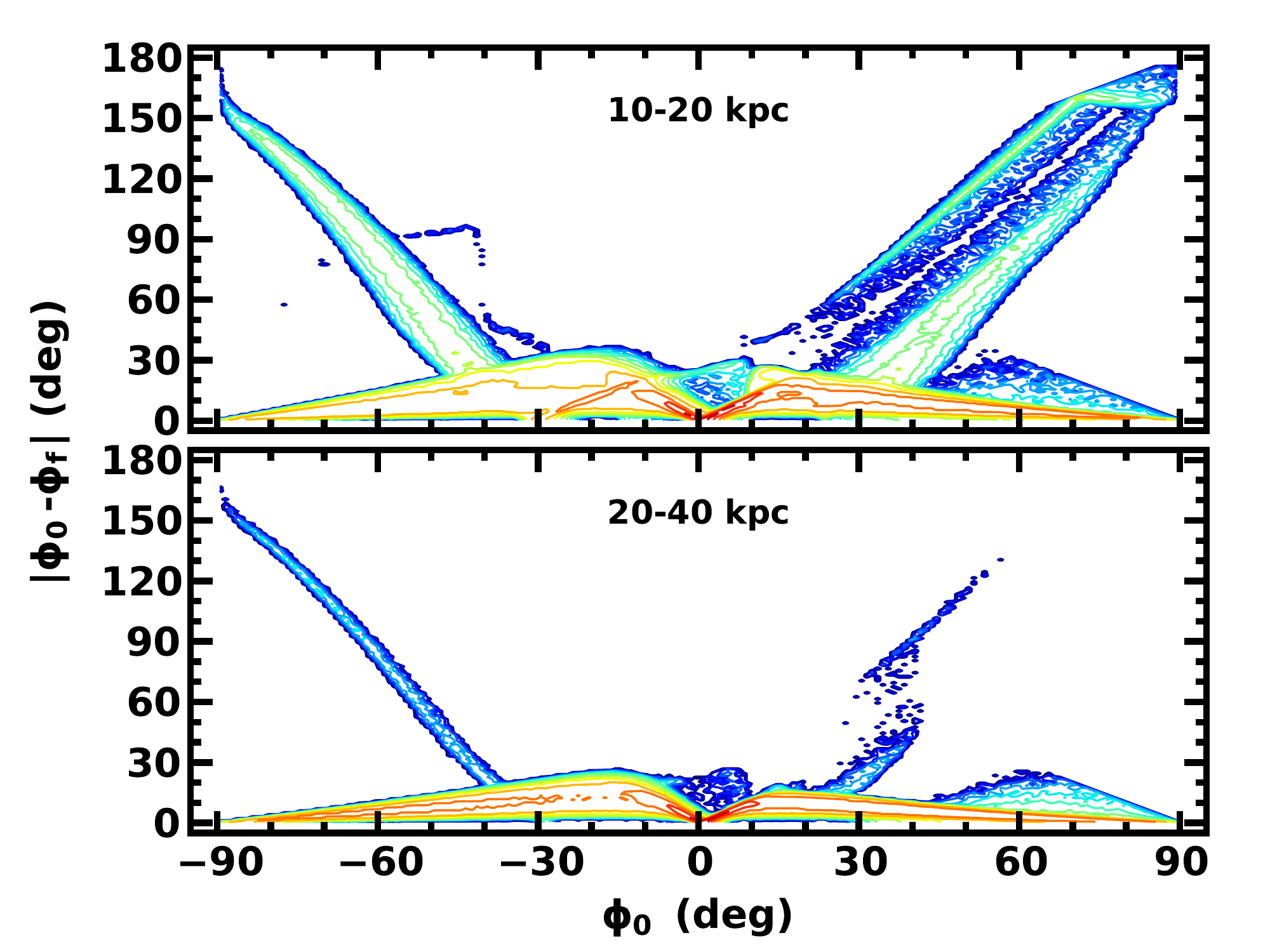}
\vskip 2ex
\caption{\label{fig: lmc4}
As in Fig.~\ref{fig: lmc3} for stars at 10--20~kpc (upper panel) and at 20--40~kpc 
(lower panel). Aside from pulling bound stars across the midplane of the disk, an
LMC along the $+z$-axis bends the trajectories of HVS ejected into the Galactic halo.
}
\end{figure}

Stars at much smaller $r$ experience a variety of deflections (Fig.~\ref{fig: lmc4}). 
Compared to results for a pure MW potential (Fig.~\ref{fig: disk4}), the overall shape 
of the $\phiz$--$\delta \phi$ relation is fairly similar: (i) most stars have modest 
deflections, (ii) there are clear peaks in $\delta \phi$ at $\phiz \approx \pm15$\deg,
and (iii) some bound stars with large $\phi_0$ undergo very large $\delta \phi$ as they
try to return to the GC. 

Among stars at 20--40~kpc, the LMC generates several new features in the 
$\phiz$--$\delta \phi$ relation (Fig.~\ref{fig: lmc4}, lower panel). 
Although a substantial `tail' of bound stars with large $\delta \phi$ at 
large \phiz\ remains for \phiz\ $\lesssim -30$\deg, there is a much weaker 
feature at \phiz\ $\gtrsim$ 30\deg. Stars with large \phiz\ accelerate towards 
the LMC and are focused towards it. With somewhat larger velocities (due to 
their smaller deceleration), fewer stars return towards the GC with the same 
maximum deflections as their counterparts with \phiz\ $\lesssim -30$\deg.
Instead, the trajectories of these stars bend towards the LMC by 20\deg\ to 
30\deg, producing an additional peak in the \phiz--$\delta \phi$ relation at 
\phiz\ $\approx$ 60\deg. 

Because the gravity of the LMC deflects HVSs with small \phiz\ (Fig.~\ref{fig: lmc2}),
there is another group of stars with $\phiz \approx$ 0\deg\ and $\delta \phi \approx$
20\deg\ to 30\deg. In a pure MW potential, stars ejected into the plane feel little 
gravity from the disk and are undeflected. With an LMC along the $+z$-axis, these stars
are deflected towards the LMC and make it into the halo at $r \approx$ 20--40~kpc.

Overall, the LMC has a modest impact on bound stars at 10--20~kpc (Fig.~\ref{fig: lmc4},
upper panel). Most stars have modest deflections. The \phiz--$\delta \phi$ relation has
(i) the standard peaks of 
$\delta \phi \approx$ 30\deg\ at \phiz\ $\approx -20$\deg\ and
$\delta \phi \approx$ 20\deg\ at \phiz\ $\approx +20$\deg, 
(ii) tails at $|\phiz| \gtrsim$ 30\deg\ with large $\delta \phi$, and
(iii) small subsets of bound stars with
$\delta \phi \approx$ 20\deg\ to 30\deg\ at \phiz\ $\approx$ 0\deg\ and at 
\phiz\ $\approx$ 60\deg.
Compared to the pure MW model, the tail at \phiz\ $\gtrsim$ 30\deg\ has a much
broader morphology and is more chaotic. With the LMC along the $+z$-axis, bound stars
traveling toward the Galactic pole are pulled towards the LMC, producing a different
set of deflections compared to the pure MW potential.

\begin{figure}
\includegraphics[width=6.5in]{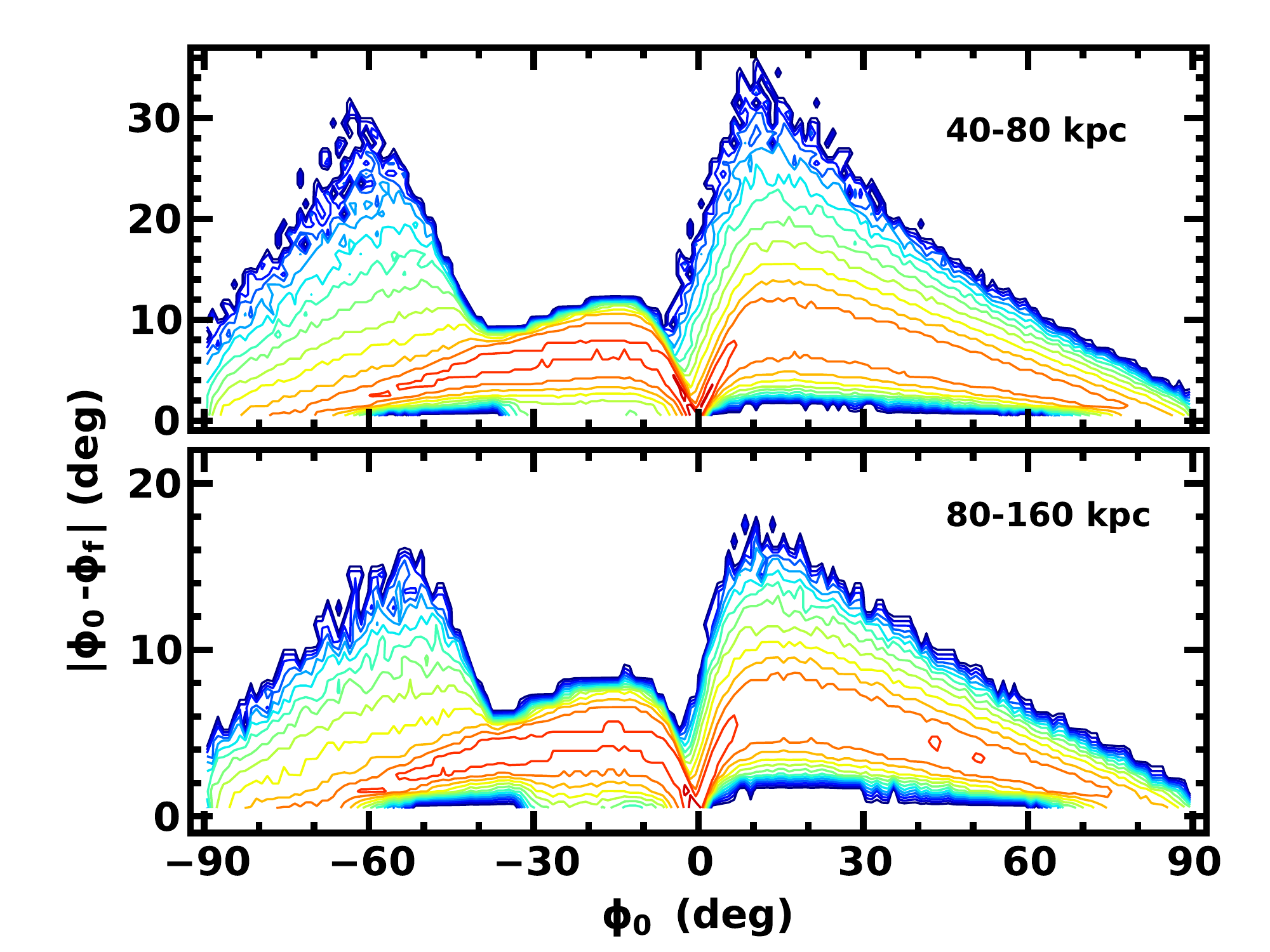}
\vskip 2ex
\caption{\label{fig: lmc5}
As in Fig.~\ref{fig: lmc3} for an LMC analog with 
$l_L$ = $-79$\degpoint5 and $b_L$ = $-33$\deg. 
}
\end{figure}

Ejecting a sample of $10^7$ HVSs into a potential with a fixed LMC at its current
position yields fairly similar \phiz--$\delta \phi$ relations. Among unbound stars
at 80--160~kpc, an LMC with $d$ = 49.66~kpc at $(l, b)$ = $(+280.5, -32.9)$ = 
$(-79.5, -32.9)$ creates somewhat larger extreme deflections, with peaks 
at $\phiz \approx -50$\deg, $\phiz \approx -15$\deg, and 
$\phiz \approx +10$\deg\ (Fig.~\ref{fig: lmc5}, lower panel).  Stars 
injected into the Galactic plane have typical $\delta \phi \approx$ 
4\deg\ to 8\deg.  Compared to a calculation with the LMC on the $+z$-axis, these
deflections are either a degree or two smaller ($\phiz \gtrsim -30$\deg) or 
a degree or two larger
($\phiz \lesssim +30$\deg). Stars ejected into the Galactic pole either have small
$\delta \phi$ ($\phiz \gtrsim$ 60\deg) or large $\delta \phi$ ($\phiz \lesssim$ 
$-60$\deg). 

An LMC in the southern hemisphere has a more dramatic impact on stars at smaller 
distances, $r \approx$ 40--80~kpc (Fig.~\ref{fig: lmc5}, upper panel). Bound stars
ejected just above the disk midplane ($\phi_0 \approx$ 10\deg) and into the southern
Galactic halo ($\phiz \approx -60$\deg) then have large $\delta \phi \approx$ 30\deg.
These deflections are roughly 50\% larger than in a system with the LMC along the
$+z$-axis. Unbound stars injected with $-30$\deg\ $\lesssim \phiz \lesssim +30$\deg\ typically
have modest $\delta \phi \approx$ 5--10\deg. HVS traveling above the plane ($z \gtrsim$ 0)
are deflected more than HVSs below the plane. Those with much larger $\phiz$ have much
smaller $\delta \phi \approx $ 1\deg\ to 2\deg.

\begin{figure}
\includegraphics[width=6.5in]{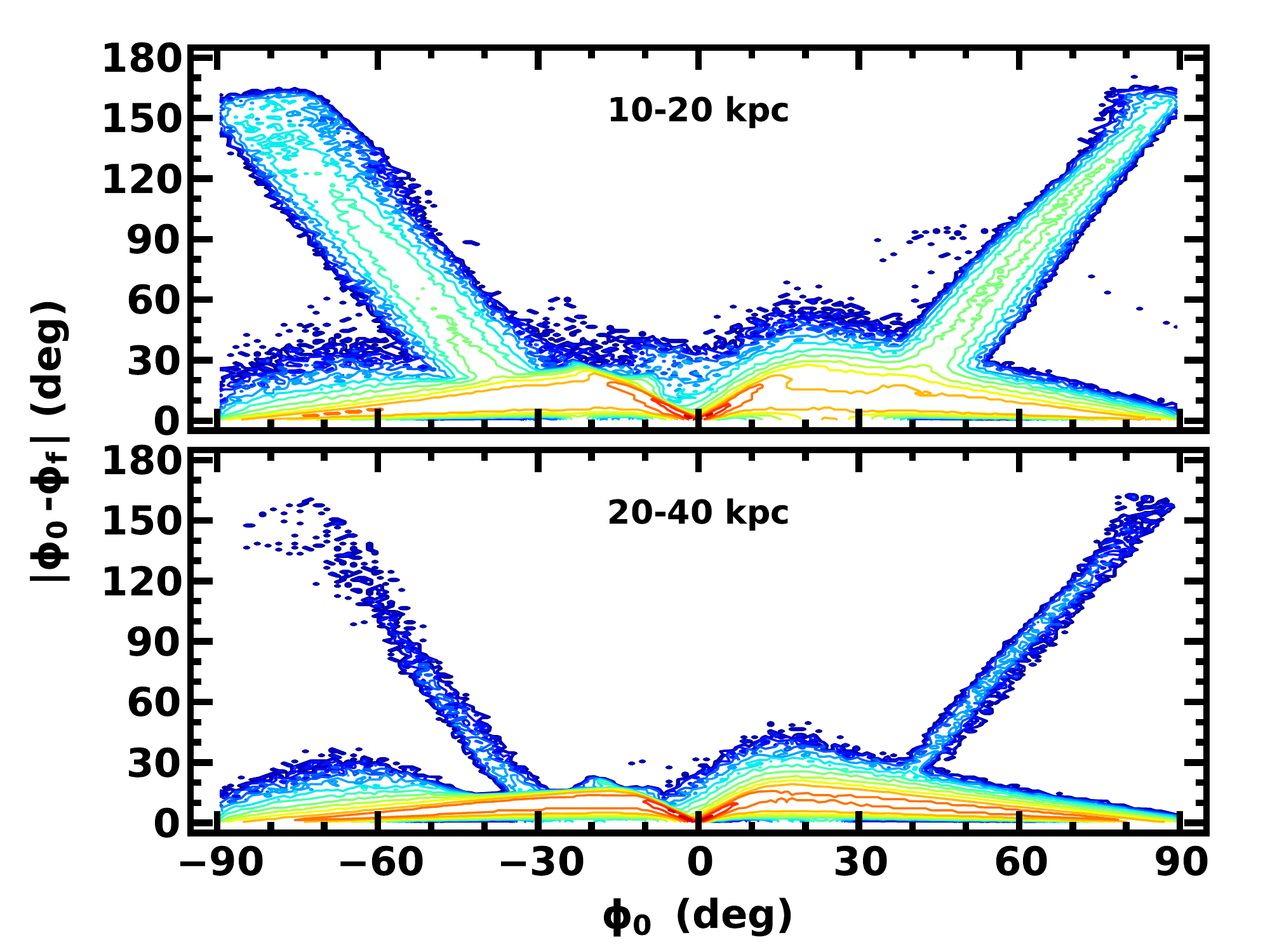}
\vskip 2ex
\caption{\label{fig: lmc6}
As in Fig.~\ref{fig: lmc4} for an LMC analog with 
$l_L$ = $-79$\degpoint5 and $b_L$ = $-33$\deg. 
}
\end{figure}

Bound stars with lower \v0\ have even larger deflections. At 20--40~kpc (Fig.~\ref{fig: lmc6},
lower panel), stars typically have $\delta \phi \approx$ 10\deg\ to 20\deg\ at low $\phiz$
and a few deg at large $\phiz$. However, groups of bound stars have large $\delta \phi$
for all $\phiz$. As in previous examples, bound stars that travel out through the galaxy,
turn around, and head back towards the GC often have $\delta \phi \approx$ 90\deg\ to
150\deg. This group grows considerably among stars with $r$ = 10--20~kpc. 

\subsection{Full LMC Model}
\label{sec: res-full}

Although the toy models illustrate how a stationary LMC impacts the trajectories of HVSs, the 
real LMC travels many kpc during the 100+ Myr flight time for stars ejected from the GC into 
the Galactic halo.  To consider the impact of a more realistic LMC, we add a moving LMC into
our potential model. Starting with an LMC at its current position, we calculate the acceleration
of the LMC (MW) due to the MW (LMC).  Relative to a fixed center-of-mass, we then integrate the 
orbits of the LMC and the MW backwards in time using the same procedure as for HVSs ejected from 
the GC. After an evolutionary time of 10~Gyr, we adopt the endpoint of the `backwards' integration 
as the starting point for a second integration, where we allow the LMC and the MW to fall back 
towards the center-of-mass.  The differences between the endpoint of this second integration and 
the current LMC position are smaller than $\pm$0.05~kpc in each cartesian position coordinate and 
less than $\pm$0.05~\kms\ in each cartesian velocity.  Although we could achieve higher accuracy 
with shorter timesteps, this agreement is satisfactory.

In this exercise, the masses, gravitational potentials, and other structural parameters of the LMC 
and MW are fixed in time\footnote{Although the Local Group is embedded in diffuse gas, observational
and theoretical analyses suggest a total mass comparable to the Galactic disk and a typical accretion
rate of 1--10~\msunyr\ \citep[e.g.,][and references therein]{nuza2014,lehner2015,richter2017}.  Tidal 
stripping likely reduces the mass of the LMC over time \citep[e.g.,][]{fox2013,fox2014}.  During the 
200--300~Myr of a typical simulation, the mass added to the MW or lost by the LMC makes a negligible 
contribution to the potential of either galaxy.}. To treat dynamical friction by the MW on the LMC, 
we follow previous studies and adopt a simple formula \citep[e.g.,][]{besla2007,gomez2015,jethwa2016}:
\begin{equation}
{d \vec{v}_L \over dt} = -4 \pi G^2 ~ M_L ~ \rho_{MW} ~ {\rm ln} \Lambda ~ 
\left | \int_0^{v_L} v^2 f_{MW} dv \right | ~
{\vec{v}_L \over v_L^3 } ~ ,
\label{eq: dvdt}
\end{equation}
where $\rho_{MW}$ is the mass density of the MW at the position of the LMC, $\Lambda = r/4800$ is 
the Coulomb logarithm, $f(v)$ is the velocity distribution function, and $v_L$ is the velocity of 
the LMC relative to GC. 

Typically, this acceleration from dynamical friction is 10\% to 15\% of the acceleration from the
MW on the LMC.  Our approach ignores the factor of 100 smaller acceleration from dynamical friction 
on the MW by the LMC.

It is standard to approximate the integral in eq.~\ref{eq: dvdt} by
\begin{equation}
\int_0^{v_L} v^2 f_{MW} dv = {\rm erf(\xi)} - { 2 \xi \over \sqrt{\pi} } e^{-\xi^2} ~ ,
\label{eq: int}
\end{equation}
where $\xi = v_L / (\sqrt{2} \sigma)$ and $\sigma$ is the one-dimensional velocity dispersion of
the MW halo \citep{gomez2015,jethwa2016}. For an NFW profile, $\rho_{MW}$ and $\sigma$ can be 
expressed as:
\begin{equation}
\rho_{MW} = {\rho_h \over x_h (1 + x_h)^2 } 
\label{eq: rho}
\end{equation}
and
\begin{equation}
\sigma = 1.4393 ~ v_{max} \left ( { x_c^{0.354} \over 1 + 1.1756 ~ x_c^{0.725} } \right ) ~ 
\label{eq: sigma}
\end{equation}
where $\rho_h = M_h / 4 \pi r_h^3$, $x_h = r / r_h$, $x_c = r_{c,max} / r_h$, 
$r_{c,max} = 2.16258 ~ r_h$ is the radius of maximum circular velocity, 
$v_{max} = (G M(r_{c,max}) / r_{c,max})^{1/2} $ is the maximum circular 
velocity at $r = r_{c,max}$, and $M(r_{c,max})$ is the mass contained within $r_{c,max}$ 
\citep[e.g.,][]{zentner2003,gomez2015,jethwa2016}. Setting
\begin{equation}
M(r) = M_h \left [ {\rm ln} \left ( { r + r_h \over r_h } \right ) - \left ( { r \over r + r_h } \right ) \right ] ~ 
\label{eq: mhalo}
\end{equation}
yields the mass contained inside $r$ for an adopted $M_h$ and $r_h$ \citep{nav1996,nav1997}. 

To test our algorithm, we conducted a series of tests designed to reproduce published results.
Our solutions for the separation of the LMC and MW, $r_{LMC}(t)$, as a function of the masses 
and structural parameters follow the trends in Fig.~1 of \citet{gomez2015}, who derive the 
motion of the LMC for a broad range of MW and LMC masses. For our choice of $M_h$ and $M_L$,
the trajectory of the LMC across the sky in $(l, b)$ matches the trajectory in Fig.~1 of 
\citet{boubert2016}, who adopt an LMC orbit from the calculations of \citet{jethwa2016}. 

For an ensemble of $10^7$ HVSs traveling through a time-varying MW+LMC potential, we select
stars using our standard prescription. Once a star is placed at $r$ = 1.4~pc with velocity
\v0\ and ejection angles \thz\ and \phiz, we use a look-up table to place the LMC at its
expected position at a time $t$ = $t_{ej}$. The LMC is then at a position 
$\delta t = t_{obs} - t_{ej}$ {\it earlier} than its current position relative to the GC.
As we integrate the orbit of the star through the MW+LMC potential, we update the LMC position 
every time step, until the LMC reaches its current position at $t$ = $t_{obs}$. In this way,
every star travels through a time-varying potential, with the initial LMC position set by
$t_{ej}$.

In these calculations, we ignore the changing velocity of the MW relative to the center-of-mass.
Over the 350~Myr main sequence lifetime of the B-type stars in our model, the velocity of the
MW relative to the center-of-mass changes by roughly 20~\kms. Compared to typical ejection
velocities of 700~\kms\ to more than 1000~\kms, this difference is small. Among known HVSs,
travel times from the GC to the halo are 50--250~Myr \citep{brown2014}. During this time frame,
the velocity of the MW changes by only 10~\kms. This velocity is comparable to typical errors 
in the radial velocity and much smaller than typical errors in the tangential velocity 
\citep[e.g.,][]{brown2009b,brown2012a,brown2014,brown2015a,marchetti2018b,hattori2018b}. 

Overall, the moving LMC has little impact on distributions of HVS deflections. At 80--160~kpc,
the magnitude of typical and extreme deflections as a function of $\phiz$ is similar to that
in models with a fixed LMC at its current position (Fig.~\ref{fig: lmc7}, lower panel). At any
$\phiz$, the largest $\delta \phi$ is roughly 1\deg\ smaller; typical deflections are nearly
identical. Among closer stars with $r$ = 40--80~kpc, maximum deflections are nearly 2\deg\ smaller;
typical deflections are less than 1\deg\ different from those with a stationary LMC
(Fig.~\ref{fig: lmc7}, upper panel). Within both samples, a moving LMC fills a larger volume 
throughout the simulation and therefore deflects a larger percentage of stars. Compared to 
calculations with a stationary LMC, more HVSs have typical deflections in this simulation. 
Correspondingly fewer have the minimum deflection.

This situation repeats for stars at 10--40~kpc (Fig.~\ref{fig: lmc8}). Although the maximum 
and typical deflections are smaller, more HVSs experience significant deflections. Among
bound and unbound HVSs, there is a larger percentage of clearly non-radial orbits. Despite 
these clear differences, a moving LMC has little impact on the vast majority of stars in 
these samples (within the red contours in the figure). Most stars have modest deflections
which range from a few degrees for stars ejected into the Galactic plane or into the 
Galactic pole to 10\deg--30\deg\ for stars ejected with \phiz\ $\approx$ 10\deg\ to 40\deg.
While the gravity of the LMC deflects these stars, its motion has little impact.

\subsection{Summary}
\label{sec: res-summ}

Within a MW+LMC potential, HVSs ejected from the GC deviate significantly from radial orbits.
On its own, the gravity of the disk deflects the trajectories of HVS moving near the Galactic
plane (Figs.~\ref{fig: disk1}--\ref{fig: disk2}). For HVSs with $\phiz \lesssim$ 30\deg, 
typical deflections range from $\delta \phi \approx$ 5\deg\ for unbound stars to $\delta \phi$
$\approx$ 20\deg\ to 30\deg\ for bound stars (Figs.~\ref{fig: disk3}--\ref{fig: disk4}). 
Although stars ejected into higher Galactic latitudes experience a factor of 2--3 smaller 
$\delta \phi$, all HVSs are deflected towards the disk. Within a large ensemble of HVSs,
more than 60\% have final Galactic latitude $b_f \lesssim$ 30\deg.

The LMC adds another source of asymmetry to the potential. With a nominal mass larger than the 
MW disk mass, the LMC slows down HVSs faster than the disk and generates larger deflections 
with respect to the initial, purely radial motion. Changes to the radial trajectories are 
fairly independent of the initial angle of ejection relative to the $x-z$ plane: most HVSs 
have $\delta \phi \approx$ 2\deg\ to 5\deg, a factor of 2--3 larger than in a system with 
no LMC.

In these examples, the position and large mass of the LMC are responsible for the shape of 
the $\phi_0$-$\delta \phi$ relation.  Large deflections always occur for stars ejected with 
small \phiz; moving the LMC closer to the disk tends to make these deflections larger. Stars 
with larger \phiz\ develop significant non-radial motions when the LMC lies somewhere near 
their nominal trajectory. An LMC along the $+z$-axis deflects stars with $\phiz \gtrsim$ 60\deg, 
but not those with $\phiz \lesssim -60$\deg.  Similarly, placing the LMC at its current position 
changes the trajectories of stars with $\phiz \approx$ 0\deg\ to $-60$\deg\ and 
$\thz \approx$ $-135$\deg\ to $-25$\deg\ much more than stars ejected into other quadrants of 
the Galaxy.

For any LMC position, reducing (increasing) the mass of the LMC leads to smaller 
(larger) peaks at $\phi_0 \approx$ $-15$\deg\ and at $\phi_0 \approx$ $75$\deg. 
Although changing the LMC scale length $r_L$ changes \vfinal\ and \rfinal\ for HVSs 
in the outer halo, $r_L$ has little impact on the magnitude or the shape of the 
$\phi_0$--$\delta \phi$ relation. For peaks at $-15$\deg\ and $+15\deg$, the 
distance of the LMC is large compared to the scale length. The magnitude of the 
deflections then depends only on the LMC mass. When stars are ejected along the
$+z$-axis, the scale length has a modest impact on the height and shape of the peak at
$+75$\deg. However, factor of two changes in $r_L$ produce much smaller variations in
the peak than factor of two changes in the LMC mass.

\section{OBSERVATIONAL DISCRIMINANTS}
\label{sec: obs}

To develop observational predictions from these calculations, we consider subsets of
the $10^7$ stars in calculations with and without an LMC analog. Selecting 3~\msun\ stars 
with $d \le$ 100~kpc, absolute $g$-band magnitudes $M_g \approx$ 0 \citep{bress2012}, 
and $g \le$ 20 yields a sample accessible with large ground-based optical telescopes 
\citep[e.g.,][]{brown2005,brown2009b,brown2013}. For this group of stars, we consider 
how the radial space density, sky surface density, and distributions of the deflection 
angle and the radial and tangential velocity distinguish MW potentials with an LMC from 
those without an LMC.

Among all of our calculations, the radial variation of the space density of bound and 
unbound stars is nearly independent of the presence of the LMC.  As in Fig.~\ref{fig: rho1}, 
stars outside (inside) 80~kpc are mostly unbound (bound to the Galaxy). Within 40~kpc, the 
relative numbers of unbound stars and bound outliers to halo-like stars are independent of 
the LMC mass. In models with and without the LMC, the relative density of bound outliers
(unbound stars) peaks at 70--75~kpc (100--105~kpc). 

The fraction of stars in the disk and halo is also independent of the physical properties of 
the LMC. For halo stars and bound outliers, roughly 33\% have $|b_f| \ge$ 30\deg. The fraction 
of stars with $|b_f| \ge$ 30\deg\ grows to 43\% among unbound stars, where deflections from
the disk or the LMC are smaller.  Despite isotropic ejections from the GC, most bound HVS 
lie close to the Galactic plane. Unbound HVSs are distributed more isotropically.

Despite the high concentration at low Galactic latitude, HVSs in the pure MW calculations
are otherwise distributed rather uniformly in space and velocity (Table~\ref{tab: kin-stats}). 
For the entire sample of stars, the average/median position and velocity is consistent with
zero. The dispersion and inter-quartile range is roughly 20~kpc in each coordinate.  The 
typical dispersion of 175~\kms\ (150~\kms) in $x, y$ ($z$) yields a 3D velocity dispersion 
of roughly 300~\kms, which is only twice the typical radial velocity dispersion of halo stars 
in the outer Galaxy \citep[e.g.,][]{brown2010a,kafle2012,king2015,cohen2017}. The smaller 
velocity dispersion in the $z$ direction is a measure of the influence of the disk potential.

In calculations that include the LMC potential, the distribution of stars is less isotropic. 
The positional centroid of the population shows a clear displacement of stars towards the LMC. 
The offset in velocity is smaller but still towards the adopted position of the LMC. Despite
these differences, the velocity dispersion in LMC models is nearly identical to that in pure 
MW models.

These differences are also apparent in the specific angular momenta. Defining the cartesian
components $l_x = y \cdot v_z - z \cdot v_y$, $l_y = z \cdot v_x - x \cdot v_x$, and
$l_z = x \cdot v_y - y \cdot v_x$, stars initially on radial trajectories from the GC have
a specific angular momentum close to zero. After passing through the MW potential, the ensemble 
of $10^7$ HVSs in a MW only model still has average and median $(l_x, l_y, l_z)$ close to zero.
The nearly identical large values of $|l_x|$ and $|l_y|$ result from the large distances of 
HVSs from the GC.  The much smaller $|l_z|$ is a consequence of the symmetry of the potential 
relative to the Galactic poles.

Adding the LMC into the potential significantly changes these results.  For the full set 
of HVSs, the average and median $(l_x, l_y, l_z)$ in Table~\ref{tab: kin-stats} show the
large impact of the LMC.  Considering the absolute values, the average and median specific 
angular momenta illustrate the ability of the LMC to impart a rotational component to the 
velocities of bound HVSs.

Among unbound stars with $r \gtrsim$ 80~kpc in the MW only potential, typical angular momenta 
are somewhat smaller than those listed in the Table.  Stars at larger distances in the MW+LMC 
calculations have larger specific angular momenta.  Beyond 80~kpc, average and median values 
for $(|l_x|, |l_y|, |l_z|)$ are roughly a factor of two larger than those for the entire ensemble. 
Although the trajectories of unbound stars are deflected less than those of bound stars, their 
larger distances conspire to produce rather larger specific angular momenta.

The sky surface density of unbound stars and bound outliers shows some of these features
(Fig.~\ref{fig: allsky1}). In the top panel, a map derived from models with no LMC shows 
(i) a strong concentration of stars towards the disk superimposed on (ii) a roughly 
axially symmetric distribution of stars centered on the GC.  Although HVSs are ejected 
symmetrically from the GC, the gravitational potential of the disk bends trajectories 
towards the Galactic plane.  Otherwise, stars are fairly isotropically distributed about
the GC. 

Results for calculations with the moving LMC analog are obviously different. In addition to
a strong concentration of stars towards the disk midplane and the GC, there is a clear
enhancement in the surface density towards the LMC at $(l, b)$ = ($-80$\deg, $-33$\deg). 
Although the contours in the northern Galactic hemisphere resemble those in the MW only 
map, those in the southern Galactic hemisphere are more distorted and less symmetrical relative
to $l$ = 0\deg. These differences show the large impact of gravitational focusing, where the
LMC bends the trajectories of HVSs around it. The extra concentration of stars towards the 
LMC is responsible for the shift in the median position of HVSs from the GC towards the LMC.

To examine the surface density enhancement in more detail, we consider the number of stars in 
10\deg\ $\times$ 10\deg\ boxes separated by 5\deg\ intervals (Fig.~\ref{fig: allsky2}).  In 
the left panel, the curves plot the number of stars as a function of Galactic longitude $l$ 
for boxes centered at $b$ = $+33$\deg\ and $b$ = $-33$\deg. In the MW only models (purple and
blue curves), the number of stars is independent of $b$. Remarkably, the number of stars at
$b$ = $+33$\deg\ in the MW + LMC model (green curve) follows the MW only model very closely. 
In contrast, there is a clear factor of 2--3 enhancement in stars at $l$ = $-120$\deg\ to 
$-60$\deg\ along the $b$ = $-33$\deg\ track (orange curve). The half width of the enhancement 
is roughly twice the LMC scale length. 

The right panel of Fig.~\ref{fig: allsky2} illustrates the variation of surface density 
along a line of constant $l$.  For MW only models, 
the number of stars as a function of $b$ is symmetric with $l$: the track for 
$l$ = $+80$\deg\ (purple curve) closely follows the one for $l$ = $-80$\deg\ (blue curve). 
Adding in the LMC has little impact on the track for $l$ = $+80$\deg\ (green curve). Along
$l$ = $-80$\deg\ (orange curve), however, the enhancement in stars around the LMC is clear. 
Here, the half width of the enhancement is also roughly twice the LMC scale length. 

To explore the velocity differences between the calculations, we compare the distributions of 
radial and tangential velocities for unbound stars and bound outliers along lines-of-sight 
towards the LMC and other directions through the Galaxy. Fig.~\ref{fig: vhist1} illustrates 
results for $(l, b)$ = $(-80, -30)$ in the lower panels and for $(l, b)$ = $(+80, +30)$ in 
the upper panels. Along a line-of-sight towards the LMC, the distributions of $v_r$ for a
MW only model (2000 stars, purple histogram) and a MW+LMC model (3700 stars, green histogram) 
are clearly different: models with the LMC have systematically smaller $v_r$ than those 
without the LMC. Using a K--S test \citep{press1992}, the two $v_r$ distributions have a 
small chance, $\lesssim 10^{-20}$, of arising from the same parent population.

The distributions of $v_t$ towards the LMC are also different (Fig.~\ref{fig: vhist1},
lower right panel). In the MW only model (purple histogram), the tangential velocity peaks
at a smaller value ($\sim$ 40--60~\kms) than in the MW+LMC model (green histogram, peak
at 80--100~\kms). The K--S test again predicts a small likelihood, $\lesssim 10^{-20}$,
that the two distributions have a common parent. 

The top panels of Fig.~\ref{fig: vhist1} compare distributions for a line-of-sight on the
opposite side of the GC from the LMC. The $v_r$ and $v_t$ distributions look identical;
K--S tests yield 20\% to 40\% probabilities that the samples are drawn from the same 
parent population.

For other lines-of-sight through the Galaxy, there is a clear correlation between the 
results of K--S tests and $(l, b)$. At high latitudes with $|b_f| \gtrsim$ 50\deg, the
LMC has little impact on the $v_r$ and $v_t$ distributions. For the MW only and MW+LMC 
samples, the averages/medians differ by less than 10~\kms; the corresponding dispersions
or inter-quartile ranges are typically 150--200~\kms. Based on K--S tests, the
distributions have a high probability of selection from a common parent population.

At lower latitudes, the distributions of $v_r$ and $v_t$ show significant differences.
For a specific $l$, the K--S probability systematically grows from roughly 
$10^{-5} - 10^{-4}$ at $|b_f| \approx$ 40\deg\ to $\lesssim 10^{-20}$ at 
$|b_f| \approx$ 0\deg--10\deg. When $|b_f| \approx$ 40\deg, the typical $v_t$ in models
with an LMC is a few per cent larger than models with no LMC. The typical $v_r$ is 
correspondingly smaller. Near the plane of the disk, the difference in $v_r$ and $v_t$
rivals the differences for the line-of-sight that includes the LMC shown in
Fig.~\ref{fig: vhist1}. 

Fig.~\ref{fig: vhist2} illustrates results for two lines-of-sight well away from the 
current position of the LMC. In the upper panels, there is little difference between
the distributions for MW only (purple histograms) and MW+LMC models (green histograms). 
The peaks and overall
shapes of the distributions look identical within counting statistics. A K--S test
confirms that the probability that the distributions are drawn from the same parent
population is 20\% to 30\%. In the lower panel, the distributions start to show the
tell-tale signature of the LMC potential: a somewhat smaller set of radial velocities
and a somewhat larger set of tangential velocities. In this example, the K--S test
yields a $1 - 3 \times 10^{-5}$ probability that the two sets of velocities are drawn from
the same parent. 

The physical origin for the variation of K--S probability with $b_f$ follows from
sec. 4.2. For stars ejected into high Galactic latitudes, an LMC fairly close to 
the Galactic plane only slightly deflects the trajectories of unbound stars and 
bound outliers. Once these stars reach 50--100~kpc, most of their motion is radial;
$v_t$ is negligible. Thus, any differences in the $v_t$ distributions are small and
difficult to measure quantitatively. Although high latitude HVSs traveling out of 
the MW+LMC potential typically reach somewhat smaller distances with somewhat smaller 
$v_r$, it is challenging to measure this difference.  Within an ensemble of HVSs, 
this feature of LMC models moves the lowest velocity stars among the bound outliers 
(unbound stars) into the halo-like (bound outlier) population. Because they have
the highest velocities and are decelerated the least, stars remaining among the 
bound outliers and unbound stars have fairly similar radial velocity distributions
dominated by the (rare) highest velocity ejections. Thus, the populations are
indistinguishable.

When HVS are ejected at low Galactic latitudes, the disk always deflects them from
radial orbits. The gravity of the LMC magnifies these deflections. Stars with
$|b_f| \lesssim$ 40\deg\ are first deflected towards the disk by the disk potential
and then by the LMC. Compared to stars at higher latitudes, these stars end up with 
larger $v_t$ and smaller $v_r$. With larger deflections at smaller $|b_f|$, the
differences in the distributions of $v_r$ and $v_t$ grow with decreasing $|b_f|$.

In these examples, the K--S tests require samples of a few hundred stars along most
line-of-sight to yield probabilities $\lesssim 10^{-20}$. Samples of 20--30 stars 
in areas that subtend 10\deg\ $\times$ 10\deg\ on the sky yield smaller confidence 
levels, $\lesssim 10^{-4}$.  Testing differences among different potential models 
thus requires $\sim 10^4$ unbound HVSs over the sky. 

To conclude this section, we examine the ability of 6D position and velocity information
to measure the deflections of HVS from their original radial orbits. For stars with
observed position \rvf\ and velocity \vvf, the angle between these two vectors is
\begin{equation}
{\rm cos}~\gamma = \vec{r}_f \cdot \vec{v}_f / (r_f ~ v_f) ~ .
\label{eq: gamma}
\end{equation}
Stars on purely radial orbits have $\gamma$ = 0\deg; disk stars orbiting the GC have
$\gamma \approx$ 90\deg.

Despite our inability to measure $\delta \phi = \phi_0 - \phi_f$ directly, $\gamma$
is an excellent proxy for $\delta \phi$. For nearby stars ($d$ = 10--40~kpc) on their 
first pass out through the MW only potential, $\gamma$ ranges from roughly zero at 
$b_f \approx$ 0\deg\ to a broad maximum of 10\deg\ to 20\deg\ at 
$b_f \lesssim$ 30\deg\ to roughly zero again at $b_f \approx$ 90\deg\ (see 
Figs.~\ref{fig: disk3}--\ref{fig: disk4}).  The potential of the disk deflects bound 
stars near the Galactic plane by 10\deg\ to 20\deg\ more than unbound stars ejected 
from the GC at larger velocities.  Bound stars on their way back to the GC have 
$\gamma \gtrsim$ 30\deg; many stars have $\gamma \approx$ 150\deg\ to 180\deg.  
The frequency of $\gamma$ is symmetric about the disk plane: stars with 
$b_f \approx$ 30\deg\ have the same distribution of $\gamma$ as stars with 
$b_f \approx -30$\deg.

Among more distant stars ($d$ = 40--160~kpc), the population is dominated by unbound stars
leaving the MW. With few bound stars returning to the GC, $\gamma$ is always rather small.
Close to the Galactic plane, $\gamma \lesssim$ 1\deg. At somewhat larger $b_f \approx$ 
10\deg\ to 30\deg, a few stars have $\gamma \gtrsim$ 3\deg. Near the Galactic poles,
$\gamma \lesssim$ 1\deg.

\begin{figure}
\includegraphics[width=6.5in]{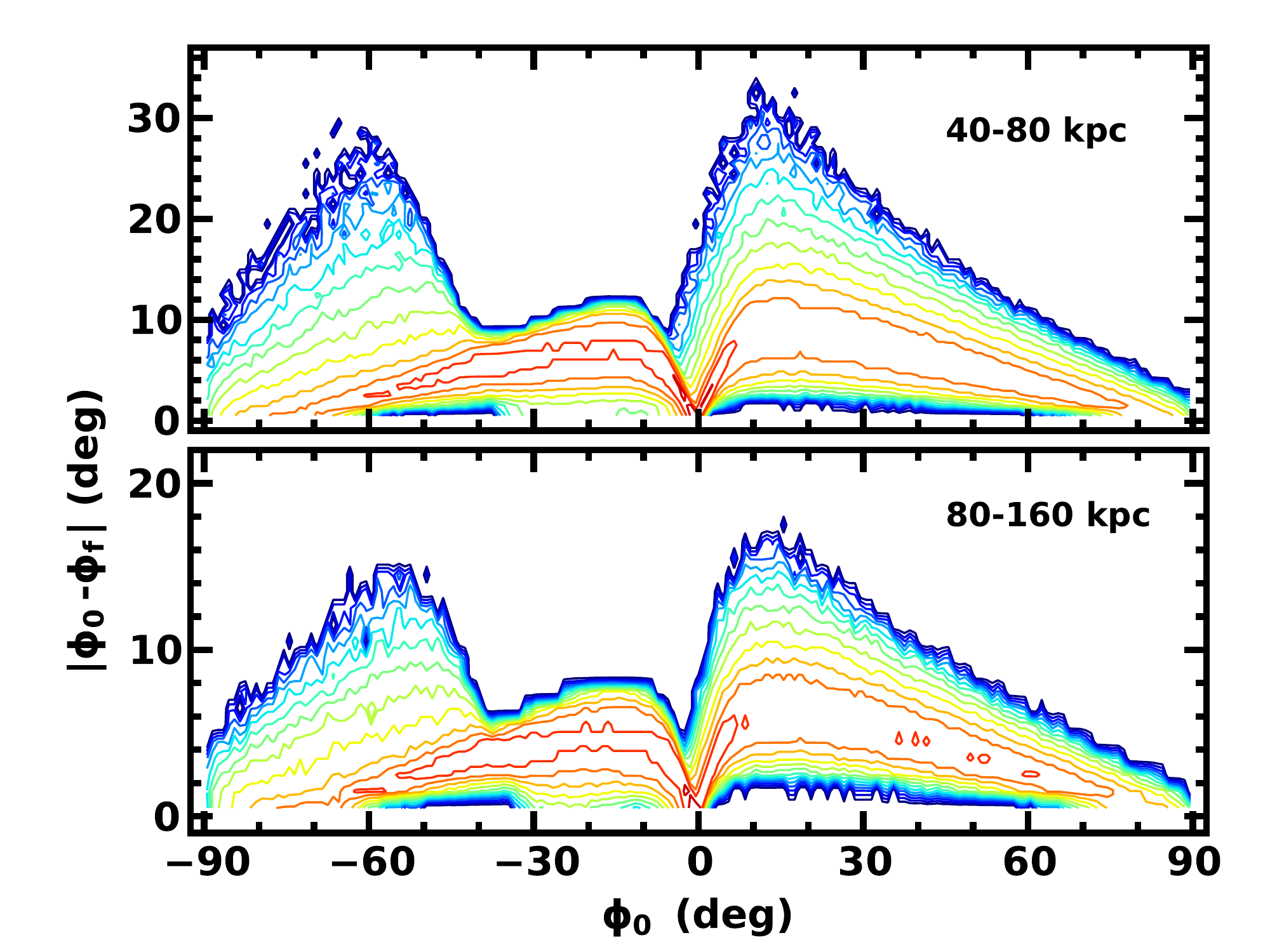}
\vskip 2ex
\caption{\label{fig: lmc7}
As in Fig.~\ref{fig: lmc5} for an LMC analog whose
distance varies in time as outlined in the text.
}
\end{figure}

\begin{figure}
\includegraphics[width=6.5in]{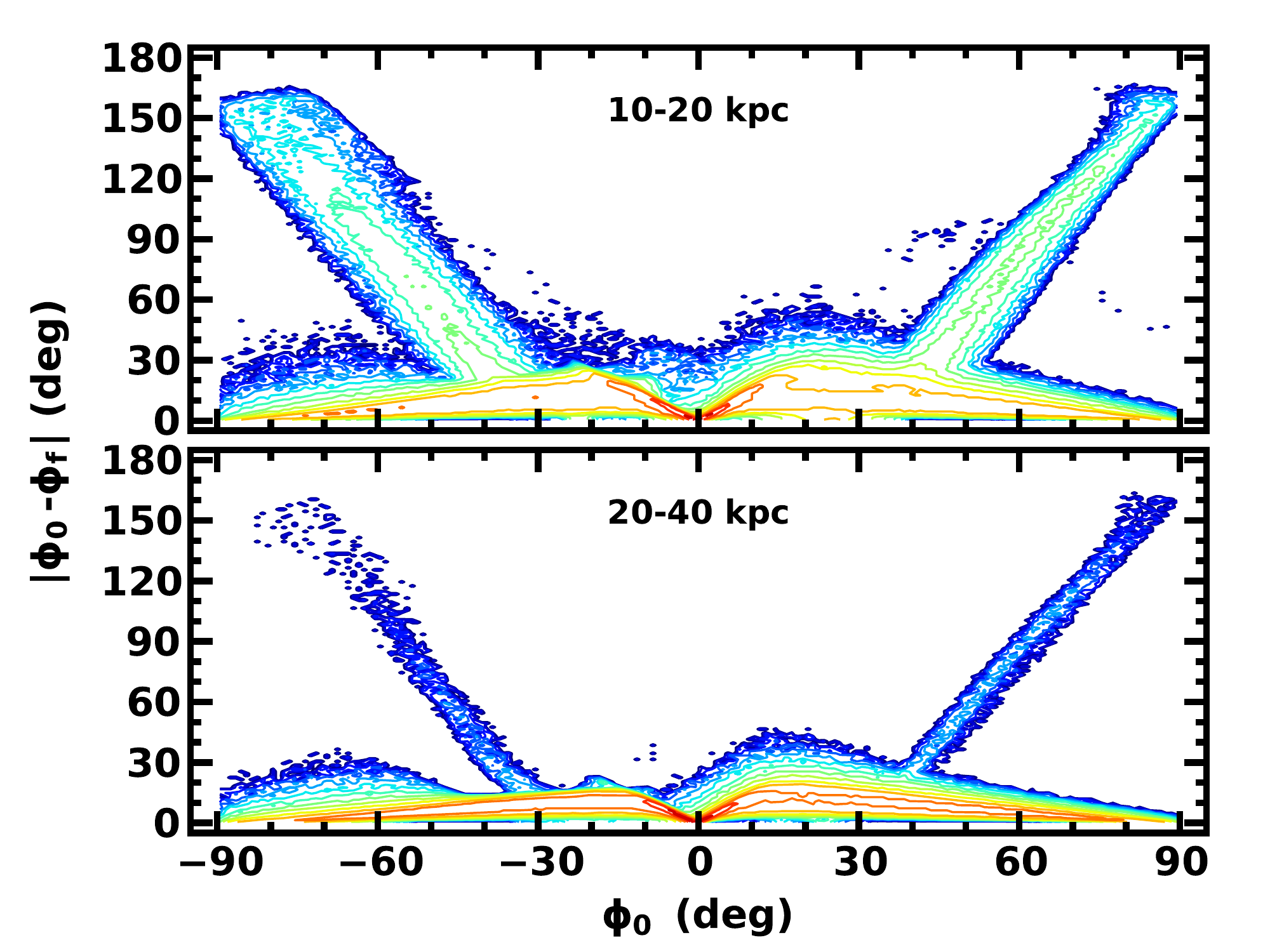}
\vskip 2ex
\caption{\label{fig: lmc8}
As in Fig.~\ref{fig: lmc6} for an LMC analog whose
distance varies in time as outlined in the text.
}
\end{figure}

\begin{figure}
\includegraphics[width=6.5in]{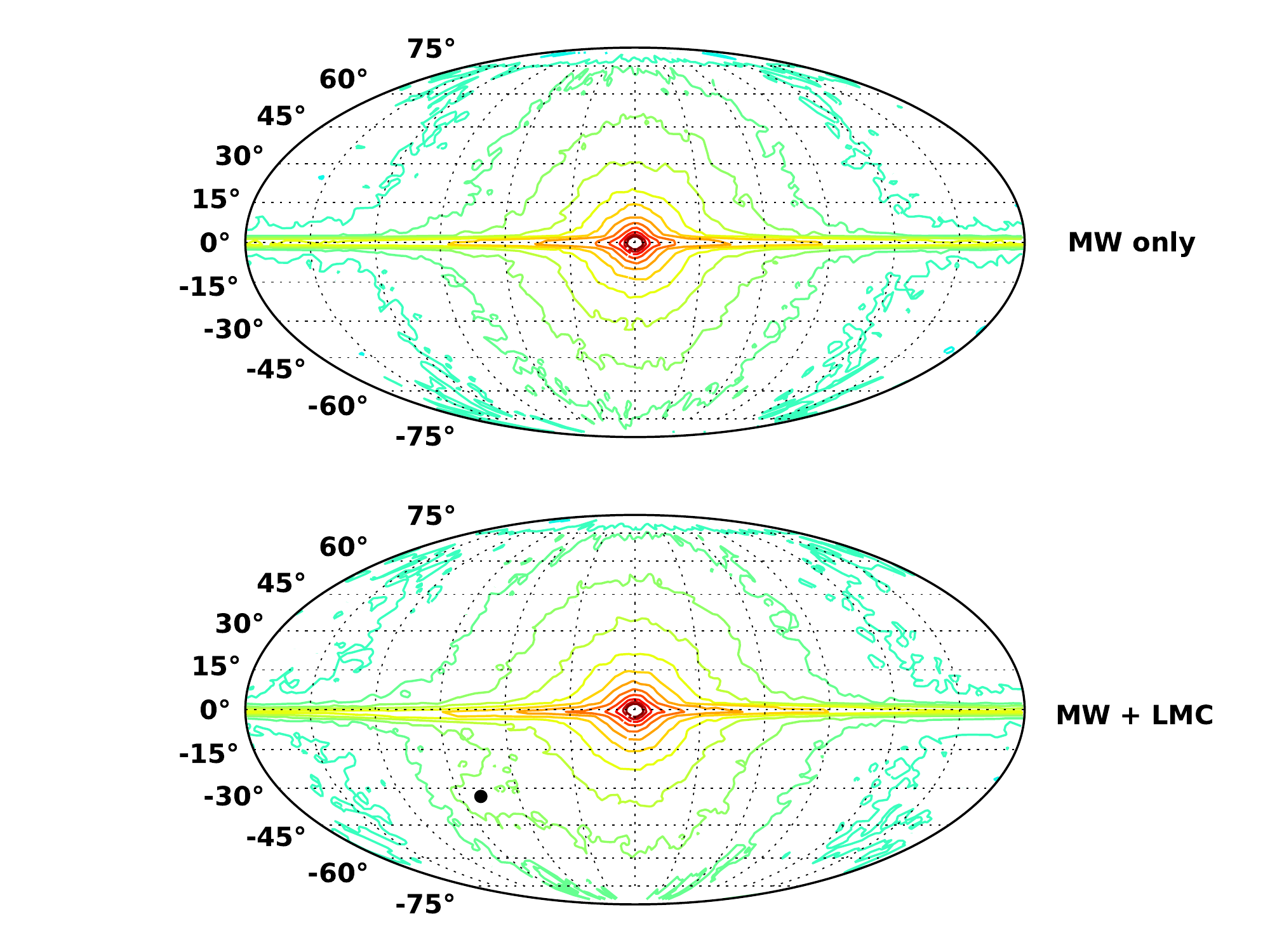}
\vskip 2ex
\caption{\label{fig: allsky1}
Contour map of surface density for unbound stars and bound outliers
in Galactic coordinates. The y-axes are labeled with Galactic latitude.
The longitude runs from $-180$\deg\ at the left to $+180$\deg\ at the right.
Upper panel: results for MW-only models.  Lower panel: results for 
MW + LMC models.  Aside from the obvious concentration of stars in the disk, 
calculations with the LMC potential show an overdensity around the current 
position of the LMC (indicated by the black dot at $(l, b)$ = 
($-$79\degpoint5,$-$32\degpoint9)).
}
\end{figure}

\begin{figure}
\includegraphics[width=6.5in]{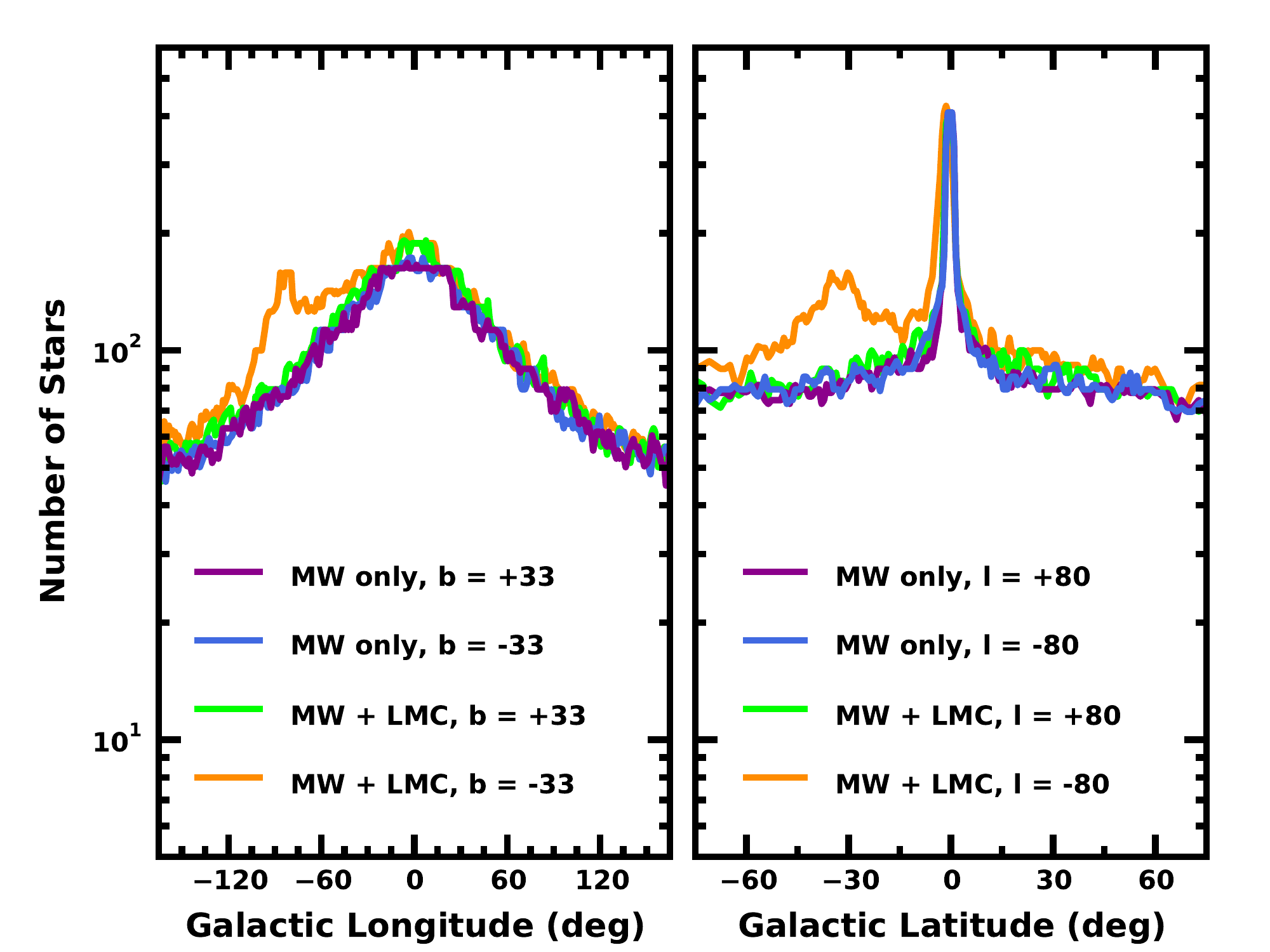}
\vskip 2ex
\caption{\label{fig: allsky2}
Stellar density in 10\deg\ $\times$ 10\deg\ regions as a function of
(a) $l$ for $b$ = $+33$\deg\ and $b$ = $-33$\deg\ (left panel) and 
(b) $b$ for $l$ = $+80$\deg\ and $l$ = $-80$\deg\ (right panel) for
MW-only and MW + LMC models as indicated in the legend. Each panel 
shows a large overdensity in the disk midplane and a more modest 
overdensity of stars in the direction of the LMC.
}
\end{figure}

\begin{figure}
\includegraphics[width=6.5in]{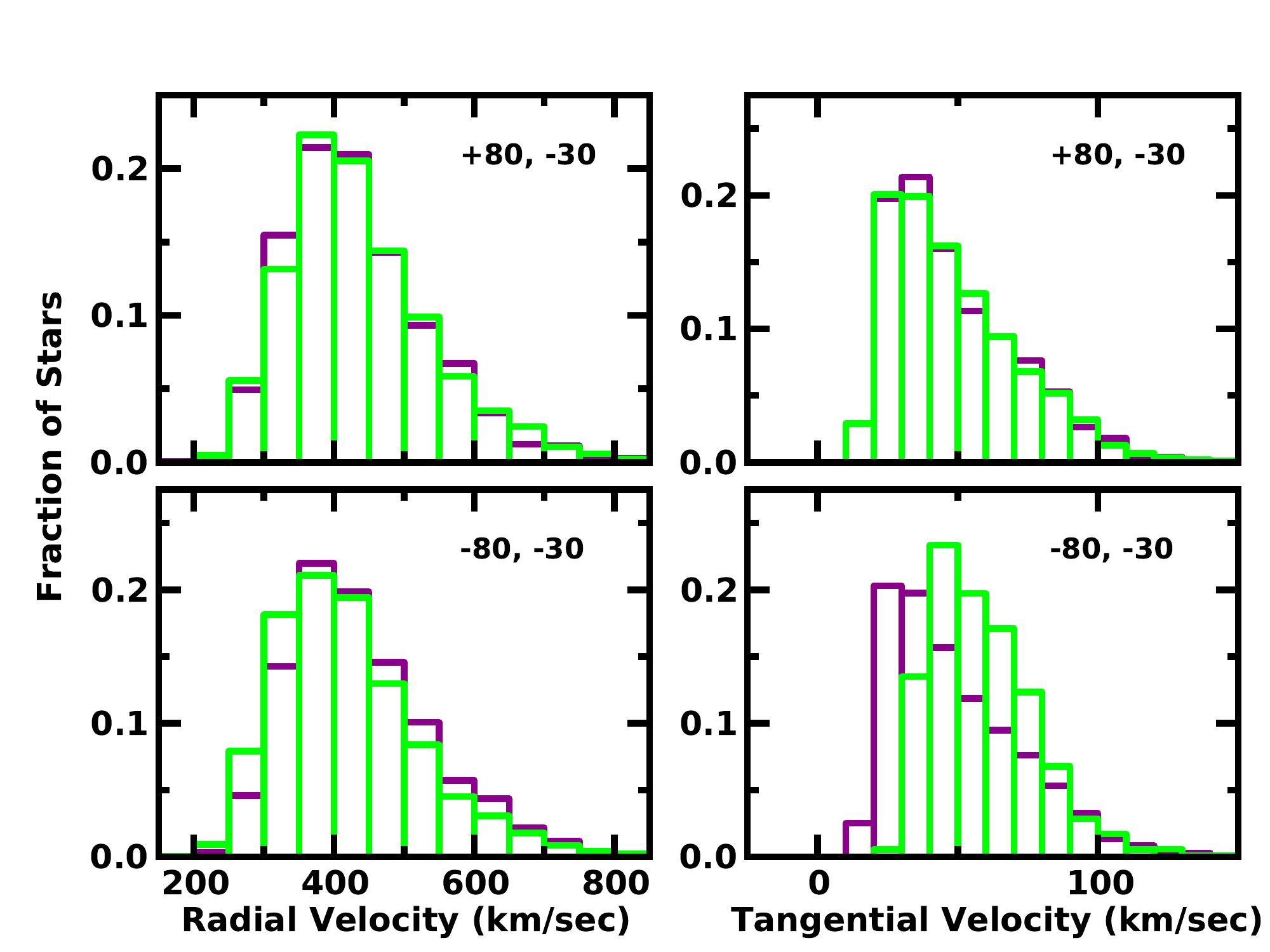}
\vskip 2ex
\caption{\label{fig: vhist1}
Distributions of $v_r$ (left panels) and $v_t$ (right panels) for unbound 
stars with $d$ = 40--160~kpc in 10\deg\ $\times$ 10\deg\ regions centered 
on the $(l, b)$ indicated in the legend for MW-only (purple) and MW + LMC
models (lime). Towards the LMC (lower panels), stars in the MW+LMC models
have smaller $v_r$ and larger $v_r$ than stars in the MW only models. On
the opposite side of the GC, MW only and MW + LMC models have identical 
distributions of $v_r$ and $v_t$. 
}
\end{figure}

\begin{figure}
\includegraphics[width=6.5in]{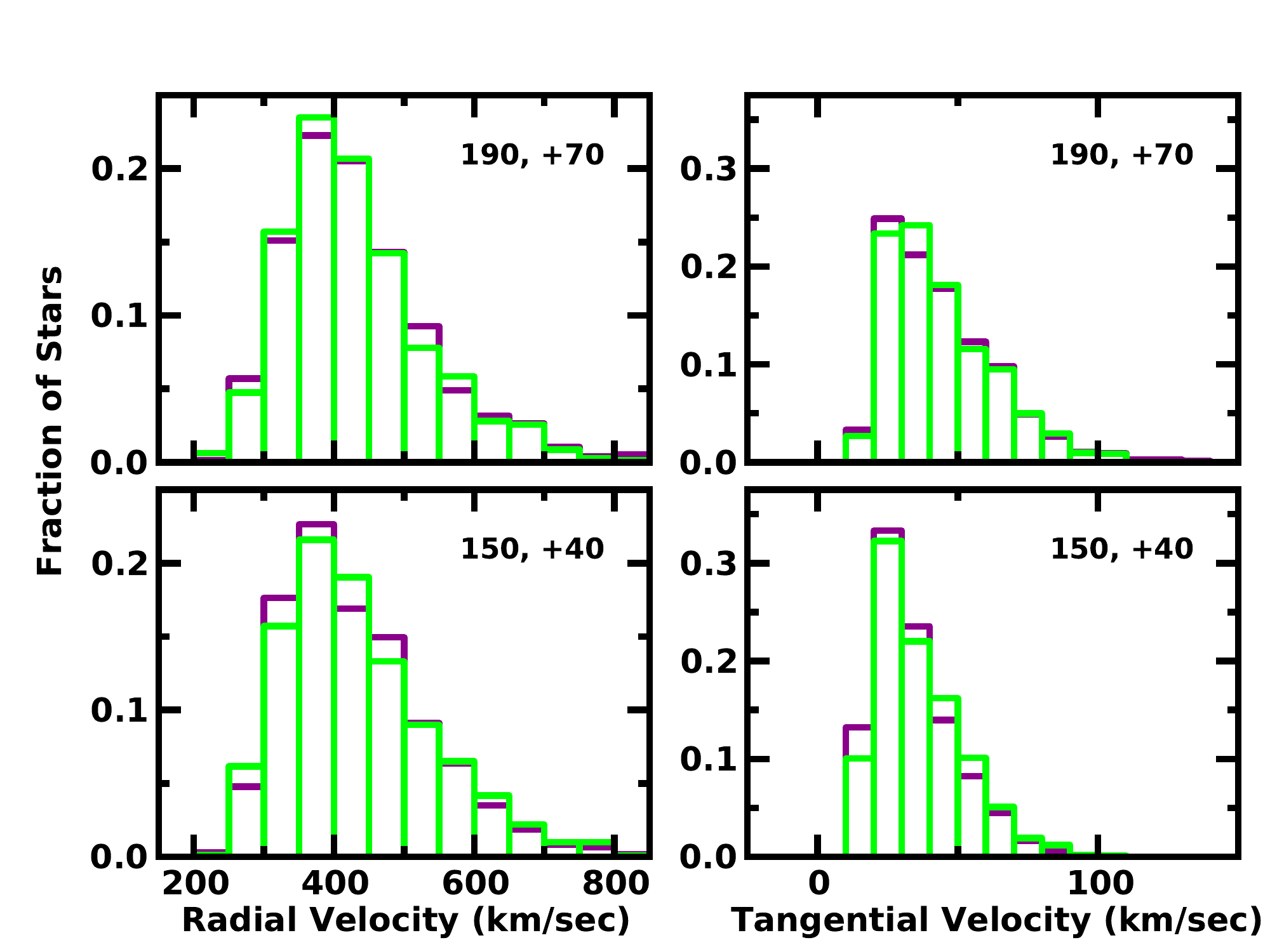}
\vskip 2ex
\caption{\label{fig: vhist2}
As in Fig.~\ref{fig: vhist1} for lines-of-sight in the direction of the
Galactic anti-center. At low latitudes (lower panels), stars in the MW+LMC
models have larger $v_t$ than stars in the MW only models; the distributions
of $v_r$ are nearly identical. At higher latitudes (upper panels), the
distributions of stars are indistinguishable. 
}
\end{figure}

In models with the LMC potential, the behavior of $\gamma$ with $b_f$ is more complicated
(Figs.~\ref{fig: lmc3}--\ref{fig: lmc8}). Nearby, the large population of bound stars 
generates a broad range of $\gamma \approx$ 0\deg\ to 180\deg, with a clear preference for
$\gamma \approx$ 10\deg\ to 30\deg\ in the direction of the LMC and close to the Galactic
plane. Near the LMC, gravitational focusing pulls stars on radial trajectories towards 
the LMC. Close to the disk midplane, the gravity of the LMC pulls stars across the disk.
After reaching apogalacticon, bound stars returning towards the GC have large 
$\gamma \approx$ 180\deg. As in the MW only models, these stars occupy the full range of 
possible $b_f$ with a modest overdensity at $|b_f| \lesssim$ 30\deg\ due to the general
overdensity of bound stars in the plane.

Compared to nearby stars in the MW + LMC potential, more distant unbound stars have 
relatively small $\gamma$. Near the Galactic poles ($b_f \approx$ 70\deg\ to 90\deg),
the gravity of the LMC simply decelerates these stars more rapidly than the MW on its own. 
Deflection angles are then small ($\gamma \lesssim$ 3\deg; see also Fig.~\ref{fig: lmc7}). 
Stars ejected towards the LMC and into the Galactic plane are deflected by much larger 
angles, $\gamma \gtrsim$ 5\deg. Few stars have much larger deflection angles.

\begin{figure}
\includegraphics[width=6.5in]{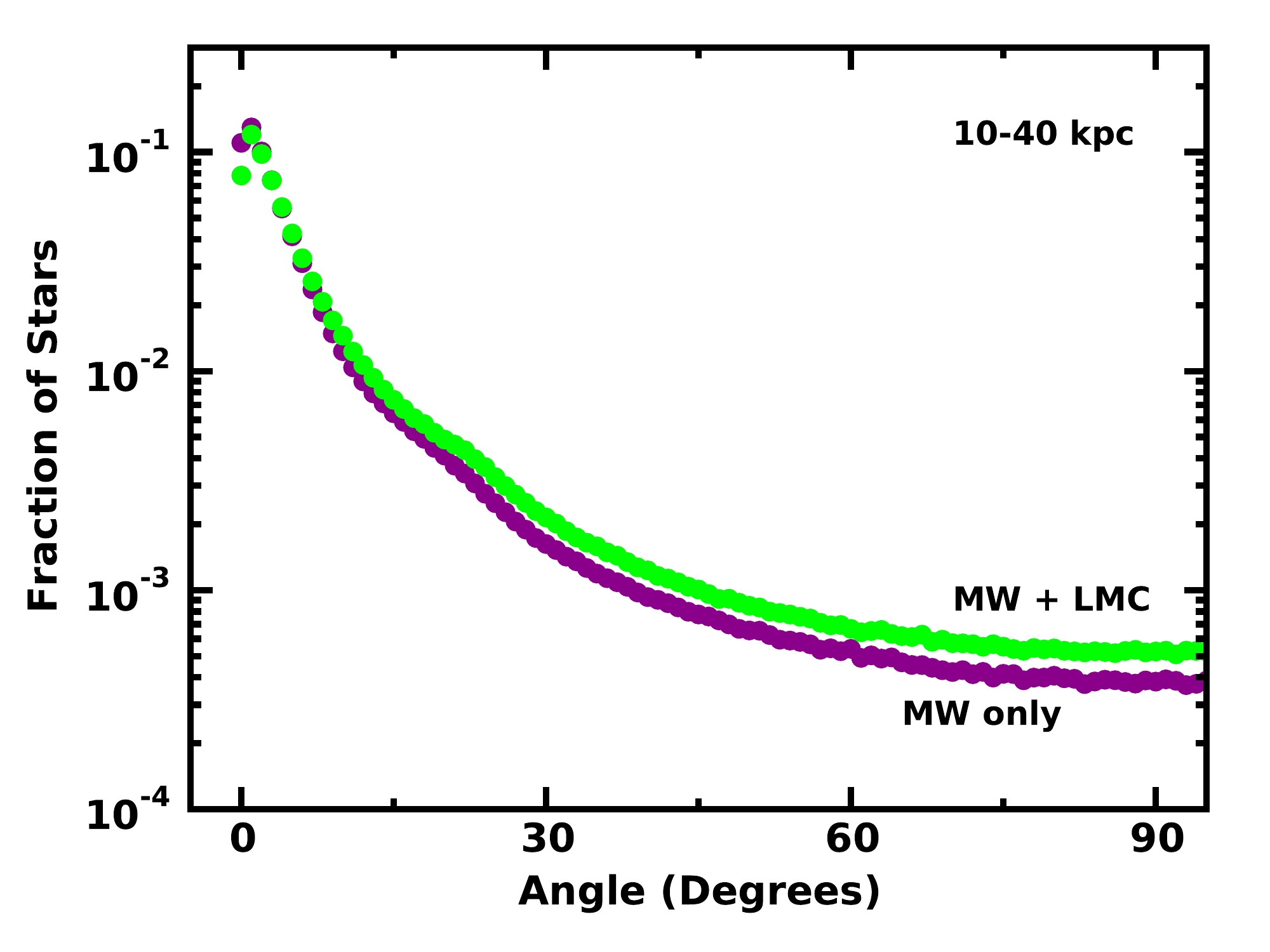}
\vskip 2ex
\caption{\label{fig: angle1}
Fraction of nearby stars ($d$ = 10--40~kpc) with an angle $\gamma$ between their final 
position and velocity vectors for MW only calculations (purple symbols) and MW + LMC
calculations (green symbols). Among nearby stars, the distribution of $\gamma$ is 
nearly independent of the underlying potential.
}
\end{figure}

\begin{figure}
\includegraphics[width=6.5in]{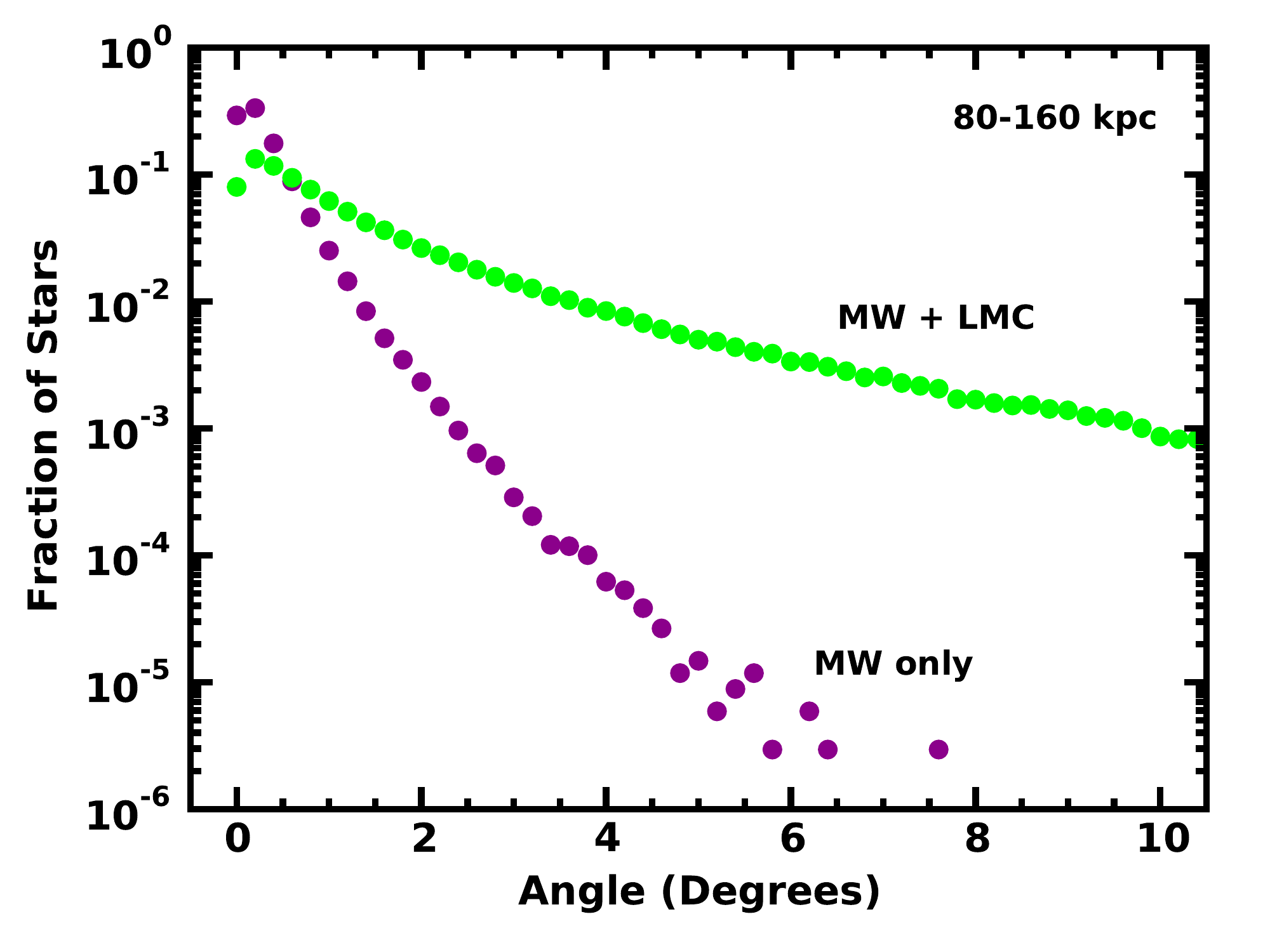}
\vskip 2ex
\caption{\label{fig: angle2}
As in Fig.~\ref{fig: angle1} for stars with $d$ = 80--160~kpc.  At large distances, 
unbound stars in a MW only potential lie on much more radial orbits than those in
a MW + LMC potential.
}
\end{figure}

To contrast results for the different potential models, 
Figs.~\ref{fig: angle1}--\ref{fig: angle2} show the distribution of $\gamma$ for the MW 
only model (purple symbols) and the MW + moving LMC model (green symbols).  Among stars
at 10--40~kpc (Fig.~\ref{fig: angle1}), the distributions are nearly identical. Results
for $\gamma \gtrsim$ 90\deg\ are approximately a mirror image of those at $\gamma \lesssim$
90\deg.  Overall, the MW only potential yields more stars with $\gamma \approx$ 0\deg\ and
180\deg; the MW+LMC models generate more stars with intermediate $\gamma$.  

At larger distances (80--160~kpc), the two distributions are clearly different 
(Fig.~\ref{fig: angle2}). Although the disk deflects a few stars traveling through the
Galactic plane by 2\deg\ or more, roughly 96\% of unbound stars in the MW only model 
have $\gamma \lesssim$ 1\deg; more than 99.5\% have $\gamma \lesssim$ 2\deg. Including
the LMC potential significantly reduces the number of unbound stars with negligible
deflections: only 56\% (75\%) have $\gamma \lesssim$ 1\deg\ (2\deg); roughly 12\% have
$\gamma \gtrsim$ 5\deg. 

Two factors cause the large differences in the distributions of $\gamma$ for unbound stars. 
In the MW + LMC potential, roughly 10\% of unbound stars pass within 2 $r_L$ (roughly 33\deg) 
of the center of the LMC and are deflected by several deg (Fig.~\ref{fig: lmc7}; 
Fig.~\ref{fig: allsky2}). Another 20\% of unbound stars are ejected within 12\deg\ of the 
disk midplane; the LMC pulls many of these across the midplane. Together, unbound stars
ejected into the midplane or passing close to the LMC comprise nearly all of the distant
stars with large $\gamma$ in Fig.~\ref{fig: angle2}.

Aside from the distinctive frequency distributions of $\gamma$ for complete ensembles 
of HVSs, there are clear differences in the distributions along most lines-of-sight through 
the Galaxy. Following the same procedure as for our analysis of the distributions for $v_r$ 
and $v_t$ in Figs.~\ref{fig: vhist1}--\ref{fig: vhist2}, we infer the median $\gamma$ as
a function of distance along various lines-of-sight. For nearby stars ($d$ = 10--40~kpc),
the median $\gamma$'s for MW only and MW+LMC models differ by 0\degpoint2 or less. Using a
K--S test, the two populations are almost always consistent with draws from the same 
underlying population. Among stars at 80--160~kpc, the median $\gamma$'s differ by as much
as 1\deg; K--S tests suggest the distributions of $\gamma$ are almost never consistent with
draws from the same parent population, with very low K--S probabilities of $\lesssim 10^{-15}$.

In these examples, all-sky samples of $\gtrsim 10^4$ unbound HVSs are required to demonstrate
that the distribution of $\gamma$ for stars along a particular line-of-sight in the MW only
potential is not drawn from the same parent population as the $\gamma$'s for stars ejected
into the MW+LMC potential. However, the distribution of $\gamma$ within smaller ($\sim 10^3$)
all-sky samples of unbound HVSs randomly selected from any of our calculations clearly differ
from one another and from samples of randomly generated stars on purely radial orbits from the
GC. Thus, it should be possible to distinguish some potential models from others with more
modest increases in the current sample of HVSs \citep[see also][]{gnedin2005,yu2007}.

Changing the LMC mass has little impact on these conclusions. In systems where the LMC mass 
is a factor of two smaller than our nominal mass, there is substantially less gravitational
focusing (Fig.~\ref{fig: lmc1}). The overdensity of HVSs in the direction of the LMC is then
roughly a factor of two smaller than shown in Fig.~\ref{fig: allsky2}. Because the LMC mass
has a limited impact on the velocities of HVSs, the overall shape of the histograms in 
Figs.~\ref{fig: vhist1}--\ref{fig: vhist2} is unchanged. However, the typical tangential
velocities are smaller and it is harder to distinguish the velocity distributions from those
with no LMC. Similarly, the typical angle $\gamma$ between the position and velocity vectors 
is smaller in calculations with a lower mass LMC. At 10--40~kpc (Fig.~\ref{fig: angle1}), the
differences are negligible. Among more distant stars (Fig.~\ref{fig: angle2}), the distribution
of $\gamma$ is roughly midway between those of the pure MW model and the nominal MW+LMC model.

Doubling the nominal mass of the LMC has a somewhat smaller impact on observables. Despite the 
larger gravitational focusing of a more massive LMC (Fig.~\ref{fig: lmc1}), stars passing 
at more than 2--3 LMC scale lengths still feel the gravity of the Galaxy more than the gravity 
of the LMC. Thus, the increase in the surface density of stars near the LMC is rather small,
$\sim$ 25\%.  In calculations with a heavy LMC, HVSs have larger deflection angles, larger
tangential velocities, and larger $\gamma$ than HVSs in the nominal MW+LMC model. Among stars
with $d$ = 10--40~kpc, these differences are negligible. For stars at larger distances, the
heavier LMC produces a shallower distribution of $\gamma$ with ejection angle 
(Fig.~\ref{fig: angle2}).

We conclude that there are robust observational measures that distinguish HVSs ejected from
MW's with and without an LMC companion. Aside from generating an overdensity of bound and
unbound stars in the general direction of the LMC, the gravity of the LMC modifies the 
distributions of $\gamma$, $v_r$, and $v_t$ along many lines-of-sight through the Galaxy. 
Although site lines towards the LMC and the Galactic plane are those most strongly affected,
the gravity of the LMC also impacts $\gamma$ the observed angle between the current position
and velocity of unbound stars at all $l$ and $b$.

\section{IDENTIFYING UNBOUND HYPERVELOCITY STARS}
\label{sec: id}

In previous studies \citep{bromley2006,kenyon2008,kenyon2014}, we have emphasized 
that the Hills mechanism ejects bound and unbound HVSs from the GC \citep[see 
also][]{rossi2014,rossi2017}.  Aside from generating unbound hyper-runaway stars, 
close interactions of massive stars and supernova explosions in close binary stars 
also primarily eject stars closely bound to the MW \citep[e.g.,][]{blaauw1961,pov1967,
leon1991,bromley2009,kenyon2014}. Nearly all of the bound stars have radial or tangential
velocities close to those of indigenous stars in the disk or halo. Identifying the
handful of extreme outliers in this population is often tedious \citep[e.g.,][]{brown2009b}.

To facilitate the development of robust observing strategies, it is useful to consider 
the ability of heliocentric observations in recovering unbound stars from one of our 
simulations. We focus on $v_r$ and $v_t$. For unbound stars with $v_f > v_e$, we derive
the fraction recovered from only one observable, either $v_r > v_e$ or $v_t > v_e$. We
also consider the fraction of unbound stars recovered from an ensemble of stars selected
by either $0.75 v_e < v_r < v_e$ or $0.75 v_e < v_t < v_e$. Our choice of the 0.75 factor 
is based on the observed velocity dispersion of bound HVS, which is only twice the local 
velocity dispersion of halo stars.  Choosing a smaller factor leads to a larger confusion 
between possible HVSs and true halo or thick disk stars \citep[for other approaches to
analyzing large sets of observed velocities, see][and references therein]{koll2009,li2012,
zhong2014,li2015,hawkins2015,favia2015,hattori2018b}.

\begin{figure}
\includegraphics[width=6.5in]{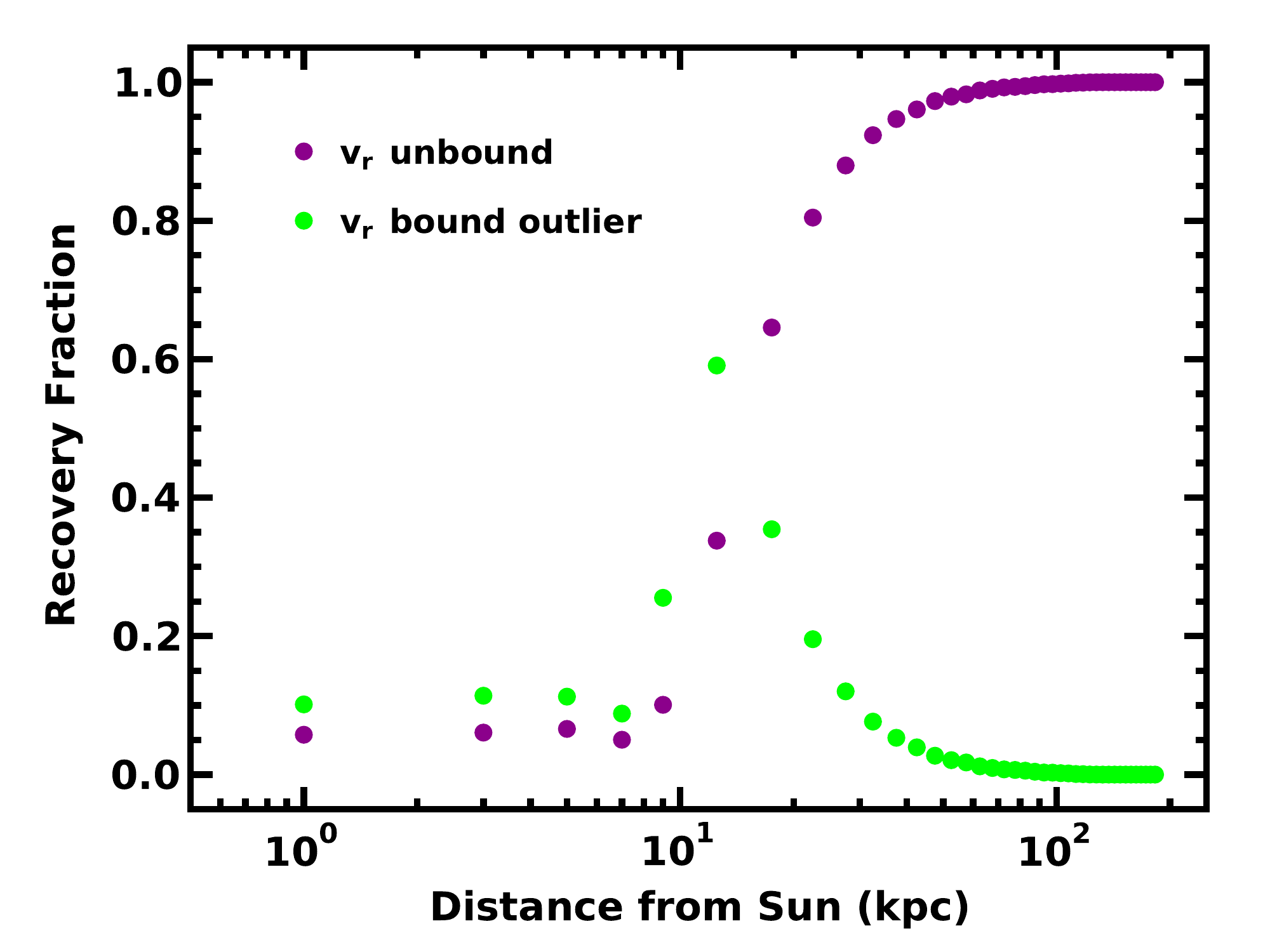}
\vskip 2ex
\caption{\label{fig: nvrvt1}
Recovery fraction as a function of distance from the Sun for unbound stars selected 
by radial velocity $v_r$.  Stars with $v_r > v_e$ (purple points) rarely reflect the 
true fraction of unbound stars at $d \lesssim$ 10--20~kpc. When $d \gtrsim$ 30--40~kpc, 
the number of stars with $v_r > v_e$ is very close to the true number of unbound 
stars. At $d \approx$ 8--20~kpc, stars identified as bound outliers ($v_r =$ 
0.75-1.00~$v_e$) include a large fraction of unbound stars. At smaller ($d \lesssim$
8~kpc) or larger ($d \gtrsim$ 20~kpc) distances, bound outliers are rarely unbound.
}
\end{figure}

For distant stars ($d \gtrsim$ 20~kpc), the radial velocity recovers the vast majority
of unbound stars (Fig.~\ref{fig: nvrvt1}). At these distances, HVSs on nearly radial 
orbits away from the GC have negligible proper motion and tangential velocity. The recovery
fraction grows from 60\%--70\% at $d \approx$ 20~kpc to nearly 100\% at $d \approx$ 100~kpc.

Requiring that the Galactic rest frame radial velocity exceed the local escape velocity 
underestimates the true fraction of unbound stars by a factor of 10--20 for nearby stars 
with $d \lesssim$ 10--15~kpc.  Among nearby HVSs on nearly radial orbits, the radial 
velocity is useful only for the fraction with motion directed right at or right away 
from the Sun (e.g., $l \approx$ 0\deg\ or 180\deg). This group is a small fraction of 
the total sample. For other stars, the radial component of the motion is a small 
fraction of the total motion; the radial velocity is then similar in magnitude to 
the typical velocity of other nearby stars.

Selecting stars with $0.75 v_e < v_r < v_e$ is an attractive way to identify nearby HVSs.
At $d \lesssim$ 8~kpc, a less restrictive constraint on $v_r$ identifies 1.5--2 times the
number of unbound stars. However, the recovery fraction is still small, $\lesssim$ 10\%.
Among stars with intermediate distances of 8--15~kpc, however, the $v_r$ selection samples 
a much larger fraction of solid angle on the sky and recovers 30\% to 60\% of unbound stars. 
At still larger distances, the motion becomes more purely radial: stars that are bound based
on $v_r$ have little tangential velocity and are truly bound to the MW.

\begin{figure}
\includegraphics[width=6.5in]{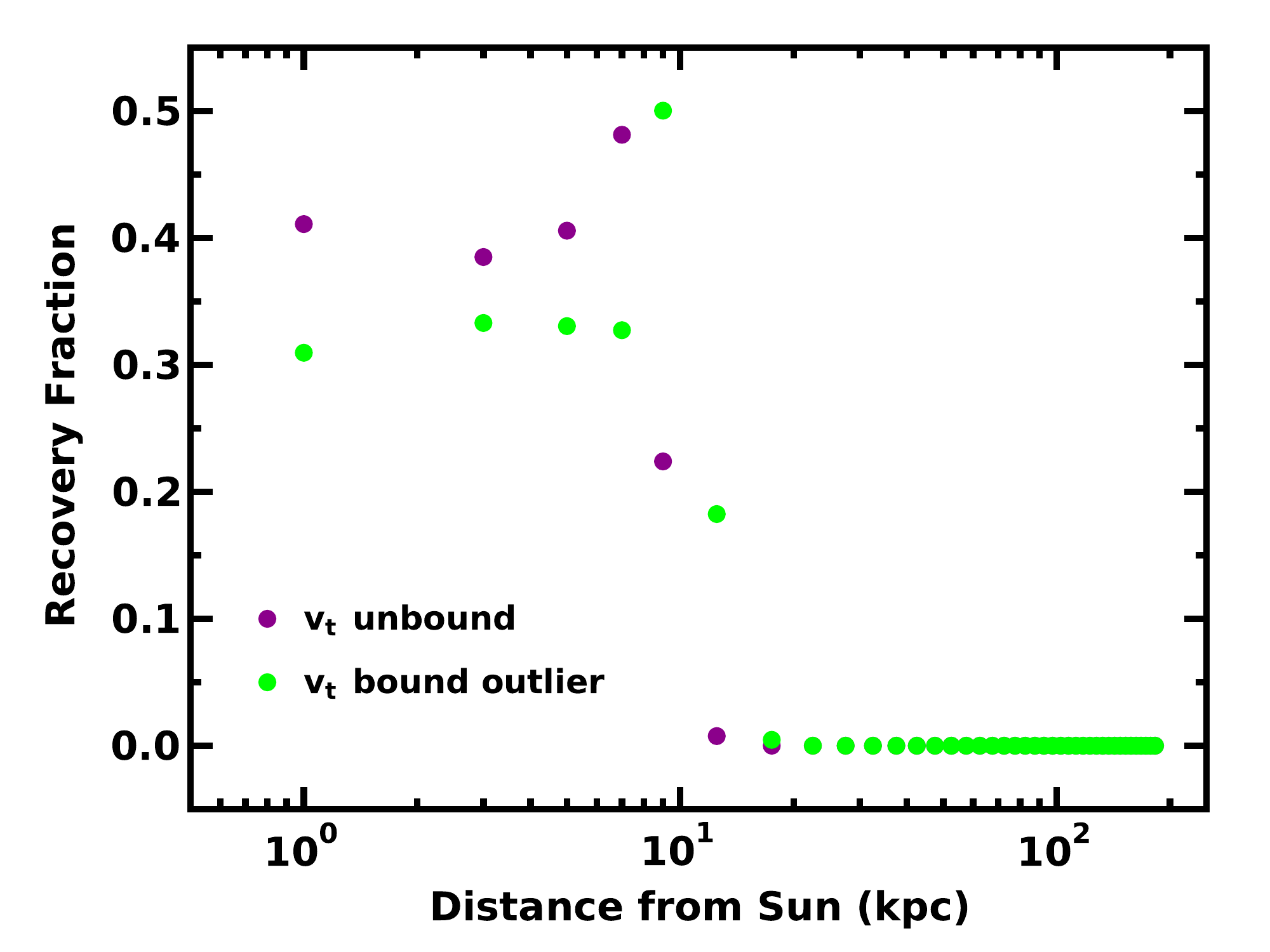}
\vskip 2ex
\caption{\label{fig: nvrvt2}
As in Fig.~\ref{fig: nvrvt1} for unbound stars selected by tangential velocity
($v_t$). Close to the Sun ($d \lesssim$ 20~kpc), $v_t$ recovers roughly 70\%
of unbound stars. Slightly more than half of these have $v_t > v_e$; the rest
have $v_t = $ 0.75--1.00 $v_e$. Beyond 20~kpc, $v_t$ does not discriminate
bound and unbound stars.
}
\end{figure}

The strengths and weaknesses of the tangential velocity are exactly opposite those of $v_r$
(Fig.~\ref{fig: nvrvt2}). Nearby ($d \lesssim$ 10~kpc), most HVSs move at large tangential 
velocities relative to the Sun. Searching for stars with $v_t > v_e$ identifies 40\% to 50\%
of all unbound stars. Adding in stars with $0.75 v_e < v_t < v_e$ selects another 30\% of
the unbound stars with $d \lesssim$ 8~kpc and close to 50\% of unbound stars at $d \approx$
10~kpc. Together, these two criteria identify 70\% to 80\% of all unbound stars. 

Beyond 10~kpc, $v_t$ provides a very poor way to identify HVSs. At these distances, the
tangential component of the motion is simply a small fraction of the total motion. 

These results demonstrate that surveys using $v_t$ for nearby stars with $d \lesssim$
10--20~kpc and $v_r$ for more distant stars with $d \gtrsim$ 20~kpc can recover 
$\gtrsim$ 80\% of unbound HVSs ejected from the GC. Because the Sun lies at a distance
$r_\odot \approx$ 8~kpc from the GC, both techniques work rather well for $d \approx r_\odot$. 

Once unbound stars are identified, it is straightforward to pinpoint their origins from 
the observed $v_r$ and $v_t$. Along lines-of-sight well away from the disk or the LMC, 
nearly all unbound stars have $\gamma \lesssim$ 2\deg--3\deg. For samples of stars with 
modest errors in $v_t$, stars ejected from the GC are easily distinguished from runaways
ejected from the disk and stars ejected from the LMC or some other location within the 
Local Group. Within a few deg of the disk, unbound HVSs on nearly radial orbits 
($\gamma \lesssim$ 5\deg) are hard to distinguish from unbound runaways ejected at small 
angles with respect to the Galactic plane. With high quality $v_t$, however, it should 
be possible to establish the rotational component of motion for any runaway and use this 
measurement to separate runaways from HVSs on more radial trajectories. Toward the LMC, 
gravitational focusing increases the typical $\gamma$ for unbound stars. Still, high 
quality $v_t$ serves to isolate HVS ejected from the GC from other types of high velocity
stars.

\section{DISCUSSION}
\label{sec: disc}

Our calculations clarify the impact of the disk and the LMC on the trajectories of HVSs
ejected from the GC. Aside from decelerating all HVSs, the disk and the LMC dramatically
change the trajectories of HVSs ejected into low Galactic latitudes ($b \lesssim$ 30\deg),
pulling stars across the plane and (sometimes) into the inner halo. At higher latitudes, 
the gravity of the disk and the LMC deflects HVSs from their original radial paths through
the halo. These deviations range from a few tenths of a degree for unbound stars to several 
tens of degrees for bound stars.

The disk and the LMC also generate overdensities of HVSs relative to the initial spatial
distribution ejected from the GC. The disk potential concentrates bound HVSs towards the 
disk. For a reasonable range of MW+LMC potential models, roughly 25\% of HVSs ejected with 
an initial Galactic latitude $b_0 \gtrsim$ 40\deg\ end their lives with $b \lesssim$ 30\deg.
Although fewer unbound stars are dragged to lower $b$, the fraction of unbound stars with
$b \lesssim$ 30\deg\ is still larger than the fraction with $b \gtrsim$ 30\deg. The
smaller fraction of unbound stars in the halo necessitates an upward revision of 20\% to
40\% in rate estimates for HVS ejected from the GC. 

Calculations that include the LMC potential produce a factor of two overdensity in the sky
surface density of stars towards the LMC. Compared to lines-of-sight on opposite sides
of the GC, the overdensity extends for 30\deg--35\deg\ from the center of the LMC, which
corresponds to two LMC scale lengths. For our adopted LMC potential model, the LMC cannot 
capture bound stars ejected from the GC: stars reaching apogalacticon within 5--10~kpc of 
the LMC center simply fall back towards the GC. If the mass of the LMC is larger or more 
concentrated than we assume, capture of bound HVS increases the overdensity relative to 
our calculations.

Although we calculate trajectories only for 3~\msun\ stars, the results are generally
applicable to other stellar masses. Ejections from the GC are fairly independent of
stellar mass \citep{bromley2006,kenyon2008,sari2010}. Although all stars respond to the 
potential in the same way, the relative mix of bound and unbound stars depends on the 
stellar lifetime \citep[e.g.,][]{bromley2006,kenyon2008,kenyon2014}. Ensembles of lower 
(higher) mass stars with longer (shorter) main sequence lifetimes have a larger (smaller) 
number of bound stars relative to the population of unbound stars. Thus, the spatial 
distribution of higher (lower) mass stars is more (less) spherically symmetric. Deviations 
from purely radial paths and the overdensity of stars in the direction of the LMC should 
be somewhat smaller (larger) for more (less) massive stars. Based on several tests of 
HVS trajectories for 1~\msun\ and 6~\msun\ stars, we expect these differences to be 
less than a factor of two.

This analysis complements proposals to infer the shape of the Galactic halo from the
proper motions of HVSs \citep[e.g.,][]{gnedin2005,yu2007}. For HVSs ejected from the GC
at 900~\kms, a triaxial halo deflects trajectories by 0\degpoint2 to 0\degpoint5 at
$r$ = 10--70~kpc \citep{gnedin2005,yu2007}. This deflection is roughly a factor of 
two larger than the typical deflection for unbound HVSs in a MW only potential 
(Fig.~\ref{fig: angle2}, purple points). Although most HVSs ejected with 
$|b_f| \gtrsim$ 70\deg\ in a MW+LMC potential also have small deflection angles, the
LMC deflects those ejected at lower $b$ by many degrees. In a large sample of 
$\sim 10^3$ HVSs at $b \gtrsim$ 60\deg, it should be possible to isolate the 
deflections of the disk, the LMC, and the triaxial halo. At lower $b$, separating 
the different contributions to $\gamma$ will require samples of several thousand
HVSs.

Improved understanding of HVS trajectories throughout the Galaxy provides additional
constraints on techniques to infer \rsun\ and \vsun\ from the space motions of HVSs
\citep[e.g.,][]{hattori2018a}. When HVSs have purely radial trajectories from the GC, 
the observed $v_r$ and $v_t$ enable direct estimates on \rsun\ and \vsun\ that depend
only on errors in the measured distance, radial velocity, and proper motion for each HVSs.
Small deflections from purely radial motion have a modest impact on the derived \rsun\ and
\vsun. The larger $\gamma$'s implied by our simulations complicate this picture. While 
it seems plausible that \rsun\ and \vsun\ can still be inferred from a set of unbound 
HVSs, it is necessary to use data only for the highest velocity stars where $\gamma$ is
relatively small.

Our results also extend recent efforts to quantify the response of the MW to the infall 
of the LMC, the Sgr dwarf, and other less massive galaxies \citep[e.g.,][and references
therein]{garcia2002,bailin2003,weinberg2006,purcell2011,gomez2013,laporte2017,laporte2018}. 
Previous efforts focused on how the gravity of infalling galaxies might shape the dynamical 
structure of low velocity stars within the disk and the halo. Our analysis demonstrates that 
the dynamics of the highest velocity stars in the MW also have an imprint from the 
gravitational fields of nearby galaxies.

Deriving reliable properties of the MW (and Local Group) from HVSs requires samples much
larger than the 20--40 currently available \citep[e.g.,][]{brown2015b,boubert2018,brown2018}. 
Despite recent attempts to identify nearby HVSs in Gaia DR2 \citep{marchetti2018b,hattori2018b},
samples of likely unbound stars remain small. Fortunately, it seems plausible that many more
HVSs can be discovered among A-type to G-type stars in the outer halo \citep[e.g.,][]{koll2007,
kenyon2008,koll2010,kenyon2014,rossi2017,marchetti2018a}. If future surveys reveal these stars,
then HVSs can provide unique insights into the dynamics of the Milky Way system.

\section{SUMMARY}
\label{sec: summ}

Two components of the gravitational potential -- the Galactic disk and the LMC at roughly 
50~kpc from the Sun -- modify the radial trajectories of HVSs ejected from the GC. Close 
to the disk, HVS trajectories bend by as much as 30\deg\ relative to the original path. 
Towards the Galactic poles, deflections are a factor of 2--3 smaller. Bound HVSs suffer 
much larger deflections than unbound stars. Including the LMC in the potential produces 
larger deflections.  

Among large ensembles of HVSs, it is possible to distinguish the purely radial trajectories
predicted for a spherically symmetric potential and the deflected trajectories of HVSs in a
more realistic potential \citep[see also][]{gnedin2005,yu2007}. In principle, variations in 
the bending with Galactic latitude and longitude provide a way to isolate the contributions 
from the disk and the LMC. In practice, detecting these differences requires samples of 
$\gtrsim 10^3$ stars.

Aside from bent trajectories, gravitational focusing generates a factor of two overdensity of 
stars in the direction of the LMC. Although we limit our discussion to predictions for the 
overdensity for models of HVSs, the extra gravity of the LMC should also attract indigenous 
halo stars and runaway stars ejected from the disk. The scale of the overdensity on the sky 
is related to the scale length of the LMC potential.

In any potential model, the Galactic rest-frame radial ($v_r$) and tangential ($v_t$) velocities 
separately provide a robust way to identify unbound HVSs. Close the Sun ($d \lesssim$ 10~kpc), 
the likelihood of finding an HVS moving directly towards or away from us is small. Selecting stars
with $v_t$ larger than 75\% of the local escape speed robustly finds from 70\% to 90\% of unbound
stars. Despite deflections by the disk or the LMC, tangential motions among more distant HVSs 
($d \gtrsim$ 15--20~kpc) are much smaller than the local escape speed. However, requiring that
$v_r$ exceeds 75\% of the local escape speed then recovers more than 80\% of the unbound HVSs.
For the highest velocity stars where $\gamma$ is small, accurate measurements of $v_r$ and $v_t$ 
are capable of isolating stars ejected from the GC.

The success of either $v_r$ or $v_t$ in selecting unbound HVSs points to a two-pronged approach
for identifying the few HVSs likely to be found within a much, much larger sample of indigenous 
halo or disk stars. Nearby, accurate distances and proper motions provided by Gaia (for example)
can yield robust samples of high velocity stars for future study \citep[e.g.,][]{marchetti2018b,
hattori2018b}. At larger distances, radial velocity measurements of stars isolated from the disk
or halo by optical colors (for example) return physically distinct groups of high velocity stars 
\citep[e.g.,][]{brown2006a,brown2007a,brown2009b,brown2014}. If these techniques discover enough
high velocity stars, they provide unique constraints on the Galactic potential despite the presence
of the disk and the LMC.

\vskip 6ex

We thank the referee for a timely and useful report.  Resources supporting this work on the
discover cluster were provided by the NASA High-End Computing (HEC) Program through the NASA 
Center for Climate Simulation (NCCS) at Goddard Space Flight Center.

\bibliography{ms.bbl}

\begin{deluxetable}{lc}
\tablecolumns{2}
\tablewidth{5.5in}
\tablecaption{Summary of selected HVS variables}
\tabletypesize{\normalsize}
\tablehead{
{Description} & {Variable(s)}}
\startdata
Gravitational potential & $\Phi$ \\
Cartesian system centered on GC, disk midplane has $z$ = 0 & $(x, y, z)$ \\
Spherical system centered on GC & $(r, \theta, \phi)$ \\ 
Cylindrical system centered on GC ($\varrho^2 = x^2 + y^2$) & $(\varrho, \theta, z)$ \\ 
GC longitude ($x = \varrho~{\rm cos}~\theta$, $y = \varrho~{\rm sin}~\theta$) & $\theta$ \\
GC latitude ($z = r~{\rm sin}~\phi$) & $\phi$ \\
Velocity in the GC frame & $v$ \\
Heliocentric distance & $d$ \\
Heliocentric Galactic longitude & $l$ \\
Heliocentric Galactic latitude & $b$ \\
Galactic rest frame radial velocity & $v_r$ \\
Galactic rest frame tangential velocity & $v_t$ \\
Heliocentric radial velocity & $v_{r, \odot}$ \\
Heliocentric tangential velocity & $v_{t, \odot}$ \\
Initial GC distance of HVS & $r_0$ \\
Final GC distance of HVS & $r_f$ \\
Initial GC longitude and latitude of HVS & $(\theta_0, \phi_0)$ \\
Final GC longitude and latitude of HVS & $(\theta_f, \phi_f)$ \\
Initial heliocentric longitude and latitude of HVS & $(l_0, b_0)$ \\
Final heliocentric longitude and latitude of HVS & $(l_f, b_f)$ \\
Initial velocity of HVS & $v_0$ \\
Final velocity of HVS & $v_f$ \\
Main sequence lifetime & $t_{ms}$ \\
Time of ejection from GC & $t_{ej}$ \\
Time of observation & $t_{obs}$ \\
\enddata
\label{tab: vars}
\end{deluxetable}
\clearpage

\begin{deluxetable}{lcc}
\tablecolumns{3}
\tablewidth{5.0in}
\tablecaption{Summary of parameters}
\tabletypesize{\normalsize}
\tablehead{
{Parameter} & {~~~~Symbol~~~~} & {~~~~~~Value~~~~~~} }
\startdata
Mass of central MW black hole & $M_{bh}$ & $3.5 \times 10^6$~\msun \\
Mass of MW bulge & $M_b$ & $3.75 \times 10^9$~\msun \\ 
Mass of MW disk & $M_d$ & $6 \times 10^{10}$~\msun \\ 
Mass of MW halo & $M_h$ & $1 \times 10^{12}$~\msun \\ 
Mass of LMC  & $M_L$ & $1 \times 10^{11}$~\msun \\ 
Virial mass for MW & $M_{vir}$ & 1.7 $M_h$ \\
Scale length of MW bulge & $r_b$ & 105 pc \\
Radial scale length of MW disk & $a_d$ & 2750 pc \\
Vertical scale length of MW disk & $b_d$ & 300 pc \\
Scale length of MW halo & $r_b$ & 20 kpc \\
Scale length of LMC halo & $r_L$ & 15 kpc \\
Concentration parameter for MW & $c$ & 12.5 \\
Distance of Sun from GC & $r_\odot$ & 8 kpc \\
Orbital velocity of Sun around GC & $v_\odot$ & 235 \kms\ \\
Distance of LMC from Sun & $d_L$ & 49.66 kpc \\
\enddata
\label{tab: pars}
\end{deluxetable}
\clearpage

\begin{deluxetable}{lccccccc}
\tablecolumns{5}
\tablewidth{5.5in}
\tablecaption{Results for test calculations}
\tabletypesize{\normalsize}
\tablehead{
   &    & \multicolumn{3}{c}{$t(r)$ (Myr)} & \multicolumn{3}{c}{$v(r)$ (\kms) } \\
{Model} & {$r_0$ (kpc)} & {100~kpc} & {~8~kpc~} & {~~1~pc~~} & {100~kpc} & {~8~kpc~} & {~~1~pc~~} }
\startdata
HVz & 250 & 1415 & 1668 & 1680 & 261 & 568 & 918 \\
HVz & 500 & 3786 & 4007 & 4019 & 320 & 598 & 937 \\
HVz & 1000& 9726 & 9932 & 9944 & 354 & 616 & 949 \\
\\
HVx & 250 & 1415 & 1667 & 1679 & 261 & 579 & 918 \\
HVx & 500 & 3785 & 4006 & 4017 & 320 & 608 & 938 \\
HVx & 1000& 9725 & 9931 & 9942 & 354 & 616 & 949 \\
\enddata
\tablecomments{
Within a pure MW potential, particles are released at rest from a distance $r_0$
and fall toward the GC. The columns list the time $t(r)$ to reach a distance $r$ 
and the velocity $v(r)$ at $r$ for models where infall is along the $z$-axis (HVz) 
or the $x$-axis (HVx).
}
\label{tab: tests}
\end{deluxetable}
\clearpage

\begin{deluxetable}{lcccc}
\tablecolumns{5}
\tablewidth{5.0in}
\tablecaption{Predicted relative density of HVSs}
\tabletypesize{\normalsize}
\tablehead{
{Model} &
{~~~~~~d (kpc)~~~~~~} &
{~~~~~~$f_h$~~~~~~} &
{~~~~~~$f_o$~~~~~~} &
{~~~~~~$f_u$~~~~~~} }
\startdata
HVS3a & ~10--20  & 0.919 & 0.058 & 0.023 \\
HVS3a & ~20--40  & 0.852 & 0.089 & 0.058 \\
HVS3a & ~40--80  & 0.626 & 0.164 & 0.209 \\
HVS3a & ~80--160 & 0.112 & 0.147 & 0.742 \\
HVS3b & ~10--20  & 0.920 & 0.057 & 0.023 \\
HVS3b & ~20--40  & 0.853 & 0.089 & 0.058 \\
HVS3b & ~40--80  & 0.634 & 0.162 & 0.204 \\
HVS3b & ~80--160 & 0.125 & 0.151 & 0.724 \\
HVS3c & ~10--20 & 0.919 & 0.057 & 0.023 \\
HVS3c & ~20--40 & 0.853 & 0.089 & 0.058 \\
HVS3c & ~40--80 & 0.635 & 0.162 & 0.203 \\
HVS3c & ~80--160 & 0.126 & 0.151 & 0.723 \\
HVS3d & ~10--20 & 0.920 & 0.057 & 0.023 \\
HVS3d & ~20--40 & 0.852 & 0.089 & 0.059 \\
HVS3d & ~40--80 & 0.634 & 0.162 & 0.204 \\
HVS3d & ~80--160 & 0.125 & 0.152 & 0.722 \\
\enddata
\tablecomments{
The columns list the fraction of stars defined as 
`halo-like' ($f_h$), `bound outliers' ($f_o$), and
`unbound' ($f_u$) as a function of distance $r$ from
the \GC\ for different HVS models with 3~\msun\ ejected stars:
HVS3a: no LMC;
HVS3b: stationary LMC at $z$ = 49.01~kpc;
HVS3c: stationary LMC at $x$ = -0.425~kpc, $y$ = $-41.007$~kpc, and $z = -26.965$~kpc;
HVS3d: moving LMC model summarized in the main text.
}
\label{tab: den-stats}
\end{deluxetable}

\begin{deluxetable}{lcccc}
\tablecolumns{5}
\tablewidth{4.75in}
\tablecaption{Predicted kinematic parameters of HVSs}
\tabletypesize{\footnotesize}
\tablehead{
{Parameter} & {~~~HVS3a~~~} & {~~~HVS3b~~~} & {~~~HVS3c~~~} & {~~~HVS3d~~~} }
\startdata
$x_{avg}$ (pc) & 0.7 & -18 & -22 & -4 \\
$y_{avg}$ (pc) & -0.8 & -4.5 & -819 & -280 \\
$z_{avg}$ (pc) & 0.1 & 855 & -512 & -225 \\
$v_{x,avg}$ (\kms) & -0.05 & -0.1 & -0.1 & -0.3 \\
$v_{y,avg}$ (\kms) & -0.09 & 0.0 & -6.5 & -8.8 \\
$v_{z,avg}$ (\kms) & 0.00  & 6.4 & -4.1 & -5.5 \\
$\sigma(v_x)$ (\kms) & 175 & 173 & 173 & 173 \\
$\sigma(v_y)$ (\kms) & 175 & 173 & 175 & 175 \\
$\sigma(v_z)$ (\kms) & 147 & 149 & 147 & 147 \\
$l_{x,avg}$ (pc~\kms) & 54 & -266 & 23290 & 22300 \\
$l_{y,avg}$ (pc~\kms) & 41 & 54 & -184 & -452 \\
$l_{z,avg}$ (pc~\kms) & 0.0006 & 0.005 & -17.4 & -153 \\
$|l_x|_{avg}$ (pc~\kms) & 153840  & 223520 & 262540 & 259770 \\
$|l_y|_{avg}$ (pc~\kms) & 153830  & 223520 & 185510 & 185960 \\
$|l_z|_{avg}$ (pc~\kms) & 5.3 & 5.2 & 163080 & 158430 \\
//
${x_{med}}$ (pc) & 2.1 & -4.0 & -5.2 & -14\\
${y_{med}}$ (pc) & -4.6 & -0.4 & -285 & -800 \\
${z_{med}}$ (pc) & -0.4 & 360 & -226 & -500 \\
${v_{x,med}}$ (\kms) & -0.02 & 0.0 & -0.2 & -0.2 \\
${v_{y,med}}$ (\kms) & -0.01 & 0.0 & -9.0 & -6.4 \\
${v_{z,med}}$ (\kms) & 0.00 & 7.8 & -5.5 & -4.2 \\
${l_{x,med}}$ (pc~\kms) & 0.3 & -0.02 & -234 & -2575 \\
${l_{y,med}}$ (pc~\kms) & -0.05 & -0.33 & 365 & 1014 \\
${l_{z,med}}$ (pc~\kms) & 0.00 & 0.00 & 0.02 & -0.2 \\
${|l_x|_{med}}$ (pc~\kms) &  122730 & 134500 & 156550 & 155940  \\
${|l_y|_{med}}$ (pc~\kms) & 122820 & 134510 & 124990 & 124840 \\
${|l_z|_{med}}$ (pc~\kms) & 3.2 & 3.1 & 67283 & 65405 \\
\enddata
\tablecomments{Results for various observables of the full ensemble of
$10^7$ ejected stars in four models of 3~\msun\ HVSs:
HVS3a: no LMC;
HVS3b: stationary LMC at $z$ = 49.01~kpc;
HVS3c: stationary LMC at $x$ = -0.425~kpc, $y$ = $-41.007$~kpc, and $z = -26.965$~kpc;
HVS3d: moving LMC model summarized in the main text.
}
\label{tab: kin-stats}
\end{deluxetable}

\clearpage

\clearpage

\clearpage

\clearpage

\clearpage

\clearpage

\clearpage

\clearpage

\clearpage

\clearpage

\clearpage

\clearpage

\clearpage

\clearpage

\clearpage

\clearpage

\clearpage

\clearpage

\clearpage

\clearpage

\clearpage

\end{document}